\documentclass[final,3p]{elsarticle}
\makeatletter 
\def\ps@pprintTitle{%
 \def\@oddfoot{\footnotesize\itshape
       Published in \ifx\@journal\@empty Elsevier
       \else\@journal\fi\hfill February 3, 2019}}
\makeatother

\graphicspath{{./graphics/}}

\usepackage{amsmath,amsfonts,amssymb} 
\usepackage{mathtools} 
\usepackage{upgreek} 
\usepackage{easybmat} 
\usepackage{fixmath} 
\usepackage[font=footnotesize,list=true]{subcaption} 
\usepackage{booktabs} 
\usepackage[exponent-product=\cdot]{siunitx} 
\renewcommand{\ang}[1]{{{#1}^\circ}}
\usepackage{bm} 
\usepackage{xcolor}
\usepackage{microtype}
\usepackage{lmodern}

\newcommand{\graphicsFolder}{graphics}

\renewcommand{\vec}[1]{\mathbold #1} 
\newcommand{\besselj}{\mathrm{j}} 	
\newcommand{\bessely}{\mathrm{y}} 	
\newcommand{\besselJ}{\mathrm{J}} 	
\newcommand{\besselY}{\mathrm{Y}}  	
\newcommand{\legendre}{\mathrm{P}} 	
\newcommand{\hankel}{\mathrm{h}}   	
\newcommand{\Diff}{\upDelta} 

\newcommand{\diff}{\mathrm{d}} 
\newcommand{\idiff}{\, \mathrm{d}} 
\newcommand{\zerovec}{\mathbf{0}}
\newcommand{\transpose}{\top}

\newcommand{\R}{\mathbb{R}}
\providecommand{\C}{\mathbb{C}}
\newcommand{\Z}{\mathbb{Z}}
\newcommand{\N}{\mathbb{N}}
\newcommand{\PI}{\uppi}
\newcommand{\euler}{\mathrm{e}}
\newcommand{\imag}{\mathrm{i}}
\newcommand{\bigoh}{\mathcal{O}}
\providecommand*{\pderiv}[3][]{\frac{\partial^{#1}#2}{\partial #3^{#1}}}
\providecommand*{\deriv}[3][]{\frac{\diff^{#1}#2}{\diff #3^{#1}}}

\renewcommand{\leq}{\leqslant}

\DeclareMathSymbol{\GAMMA}{\mathalpha}{operators}{0}
\DeclareMathOperator{\sinc}{sinc}

\DeclareMathOperator{\fourier}{\mathcal{F}}
\DeclareMathOperator{\TS}{TS}

\renewcommand\Re{\operatorname{Re}}

\renewcommand{\Xi}{{\vec{t}_1}}

\newcommand{\energyNorm}[2]{%
  {\left\vert\kern-0.25ex\left\vert\kern-0.25ex\left\vert #1 
    \right\vert\kern-0.25ex\right\vert\kern-0.25ex\right\vert}_{#2}
}

\let\originalleft\left
\let\originalright\right
\renewcommand{\left}{\mathopen{}\mathclose\bgroup\originalleft}
\renewcommand{\right}{\aftergroup\egroup\originalright}

\usepackage{xspace}

\usepackage{hyperref} 
\hypersetup{hidelinks,colorlinks=true,citecolor=blue,linkcolor=blue,urlcolor=blue} 
\usepackage[noabbrev]{cleveref} 
\usepackage{longtable}
\usepackage{enumitem}

\journal{Journal of Sound and Vibration}

\begin{document}
\begin{frontmatter}

\title{Exact 3D scattering solutions for spherical symmetric scatterers}

\author[venas]{Jon Vegard Ven{\aa}s\texorpdfstring{\corref{cor}}{}}
\ead{Jon.Venas@ntnu.no}

\author[jenserud]{Trond Jenserud}
\ead{Trond.Jenserud@ffi.no}

\address[venas]{Department of Mathematical Sciences, Norwegian University of Science and Technology, \\Alfred Getz' vei 1, 7034 Trondheim, Norway}
\address[jenserud]{Department of Marine Systems, Norwegian Defence Research Establishment, \\Postboks 115, 3191 Horten, Norway}
\cortext[cor]{Corresponding author.}

\begin{abstract}
In this paper, exact solutions to the problem of acoustic scattering by elastic spherical symmetric scatterers are developed. The scatterer may consist of an arbitrary number of fluid and solid layers, and scattering with single Neumann conditions (replacing Neumann-to-Neumann conditions) is added. The solution is obtained by separation of variables, resulting in an infinite series which must be truncated for numerical evaluation. The implemented numerical solution is exact in the sense that numerical error is solely due to round-off errors, which will be shown using the symbolic toolbox in MATLAB. A system of benchmark problems is proposed for future reference. Numerical examples are presented, including comparisons with reference solutions, far-field patterns and near-field plots of the benchmark problems, and time-dependent solutions obtained by Fourier transformation.
\end{abstract}


\end{frontmatter}
\section{Introduction}
Acoustic scattering by elastic objects is a continuing area of study. Most phenomena in the scattering process can be adequately described by linear elasticity theory, and by further restricting the analysis to homogeneous, isotropic bodies of simple geometries, the mathematical formalism becomes simple enough to be handled by conventional analytic methods. 

The problems fall into mainly three categories: scattering of acoustic waves from elastic objects, scattering of elastic waves from fluid-filled cavities and solid inclusions, and inverse scattering, i.e., obtaining properties of a scattering object from the remotely sensed field. In the first category, the classical problems include scattering by spheres and infinite cylinders: fluid spheres ~\cite{Anderson1950ssf}, solid spheres and cylinders~\cite{Faran1951ssb, Anderson1955soa, Hickling1962aoe, Doolittle1968ssb, Flax1978toe, Gaunaurd1983rao}, and spherical and cylindrical shells with various combinations of material properties~\cite{Hickling1964aoe, Doolittle1966ssb, Gaunaurd1987lac, Gaunaurd1991ssb, Kaduchak1998rbm, Chang1994voa, Chang1994soa, Fender1972sfa}. 
Much of the work in this field up to around 1980, is summarized in Flax et al.~\cite{Flax1981pa}.

The surrounding medium is usually considered to be a lossless fluid, but viscous fluids~\cite{Lin1983asb} and viscoelastic media and materials~\cite{Hasheminejad2005asf} are also considered. 

The acoustic illumination is often taken to be a plane wave which is relevant for far-field sources, otherwise point sources are applied in the near-field. For the infinite cylinder, the incident field is in most cases applied normal to the cylinder, but obliquely incident fields are also considered~\cite{Bao1990ras, Daneshjou2017aes}. More recently, the problem of scattering of beams has received much attention~\cite{Marston2007abs, Gong2016aso}. 

Solutions to some non-symmetric problems are also given; e.g. partially fluid filled spheres~\cite{Fawcett2001sfa}, spheres with eccentric cavities~\cite{Hasheminejad2005asf}, and open spheres with internal point sources~\cite{Elias1991sba}.

The studies mentioned above consider a single object in the free field. It is also of interest to study interactions between objects, and between an object and a boundary. The problem of multiple scattering is studied in e.g.~\cite{Gabrielli2001asb} for two elastic spheres, and in~\cite{Wu2006mso} for many fluid spheres, while the scattering by objects close to boundaries, and by  partially buried objects is adressed in~\cite{Zampolli2009bpf}.


Applications of the theory are numerous, and include scattering from marine life~\cite{Anderson1950ssf, Stanton1998dbs, Stanton2000asb}, various aspects of sonar, nondestructive testing, seismology, detection of buried objects~\cite{Sessarego1998sba}, medical imaging~\cite{Wells2006ui}, determination of material properties by inverse scattering~\cite{Ayres1987ias}, and acoustic cloaking. Acoustic cloaking, i.e., making an object acoustically 'invisible', requires acoustic metamaterials and is difficult to realize in practice, but reducing the backcattering strength of an object is an important issue, and can be realized either passively by coating or actively as suggested in e.g. \cite{Avital2015ssa}. 
A recent area of research is noise control in aerospace- and automotive engineering, where sound transmission through cylindrical shells constructed from new composite materials \cite{Talebitooti2016att} and functionally graded materials \cite{Daneshjou2017aes} are studied in order to reduce noise level inside the cabin. The latter problem requires a full 3D solution. 


The method referred to as classical scattering theory starts with the linearized elasto-dynamic equation of motion (also called Naviers equation). For the intended applications, nonlinear effects are negligible, which justifies the use of the linear approximation. For a certain class of coordinate systems, the field can be expressed in terms of three scalar potentials, which satisfy scalar Helmholtz equations, and admit solutions in the form of infinite series, termed normal modes or partial waves. The formal series expansions contain all the physical features of the solution, i.e., the reflected, transmitted and circumferential (or creeping) waves. The most general problems on finite scatterers in free space are scattering by the spherical shells which requires all three potentials and give solutions in terms of double sums. However, assuming axisymmetric illumination there is no loss of generality in aligning the coordinate axis of the sphere with the axis of the incident field, resulting in an axisymmetric problem. This results in a single infinite series which is much more computational efficient than the general case. This is the approach taken here.
 
As the solution is in the form of an infinite series, it needs to be truncated at some point. The summation is terminated when the relative magnitude of the last term is less than some prescribed tolerance, such that no computational parameters are introduced if this tolerance is chosen to be the precision used in the calculations (typically double precision). It is shown, by using symbolic precision in MATLAB, that the computational errors in the implementation are due to round-off errors. This is a natural definition of a computational exact solution.

The work reviewed above solves a host of different problems, and several reference solutions are available, with complexity up to three layers. What the present work provides is the explicit solution for a fully general multilayered sphere, and with corresponding analysis of the computational residual errors. This allows easy design and modeling of reference solutions for the purpose of validating numerical methods.
More specific, the model solves the problem of scattering by an incident plane wave, or wave from a point source, by spherical objects consisting of an arbitrary number of layers. Any combinations of fluid and solid layers can be handled, and the special cases of replacing the Neumann-to-Neumann condition by a single Neumann condition is also included.



An early work on scattering from multilayered spheres and infinite cylinders is Jenserud and Tollefsen~\cite{Jenserud1990ars}. The method employed here is referred to as the global matrix method~\cite{Schmidt1985afw}, and is a systematic way of assembling local solutions for the individual layers into a global matrix for the total problem. 
The present work uses the same approach, and builds mainly upon the work of Chang and Demkowicz~\cite{Chang1994voa}, which is generalized to multilayered spherical objects.

\begin{figure}
	\centering
	\includegraphics[scale=1]{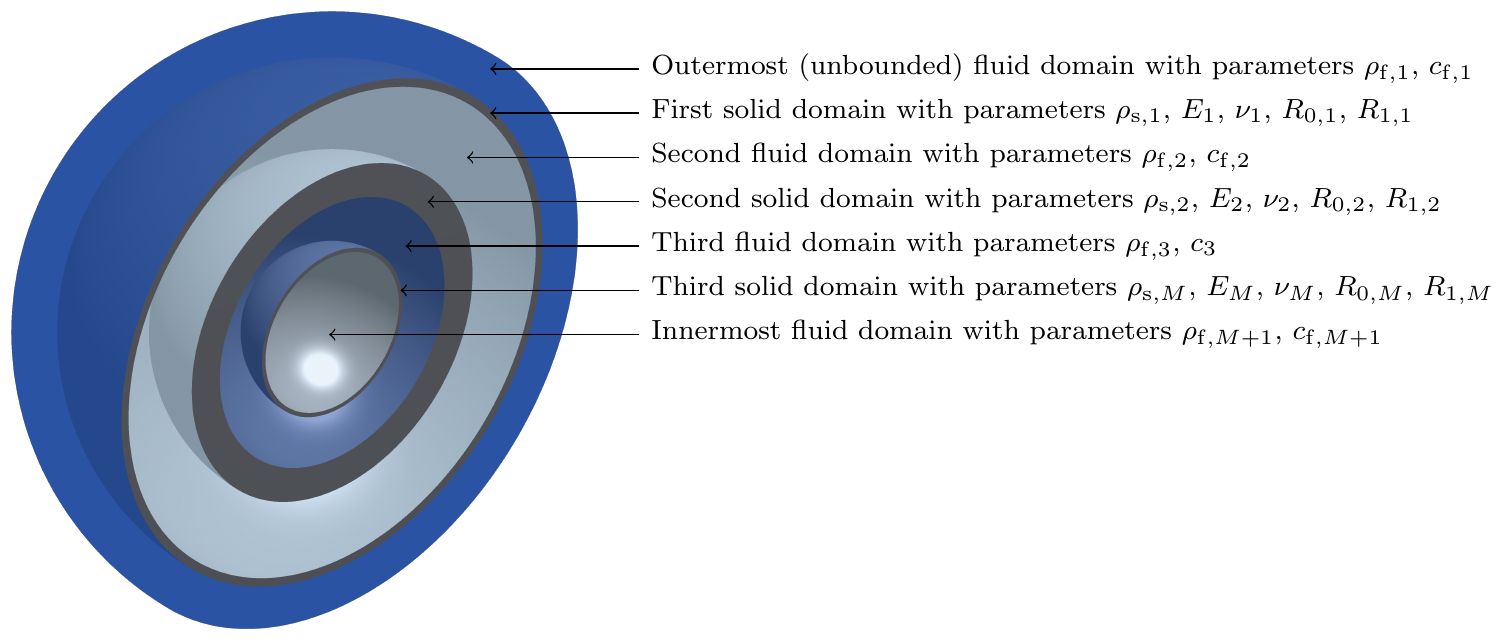}
	\caption{A model with $M=3$ steel shells with different thicknesses (clip view), illustrating the distribution of the physical parameters over the different domains.}
	\label{Fig1:illustration}
\end{figure}


\section{Governing equations}
\label{Sec1:govEquations}
In this section the governing equations for the problem at hand will be presented. In~\cite[pp. 13-14]{Ihlenburg1998fea} Ihlenburg briefly derives the governing equations for the acoustic-structure interaction problem. As the physical problem of interest is a time dependent problem, it is natural to first present the governing equations in the time-domain before presenting the corresponding equations in the frequency domain (obtained by Fourier transformation). It is noted right away that the fields described in this paper (both in the time-domain and frequency-domain) are all perturbation fields.

\subsection{Governing equations in the time domain}
Einstein's summation convention will be used throughout this work, such that repeated indices in products imply summation. For example, any vector $\vec{x}\in\R^3$ can be expressed as
\begin{equation}
	\vec{x}=\begin{bmatrix}
	x_1\\
	x_2\\
	x_3
\end{bmatrix} = \sum_{i=1}^3 x_i\vec{e}_i = x_i\vec{e}_i,
\end{equation}
where $\vec{e}_i\in\R^3$ is the standard basis vectors in a three dimensional Euclidean space.

Let $\breve{\vec{u}} = \breve{u}_i\vec{e}_i$ be the time-dependent displacement field in a given solid domain, and $\breve{\vec{\sigma}}$ the corresponding stress tensor (see \Cref{Sec1:LinearElasticity} for details). Each of the components depend on the spatial variable $\vec{x}$ and the time variable $t$, such that $\breve{\vec{u}} = \breve{\vec{u}}(\vec{x},t)$. The solid domain is then governed by Navier's equation of motion~\cite{Fender1972sfa} (derived from Newton's second law) 
\begin{equation}\label{Eq1:navierTime}
	G\nabla^2\breve{\vec{u}}+\left(K+\frac{G}{3}\right)\nabla(\nabla \cdot\breve{\vec{u}}) = \rho_{\mathrm{s}}\pderiv[2]{\breve{\vec{u}}}{t},
\end{equation}
which is equivalent to~\cite[p. 223]{Slaughter2002tlt}
\begin{equation}
	\pderiv{\breve{\sigma}_{ij}}{x_j} = \rho_{\mathrm{s}}\pderiv[2]{\breve{u}_i}{t},\quad i=1,2,3.
\end{equation}
The \textit{bulk modulus}, $K$, and the \textit{shear modulus}, $G$, can be defined by the Young's modulus, $E$, and Poisson's ratio, $\nu$, as
\begin{equation}
	K = \frac{E}{3(1-2\nu)}\quad\text{and}\quad G = \frac{E}{2(1+\nu)}.
\end{equation}

Correspondingly, denote by $\breve{p}$ the time-dependent scattered pressure field in a given fluid domain, which is governed by the wave equation
\begin{equation}\label{Eq1:waveEquation}
	\nabla^2 \breve{p} = \frac{1}{c_{\mathrm{f}}^2} \pderiv[2]{\breve{p}}{t}.
\end{equation}

\subsection{Governing equations in the frequency domain}
The dimension of the governing equations may be reduced by one using a frequency-time Fourier\footnote{The sign convention in the Fourier transform differs from the classical Fourier transform~\cite{ISO2009qau}, but agrees with most literature on the subject, for example~\cite{Fender1972sfa,Ihlenburg1998fea,Jensen2011coa,Goodman1962rat}.} pair~\cite[p. 71]{Jensen2011coa}\begin{align}
	\Psi(\vec{x},\omega) = \left(\fourier\breve{\Psi}(\vec{x},\cdot)\right)(\omega) &= \int_{-\infty}^\infty \breve{\Psi}(\vec{x},t)\euler^{\imag \omega t}\idiff t\label{Eq1:Psi}\\
	\breve{\Psi}(\vec{x},t) = \left(\fourier^{-1}\Psi(\vec{x},\cdot)\right)(t)  &= \frac{1}{2\PI}\int_{-\infty}^\infty \Psi(\vec{x},\omega)\euler^{-\imag \omega t}\idiff \omega\label{Eq1:Psit}
\end{align}
where $\Psi$ represents the scattered pressure field $p$ or the displacement field $\vec{u}$. The frequency $f$ and the angular frequency $\omega$ is related by $\omega = 2\PI f$, and the angular wave number is given by $k=\omega/c_{\mathrm{f}}$.

Consider first the scattered pressure. By differentiating \Cref{Eq1:Psit} twice with respect to time, such that
\begin{equation}
	\pderiv[2]{}{t}\breve{p}(\vec{x},t) = -\omega^2\breve{p}(\vec{x},t),
\end{equation}
the following is obtained (using \Cref{Eq1:waveEquation})
\begin{align*}
	\nabla^2 p(\vec{x},\omega) + k^2 p(\vec{x},\omega) &= \int_{-\infty}^\infty \nabla^2\breve{p}(\vec{x},t)\euler^{\imag \omega t}\idiff t + \int_{-\infty}^\infty k^2 \breve{p}(\vec{x},t)\euler^{\imag \omega t}\idiff t\\
	&= \int_{-\infty}^\infty \left[\nabla^2\breve{p}(\vec{x},t)-\frac{1}{c_{\mathrm{f}}^2}\pderiv[2]{}{t}\breve{p}(\vec{x},t)\right]\euler^{\imag \omega t}\idiff t = 0.
\end{align*}
That is, $p(\vec{x},\omega)$ satisfies the Helmholtz equation
\begin{equation}\label{Eq1:helmholtz}
	\nabla^2 p + k^2p = 0.
\end{equation}
A corresponding argument shows that the displacement field $\vec{u}(\vec{x},\omega)$ satisfies
\begin{equation}\label{Eq1:navier}
	G\nabla^2\vec{u}+\left(K+\frac{G}{3}\right)\nabla(\nabla \cdot\vec{u}) +\rho_{\mathrm{s}}\omega^2\vec{u} = \zerovec.
\end{equation}
The scattered pressure, $p$, must in addition to the Helmholtz equation satisfy the Sommerfeld radiation condition for the outermost fluid layer~\cite{Sommerfeld1949pde}
\begin{equation}\label{Eq1:Sommerfeld}
	\pderiv{p(\vec{x},\omega)}{r}-\imag k p(\vec{x},\omega) = o\left(r^{-1}\right)\quad r=|\vec{x}|
\end{equation}	
as $r\to\infty$ uniformly in $\hat{\vec{x}}=\frac{\vec{x}}{r}$.

The coupling conditions (Neumann-to-Neumann) between the solid and the fluid boundaries are given by~\cite[pp. 13-14]{Ihlenburg1998fea}
\begin{align}
	\rho_{\mathrm{f}} \omega^2 u_i n_i - \pderiv{p_{\mathrm{tot}}}{n} &= 0\\
	\sigma_{ij}n_i n_j + p_{\mathrm{tot}} &= 0
\end{align}
where $\vec{n}$ is the normal vector at the surface, and $\breve{p}_{\mathrm{tot}}$ is the total pressure\footnote{Since only perturbation fields are considered, $p_{\mathrm{tot}}$ does not include the static background pressure (and does therefore not represent the physical total pressure field).} (scattered pressure with the incident pressure field added for the outermost fluid). In addition, since the fluid is assumed to be ideal, there is no tangential traction at the surfaces. For spherical symmetric objects $\vec{n}=\vec{e}_{\mathrm{r}}$, such that the coupling equations reduces to
\begin{align}
	\rho_{\mathrm{f}} \omega^2 u_{\mathrm{r}} - \pderiv{p_{\mathrm{tot}}}{r} &= 0\label{Eq1:firstBC}\\
	\sigma_{\mathrm{rr}} + p_{\mathrm{tot}} &= 0\label{Eq1:secondBC}
\end{align}
in the spherical coordinate system (see \Cref{Sec1:sphericalCoordinates}). The tangential traction free boundary conditions becomes~\cite[p. 15]{Chang1994voa}
\begin{align}
	\sigma_{\mathrm{r}\upvartheta} &= 0\label{Eq1:traction1}\\
	\sigma_{\mathrm{r}\upvarphi} &= 0\label{Eq1:traction2}.
\end{align}
\section{General solution in the solid domain}
\label{Eq1:generalSolution}
It turns out that Navier's equation can be reduced to a set of Helmholtz equations. Since the fluid domain also is governed by the Helmholtz equation, both solid and fluid domains share the same fundamental solutions, and it thus suffices to present the general solution in the solid domain.

\subsection{Lam\'{e} solution}
Fender~\cite{Fender1972sfa} shows that the solution of \Cref{Eq1:navier} can be written in terms of a scalar potential $\phi$ and a vector potential $\vec{\psi}$ as follows
\begin{equation}\label{Eq1:LameSolution}
	\vec{u} = \nabla\phi + \nabla\times\vec{\psi}.
\end{equation}
Such a solution of Navier's equation is called a Lam\'{e} solution. The potentials $\phi$ and $\vec{\psi}$ satisfy the scalar and vector Helmholtz equation, respectively. That is,
\begin{align}
	&\nabla^2\phi + a^2\phi = 0\label{Eq1:phiHelmholtz}\\
	&\nabla^2\vec{\psi} + b^2\vec{\psi} = \zerovec\label{Eq1:PsiHelmholtz}
\end{align}
where
\begin{equation}\label{Eq1:waveNumber_a_b}
	a=\frac{\omega}{c_{\mathrm{s},1}},\quad b=\frac{\omega}{c_{\mathrm{s},2}},\quad c_{\mathrm{s},1} = \sqrt{\frac{3K+4G}{3\rho_{\mathrm{s}}}},\quad c_{\mathrm{s},2} = \sqrt{\frac{G}{\rho_{\mathrm{s}}}}.
\end{equation}
Here, the parameters $c_{\mathrm{s},1}$ and $c_{\mathrm{s},2}$ are the longitudinal and transverse (elastic) wave velocities, respectively, and $a$ and $b$ are the corresponding angular wave numbers in the solid. 

Throughout this work, axisymmetry around the $x_3$-axis is assumed. Assuming symmetry around this particular axis causes no loss of generality, as both the incident wave and the spherical shell share this symmetry property (a simple orthogonal transformation restores the generality of axisymmetry about an arbitrary axis). In the spherical coordinate system, the pressure $p$ and the displacement $\vec{u}$ are then independent of the azimuth angle $\varphi$ in the fluid and solid domains, respectively. Moreover, the solid component in the azimuth angle direction is zero, $u_\upvarphi=0$. This is a result of the axisymmetry of the problem.

\subsection{Series representation using separation of variables}
Using these assumptions Fender~\cite{Fender1972sfa} shows that $\vec{\psi}=\psi_\upvarphi\vec{e}_\upvarphi$, such that when \Cref{Eq1:phiHelmholtz,Eq1:PsiHelmholtz} are expanded in terms of spherical coordinates, the following is obtained (using \Cref{Eq1:laplaceScalarSpherical,Eq1:LapVecPotSpherical})
\begin{align}
	\pderiv{}{r}\left(r^2\pderiv{\phi}{r}\right) + \frac{1}{\sin\vartheta}\pderiv{}{\vartheta}\left(\sin\vartheta\pderiv{\phi}{\vartheta}\right) + (ar)^2\phi &= 0\\
	\pderiv{}{r}\left(r^2\pderiv{\psi_\upvarphi}{r}\right) + \frac{1}{\sin\vartheta}\pderiv{}{\vartheta}\left(\sin\vartheta\pderiv{\psi_\upvarphi}{\vartheta}\right) + \left[(br)^2-\frac{1}{\sin^2\vartheta}\right]\psi_\upvarphi &= 0.
\end{align}
Using separation of variables, each of these equations can be reduced to a couple of spherical Bessel and Legendre equations, with the associate Legendre polynomials of zero and first order (described in \Cref{subsec:legendre}) and spherical Bessel functions (described in \Cref{subsec:sphericalBesselAndHankel}) as solutions. More explicitly,
\begin{align}
	\phi(r,\vartheta) &= \sum_{n=0}^\infty \legendre_n(\cos\vartheta)\left[A_n^{(1)}\besselj_n(ar)+A_n^{(2)}\bessely_n(ar)\right]\label{Eq1:phiSolution}\\
	\psi_\upvarphi(r,\vartheta) &= \sum_{n=0}^\infty \legendre_n^1(\cos\vartheta)\left[B_n^{(1)}\besselj_n(br)+B_n^{(2)}\bessely_n(br)\right]\label{Eq1:psiSolution}
\end{align}
where the coefficients $A_n^{(i)},B_n^{(i)}\in\C$, $i=1,2$, are chosen such that the boundary conditions are satisfied.

By using \Cref{Eq1:LegendreRelation1} these functions and their partial derivatives will have their $\vartheta$-dependency contained in functions of the form (the ones relevant for this work are listed in \Cref{Eq1:Qs})
\begin{equation}
	Q_n^{(j)}(\vartheta) = \deriv[j]{}{\vartheta}\legendre_n(\cos\vartheta).
\end{equation}
That is, there is no need for the associated Legendre polynomials.

For ease of notation, the function $Z_n^{(i)}(\zeta)$, $i=1,2$, is introduced (as in~\cite{Chang1994voa, Chang1994soa}), where 
\begin{equation}
	Z_n^{(1)}(\zeta) = \besselj_n(\zeta),\quad Z_n^{(2)}(\zeta) = \bessely_n(\zeta).
\end{equation}
Moreover, the notation $\xi = \xi(r) = ar$ and $\eta = \eta(r) = br$ is used for convenience. Using the Einstein summation convention, \Cref{Eq1:phiSolution,Eq1:psiSolution} may now be rewritten as
\begin{align}
	\phi(r,\vartheta) &= \sum_{n=0}^\infty Q_n^{(0)}(\vartheta)A_n^{(i)}Z_n^{(i)}(\xi)\label{Eq1:phiSolutionSimplified}\\
	\psi_\upvarphi(r,\vartheta) &= \sum_{n=0}^\infty Q_n^{(1)}(\vartheta)B_n^{(i)}Z_n^{(i)}(\eta).\label{Eq1:psiSolutionSimplified}
\end{align}

\subsection{Expressions for the displacement and stress field}
By expanding \Cref{Eq1:LameSolution} in spherical coordinates (using \Cref{Eq1:delScalarSpherical,Eq1:CrossVecPotSpherical}) yields
\begin{equation}
	\vec{u} = \nabla\phi + \nabla\times\vec{\psi} = \pderiv{\phi}{r}\vec{e}_{\mathrm{r}} + \frac{1}{r}\pderiv{\phi}{\vartheta}\vec{e}_{\upvartheta} + \frac{1}{r\sin\vartheta}\pderiv{}{\vartheta}\left(\psi_\upvarphi\sin\vartheta\right)\vec{e}_{\mathrm{r}} - \frac{1}{r}\pderiv{}{r}\left(r\psi_\upvarphi\right)\vec{e}_{\upvartheta} 
\end{equation}
such that
\begin{equation}
	u_{\mathrm{r}} = \pderiv{\phi}{r} + \frac{1}{r}\pderiv{\psi_\upvarphi}{\vartheta} + \frac{1}{r}\psi_\upvarphi\cot\vartheta
\end{equation}
and
\begin{equation}
	u_\upvartheta = \frac{1}{r}\pderiv{\phi}{\vartheta} -\pderiv{\psi_\upvarphi}{r} -\frac{1}{r}\psi_\upvarphi.
\end{equation}
Insertion of \Cref{Eq1:phiSolutionSimplified,Eq1:psiSolutionSimplified} (using \Cref{Eq1:expandedLegendreEquationIdentity,Eq1:LegendreRelation1,Eq1:BesselDerivIdentity2})
yields
\begin{equation}\label{Eq1:u_rgen}
	u_{\mathrm{r}} = \frac{1}{r}\sum_{n=0}^\infty Q_n^{(0)}(\vartheta)\left[A_n^{(i)}S_{1,n}^{(i)}(\xi)+B_n^{(i)}T_{1,n}^{(i)}(\eta)\right]
\end{equation}
and
\begin{equation}\label{Eq1:u_tgen}
	u_\upvartheta = \frac{1}{r}\sum_{n=0}^\infty Q_n^{(1)}(\vartheta)\left[A_n^{(i)}S_{2,n}^{(i)}(\xi)+B_n^{(i)}T_{2,n}^{(i)}(\eta)\right]
\end{equation}
where
\begin{align*}
	S_{1,n}^{(i)}(\xi) &= \xi \deriv{}{\xi}Z_n^{(i)}(\xi) =  nZ_n^{(i)}(\xi)-\xi Z_{n+1}^{(i)}(\xi)\\ 
	T_{1,n}^{(i)}(\eta) &= -n(n+1)Z_n^{(i)}(\eta)\\
	S_{2,n}^{(i)}(\xi) &= Z_n^{(i)}(\xi)\\
	T_{2,n}^{(i)}(\eta) &= -Z_n^{(i)}(\eta)-\eta \deriv{}{\eta}Z_n^{(i)}(\eta) = -(n+1)Z_n^{(i)}(\eta) + \eta Z_{n+1}^{(i)}(\eta).
\end{align*}
To compute the stresses defined in \Cref{Sec1:sphericalCoordinates}, the partial derivatives of the displacement field in the spherical coordinate system are needed. These derivatives are found to be (using \Cref{Eq1:expandedLegendreEquationIdentity2,Eq1:BesselDerivIdentity1,Eq1:BesselDerivIdentity2})
\begin{align}
	\pderiv{u_{\mathrm{r}}}{r} &= \frac{1}{r^2}\sum_{n=0}^\infty Q_n^{(0)}(\vartheta)\left[A_n^{(i)}S_{3,n}^{(i)}(\xi)+B_n^{(i)}T_{3,n}^{(i)}(\eta)\right]\\
	\pderiv{u_{\upvartheta}}{r} &= \frac{1}{r^2}\sum_{n=0}^\infty Q_n^{(1)}(\vartheta)\left[A_n^{(i)}S_{4,n}^{(i)}(\xi)+B_n^{(i)}T_{4,n}^{(i)}(\eta)\right]\\
	\pderiv{u_{\mathrm{r}}}{\vartheta} &= \frac{1}{r}\sum_{n=0}^\infty Q_n^{(1)}(\vartheta)\left[A_n^{(i)}S_{1,n}^{(i)}(\xi)+B_n^{(i)}T_{1,n}^{(i)}(\eta)\right]\\
	\pderiv{u_\upvartheta}{\vartheta} &= \frac{1}{r}\sum_{n=0}^\infty Q_n^{(2)}(\vartheta)\left[A_n^{(i)}S_{2,n}^{(i)}(\xi)+B_n^{(i)}T_{2,n}^{(i)}(\eta)\right]
\end{align}
where
\begin{align*}
	S_{3,n}^{(i)}(\xi) &= \xi \deriv{}{\xi}S_{1,n}^{(i)}(\xi) - S_{1,n}^{(i)}(\xi) =  (n^2-\xi^2-n)Z_n^{(i)}(\xi) + 2\xi Z_{n+1}^{(i)}(\xi)\\ 
	T_{3,n}^{(i)}(\eta) &= \eta \deriv{}{\eta}T_{1,n}^{(i)}(\eta) -T_{1,n}^{(i)}(\eta) = -n(n+1)\left[(n-1)Z_n^{(i)}(\eta) - \eta Z_{n+1}^{(i)}(\eta)\right]\\
	S_{4,n}^{(i)}(\xi) &= \xi \deriv{}{\xi}Z_n^{(i)}(\xi) - Z_n^{(i)}(\xi) = (n-1)Z_n^{(i)}(\xi)-\xi Z_{n+1}^{(i)}(\xi)\\ 
	T_{4,n}^{(i)}(\eta) &= \eta\deriv{}{\eta}T_{2,n}^{(i)}(\eta) - T_{2,n}^{(i)}(\eta) = (\eta^2-n^2+1)Z_n^{(i)}(\eta) -\eta Z_{n+1}^{(i)}(\eta).
\end{align*}
Using \Cref{Eq1:constitutiveRelationSpherical,Eq1:strainsInSpherical}, and the relation\footnote{This relation is obtained by inserting the definition of the angular wave numbers $a$ and $b$ (\Cref{Eq1:waveNumber_a_b}) into the left hand side.}
\begin{equation}
	\frac{1}{2}\left(\frac{b}{a}\right)^2 = \frac23+\frac{K}{2G}
\end{equation}
the following formulas for the stress field components are obtained\footnote{One can save some work by observing the similarities between $\sigma_{\upvartheta\upvartheta}$ and $\sigma_{\upvarphi\upvarphi}$
\begin{align*}
	\sigma_{\upvartheta\upvartheta} &= \frac{2}{r}\left(K+\frac{G}{3}\right)u_{\mathrm{r}} + \left(K-\frac{2G}{3}\right)\pderiv{u_{\mathrm{r}}}{r} + \frac{3K-2G}{3r}\left(u_\upvartheta\cot\vartheta + \pderiv{u_\upvartheta}{\vartheta}\right) + \frac{2G}{r}\pderiv{u_\upvartheta}{\vartheta}\\
	\sigma_{\upvarphi\upvarphi} &= \frac{2}{r}\left(K+\frac{G}{3}\right)u_{\mathrm{r}} + \left(K-\frac{2G}{3}\right)\pderiv{u_{\mathrm{r}}}{r} + \frac{3K-2G}{3r}\left(u_\upvartheta\cot\vartheta + \pderiv{u_\upvartheta}{\vartheta}\right) + \frac{2G}{r} u_\upvartheta\cot\vartheta.
\end{align*}}
\begin{align}\label{Eq1:stressFieldComponents1}
	\sigma_{\mathrm{r}\mathrm{r}} &= \frac{2G}{r^2}\sum_{n=0}^\infty Q_n^{(0)}(\vartheta)\left[A_n^{(i)} S_{5,n}^{(i)}(\xi) + B_n^{(i)} T_{5,n}^{(i)}(\eta)\right]\\\label{Eq1:stressFieldComponents2}
	\sigma_{\upvartheta\upvartheta} &= \frac{2G}{r^2}\sum_{n=0}^\infty\left\{Q_n^{(0)}(\vartheta)\left[A_n^{(i)} S_{6,n}^{(i)}(\xi) + B_n^{(i)} T_{6,n}^{(i)}(\eta)\right] +  Q_n^{(2)}(\vartheta)\left[A_n^{(i)} S_{2,n}^{(i)}(\xi) + B_n^{(i)} T_{2,n}^{(i)}(\eta)\right]\right\}\\\label{Eq1:stressFieldComponents3}
	\sigma_{\upvarphi\upvarphi} &= \frac{2G}{r^2}\sum_{n=0}^\infty\left\{Q_n^{(0)}(\vartheta)\left[A_n^{(i)} S_{6,n}^{(i)}(\xi) + B_n^{(i)} T_{6,n}^{(i)}(\eta)\right] +   Q_n^{(1)}(\vartheta)\cot(\vartheta)\left[A_n^{(i)} S_{2,n}^{(i)}(\xi) + B_n^{(i)} T_{2,n}^{(i)}(\eta)\right]\right\}\\\label{Eq1:stressFieldComponents4}
	\sigma_{\upvartheta\upvarphi} &= 0\\\label{Eq1:stressFieldComponents5}
	\sigma_{\mathrm{r}\upvarphi} &= 0\\\label{Eq1:stressFieldComponents6}
	\sigma_{\mathrm{r}\upvartheta} &= \frac{2G}{r^2}\sum_{n=0}^\infty Q_n^{(1)}(\vartheta)\left[A_n^{(i)} S_{7,n}^{(i)}(\xi) + B_n^{(i)} T_{7,n}^{(i)}(\eta)\right]
\end{align}
where
\begin{align}\label{Eq1:stress}
\begin{split}
	S_{5,n}^{(i)}(\xi) &= \frac{1}{2G}\left[\left(K+\frac{4G}{3}\right)S_{3,n}^{(i)}(\xi) - \left(K-\frac{2G}{3}\right) n(n+1)Z_n^{(i)}(\xi) + 2\left(K-\frac{2G}{3}\right) S_{1,n}^{(i)}(\xi)\right] \\
	&= \left[n^2-n-\frac{1}{2}\left(\frac{b}{a}\right)^2\xi^2\right] Z_n^{(i)}(\xi) + 2\xi Z_{n+1}^{(i)}(\xi)\\
	T_{5,n}^{(i)}(\eta) &= \frac{1}{2G}\left[\left(K+\frac{4G}{3}\right)T_{3,n}^{(i)}(\eta) - \left(K-\frac{2G}{3}\right) n(n+1)T_{2,n}^{(i)}(\eta) + 2\left(K-\frac{2G}{3}\right) T_{1,n}^{(i)}(\eta)\right] \\
	&= -n(n+1)\left[(n-1)Z_n^{(i)}(\eta) - \eta Z_{n+1}^{(i)}(\eta)\right]\\
	S_{6,n}^{(i)}(\xi) &= -\left(\frac{K}{2G}-\frac13\right)n(n+1)S_{2,n}^{(i)}(\xi) +\left(\frac13+\frac{K}{G} \right)S_{1,n}^{(i)}(\xi) + \left(\frac{K}{2G}-\frac13\right)S_{3,n}^{(i)}(\xi)\\
	&= \left[n-\frac{1}{2}\left(\frac{b}{a}\right)^2\xi^2+\xi^2\right] Z_n^{(i)}(\xi) - \xi Z_{n+1}^{(i)}(\xi)\\
	T_{6,n}^{(i)}(\eta) &=  -\left(\frac{K}{2G}-\frac13\right)n(n+1)T_{2,n}^{(i)}(\eta) +\left(\frac13+\frac{K}{G} \right)T_{1,n}^{(i)}(\eta) + \left(\frac{K}{2G}-\frac13\right)T_{3,n}^{(i)}(\eta)\\
	&= -n(n+1)Z_n^{(i)}(\eta)\\
	S_{7,n}^{(i)}(\xi) &= \frac{1}{2}\left[S_{1,n}^{(i)}(\xi) + S_{4,n}^{(i)}(\xi) - S_{2,n}^{(i)}(\xi) \right] \\
	&= (n-1)Z_n^{(i)}(\xi) -\xi Z_{n+1}^{(i)}(\xi)\\
	T_{7,n}^{(i)}(\eta) &= \frac{1}{2}\left[T_{1,n}^{(i)}(\eta) + T_{4,n}^{(i)}(\eta) - T_{2,n}^{(i)}(\eta) \right] \\
	&= -\left(n^2-1-\frac{1}{2}\eta^2\right)Z_n^{(i)}(\eta) - \eta Z_{n+1}^{(i)}(\xi).
	\end{split}
\end{align}

\subsection{Validation of the displacement and stress formulas}
The correctness of the formulas may be controlled by considering Navier's equation (\Cref{Eq1:navier}) in spherical coordinates. The three components of Navier's equation in spherical coordinates are given in \Cref{Eq1:navierSpherical1,Eq1:navierSpherical2,Eq1:navierSpherical3}, the last of which is automatically satisfied due to the symmetry assumptions. The first two equations simplify to
\begin{align}
\pderiv{\sigma_{\mathrm{rr}}}{r} + \frac{1}{r}\pderiv{\sigma_{\mathrm{r}\upvartheta}}{\vartheta} + \frac{1}{r}\left(2\sigma_{\mathrm{r}\mathrm{r}} - \sigma_{\upvartheta\upvartheta} - \sigma_{\upvarphi\upvarphi} + \sigma_{\mathrm{r}\upvartheta}\cot\vartheta\right) +\omega^2\rho_{\mathrm{s}}u_{\mathrm{r}} &= 0\label{Eq1:navierSphericalSimplified1}\\
	\pderiv{\sigma_{\mathrm{r}\upvartheta}}{r} + \frac{1}{r}\pderiv{\sigma_{\upvartheta\upvartheta}}{\vartheta} + \frac{1}{r}\left[(\sigma_{\upvartheta\upvartheta} - \sigma_{\upvarphi\upvarphi})\cot\vartheta + 3\sigma_{\mathrm{r}\upvartheta} \right] +\omega^2\rho_{\mathrm{s}}u_\upvartheta &= 0.\label{Eq1:navierSphericalSimplified2}
\end{align}
Differentiation of the stress field components yields
\begin{align*}
	\pderiv{\sigma_{\mathrm{r}\mathrm{r}}}{r} &= \frac{2G}{r^3}\sum_{n=0}^\infty Q_n^{(0)}(\vartheta)\left[A_n^{(i)} S_{8,n}^{(i)}(\xi) + B_n^{(i)} T_{8,n}^{(i)}(\eta)\right]\\
	\pderiv{\sigma_{\upvartheta\upvartheta}}{\vartheta} &= \frac{2G}{r^2}\sum_{n=0}^\infty\left\{Q_n^{(1)}(\vartheta)\left[A_n^{(i)} S_{6,n}^{(i)}(\xi) + B_n^{(i)} T_{6,n}^{(i)}(\eta)\right] +  Q_n^{(3)}(\vartheta)\left[A_n^{(i)} S_{2,n}^{(i)}(\xi) + B_n^{(i)} T_{2,n}^{(i)}(\eta)\right]\right\}\\
	\pderiv{\sigma_{\mathrm{r}\upvartheta}}{r} &= \frac{2G}{r^3}\sum_{n=0}^\infty Q_n^{(1)}(\vartheta)\left[A_n^{(i)} S_{9,n}^{(i)}(\xi) + B_n^{(i)} T_{9,n}^{(i)}(\eta)\right]\\
	\pderiv{\sigma_{\mathrm{r}\upvartheta}}{\vartheta} &= \frac{2G}{r^2}\sum_{n=0}^\infty Q_n^{(2)}(\vartheta)\left[A_n^{(i)} S_{7,n}^{(i)}(\xi) + B_n^{(i)} T_{7,n}^{(i)}(\eta)\right]
\end{align*}
where
\begin{align*}
	S_{8,n}^{(i)}(\xi) &= -2S_{5,n}^{(i)}(\xi) + \xi \deriv{}{\xi}S_{5,n}^{(i)}(\xi) \\
	&= \left[n^3-3n^2+2n-\frac{n}{2}\left(\frac{b}{a}\right)^2\xi^2 + 2\xi^2\right]Z_n^{(i)}(\xi) + \left[-n^2-n-6+\frac{1}{2}\left(\frac{b}{a}\right)^2\xi^2\right]\xi Z_{n+1}^{(i)}(\xi)\\
	T_{8,n}^{(i)}(\eta) &= -2T_{5,n}^{(i)}(\eta) + \eta \deriv{}{\eta}T_{5,n}^{(i)}(\eta) \\
	&= n(n+1)\left[\left(-n^2+3n-2+\eta^2\right)Z_n^{(i)}(\eta) - 4\eta Z_{n+1}^{(i)}(\eta)\right]\\
	S_{9,n}^{(i)}(\xi) &= -2S_{7,n}^{(i)}(\xi) + \xi \deriv{}{\xi}S_{7,n}^{(i)}(\xi) \\
	&= \left[n^2-3n+2-\xi^2\right]Z_n^{(i)}(\xi) + 4\xi Z_{n+1}^{(i)}(\xi) \\
	T_{9,n}^{(i)}(\eta) &= -2T_{7,n}^{(i)}(\eta) + \eta \deriv{}{\eta}T_{7,n}^{(i)}(\eta) \\
	&= \left(-n^3+2n^2+n-2+\frac{n}{2}\eta^2-\eta^2\right)Z_n^{(i)}(\eta) +\left(n^2+n+2-\frac{1}{2}\eta^2\right)\eta Z_{n+1}^{(i)}(\eta).
\end{align*}
Inserting these expressions (alongside the stress components in \Cref{Eq1:stressFieldComponents1,Eq1:stressFieldComponents2,Eq1:stressFieldComponents3,Eq1:stressFieldComponents4,Eq1:stressFieldComponents5,Eq1:stressFieldComponents6}) into \Cref{Eq1:navierSphericalSimplified1,Eq1:navierSphericalSimplified2} and using \Cref{Eq1:expandedLegendreEquationIdentity2,Eq1:expandedLegendreEquationIdentity3}, and observing that
\begin{align*}
	\pderiv{\sigma_{\upvartheta\upvartheta}}{\vartheta} +(\sigma_{\upvartheta\upvartheta} - \sigma_{\upvarphi\upvarphi})\cot\vartheta= \frac{2G}{r^2}\sum_{n=0}^\infty Q_n^{(1)}(\vartheta)&\left\{A_n^{(i)} S_{6,n}^{(i)}(\xi) + B_n^{(i)} T_{6,n}^{(i)}(\eta)\right. \\
	&+\left. \left(-n^2-n+1\right)\left[A_n^{(i)} S_{2,n}^{(i)}(\xi) + B_n^{(i)} T_{2,n}^{(i)}(\eta)\right]\right\},
\end{align*}
the left hand side of \Cref{Eq1:navierSphericalSimplified1} and \Cref{Eq1:navierSphericalSimplified2} are indeed equal to zero.

\section{Establishing constraints from boundary conditions}
\label{Sec1:estabConstraints}
\begin{figure}
	\centering
	\includegraphics[scale=1]{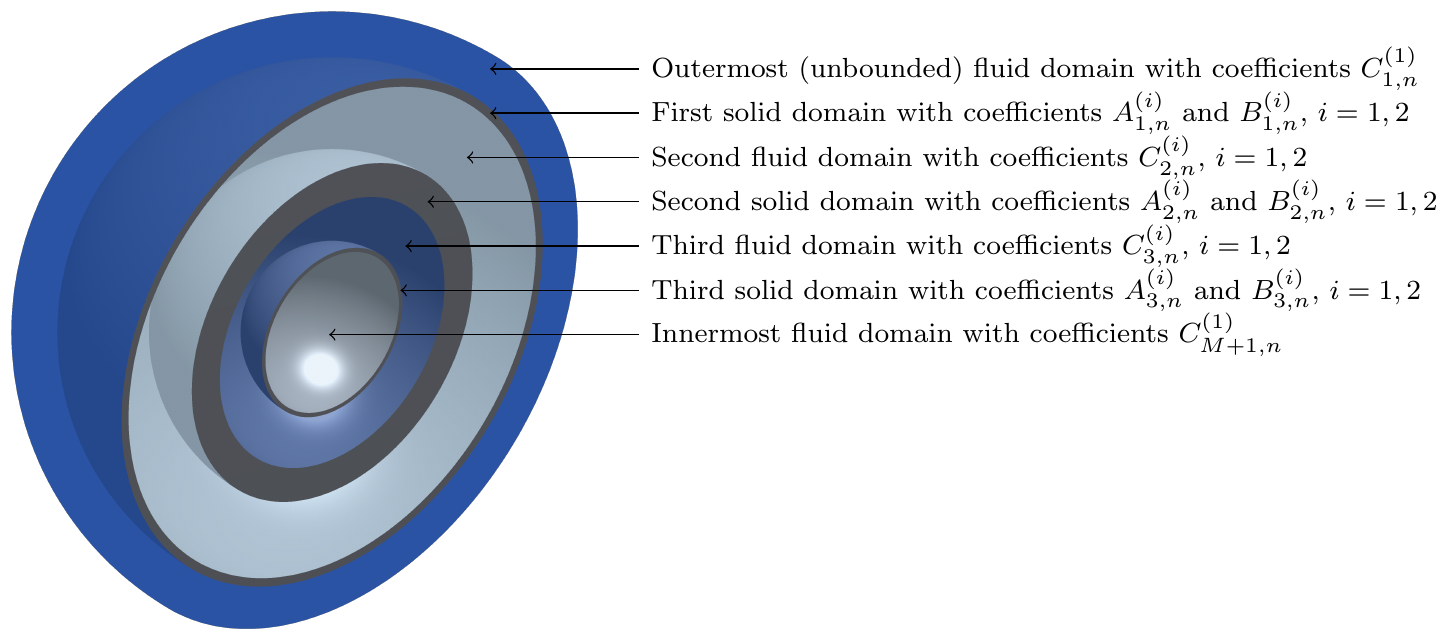}
	\caption{A model with $M=3$ steel shells with different thicknesses (clip view), illustrating the distribution of the coefficients $A_{m,n}^{(i)}$, $B_{m,n}^{(i)}$ and $C_{m,n}^{(i)}$ over the different domains.}
	\label{Fig1:coefficients}
\end{figure}

As the solution is represented as an infinite sum, the coefficients $A_{m,n}^{(i)}$, $B_{m,n}^{(i)}$ and $C_{m,n}^{(i)}$ (coefficients from the fluid domains described below) must be computed for each $n$ (see \Cref{Fig1:coefficients}). By enforcing the boundary conditions in \Cref{Eq1:firstBC,Eq1:secondBC} at each surface, constraints are developed to establish expressions for these coefficients.

\subsection{Notation for the solution in layered domains}
For the $m^{\mathrm{th}}$ solid shell the displacement field from \Cref{Eq1:u_rgen,Eq1:u_tgen} is written as
\begin{equation}
	\vec{u}_m = u_{\mathrm{r},m} \vec{e}_{\mathrm{r}} + u_{\upvartheta,m} \vec{e}_\upvartheta
\end{equation}
where
\begin{align}
	u_{\mathrm{r},m}(r,\vartheta) &= \sum_{n=0}^\infty Q_n^{(0)}(\vartheta)u_{\mathrm{r},m,n}(r) \label{Eq1:u_r}\\
	u_{\upvartheta,m}(r,\vartheta) &= \sum_{n=0}^\infty Q_n^{(1)}(\vartheta)u_{\upvartheta,m,n}(r)
\end{align}
and 
\begin{align}
	u_{\mathrm{r},m,n}(r) &= \frac{1}{r}\left[A_{m,n}^{(i)}S_{1,n}^{(i)}(a_m r)+B_{m,n}^{(i)}T_{1,n}^{(i)}(b_m r)\right]\\
	u_{\upvartheta,m,n}(r) &= \frac{1}{r}\left[A_{m,n}^{(i)}S_{2,n}^{(i)}(a_m r)+B_{m,n}^{(i)}T_{2,n}^{(i)}(b_m r)\right].
\end{align}
Corresponding expressions for the stress field in \Cref{Eq1:stress} are obtained as
\begin{align}
	\sigma_{\mathrm{rr},m}(r,\vartheta) &= \sum_{n=0}^\infty Q_n^{(0)}(\vartheta)\sigma_{\mathrm{rr},m,n}(r)\label{Eq1:sigma_rr}\\
	\sigma_{\upvartheta\upvartheta,m}(r,\vartheta) &= \sum_{n=0}^\infty Q_n^{(0)}(\vartheta)\sigma_{\upvartheta\upvartheta,m,n}^{(1)}(r) + Q_n^{(2)}(\vartheta)\sigma_{\upvartheta\upvartheta,m,n}^{(2)}(r)\\
	\sigma_{\upvarphi\upvarphi,m}(r,\vartheta) &= \sum_{n=0}^\infty Q_n^{(0)}(\vartheta)\sigma_{\upvarphi\upvarphi,m,n}^{(1)}(r) +  Q_n^{(1)}(\vartheta)\cot(\vartheta)\sigma_{\upvarphi\upvarphi,m,n}^{(2)}(r)\\
	\sigma_{\mathrm{r}\upvarphi,m}(r,\vartheta) &= 0\\
	\sigma_{\upvartheta\upvarphi,m}(r,\vartheta) &= 0\\
	\sigma_{\mathrm{r}\upvartheta,m}(r,\vartheta) &= \sum_{n=0}^\infty Q_n^{(1)}(\vartheta)\sigma_{\mathrm{r}\upvartheta,m,n}(r)\label{Eq1:sigma_rt}
\end{align}
where
\begin{align*}
	\sigma_{\mathrm{rr},m,n}(r) &= \frac{2G_m}{r^2}\left[A_{m,n}^{(i)} S_{5,n}^{(i)}(a_m r) + B_{m,n}^{(i)} T_{5,n}^{(i)}(b_m r)\right]\\
	\sigma_{\upvartheta\upvartheta,m,n}^{(1)}(r) &= \frac{2G_m}{r^2}\left[A_{m,n}^{(i)} S_{6,n}^{(i)}(a_m r) + B_{m,n}^{(i)} T_{6,n}^{(i)}(b_m r)\right]\\
	\sigma_{\upvartheta\upvartheta,m,n}^{(2)}(r) &= \frac{2G_m}{r^2}\left[A_{m,n}^{(i)} S_{2,n}^{(i)}(a_m r) + B_{m,n}^{(i)} T_{2,n}^{(i)}(b_m r)\right]\\ 
	\sigma_{\upvarphi\upvarphi,m,n}^{(1)}(r) &= \frac{2G_m}{r^2}\left[A_{m,n}^{(i)} S_{6,n}^{(i)}(a_m r) + B_{m,n}^{(i)} T_{6,n}^{(i)}(b_m r)\right]\\
	\sigma_{\upvarphi\upvarphi,m,n}^{(2)}(r) &= \frac{2G_m}{r^2}\left[A_{m,n}^{(i)} S_{2,n}^{(i)}(a_m r) + B_{m,n}^{(i)} T_{2,n}^{(i)}(b_m r)\right]\\ 
	\sigma_{\mathrm{r}\upvartheta,m,n}(r) &= \frac{2G_m}{r^2}\left[A_{m,n}^{(i)} S_{7,n}^{(i)}(a_m r) + B_{m,n}^{(i)} T_{7,n}^{(i)}(b_m r)\right].
\end{align*}

The solution to the Helmholtz equation in the $m^{\mathrm{th}}$ fluid domain (for $2 \leq m \leq M$) has the same general form as $\phi$ in \Cref{Eq1:phiSolutionSimplified}
\begin{equation}\label{Eq1:generalSol}
	p_m(r,\vartheta) = \sum_{n=0}^\infty Q_n^{(0)}(\vartheta)C_{m,n}^{(i)} Z_n^{(i)}(k_m r)
\end{equation}
where the coefficients $C_{m,n}^{(i)}\in\C$ are chosen such that the boundary conditions are satisfied. As the spherical Hankel functions of first and second kind (described in  \Cref{subsec:sphericalBesselAndHankel}) are linear combinations of the spherical Bessel functions of first and second kind, the general solution can be written in terms of these functions. For the outer (unbounded) fluid the Hankel function of the second kind is eliminated due to the Sommerfeld radiation condition in \Cref{Eq1:Sommerfeld}~\cite[p. 26]{Ihlenburg1998fea}. Thus, for the outermost fluid, the scattered pressure field is given by
\begin{equation}\label{Eq1:outerFluid}
	p_1(r,\vartheta) = \sum_{n=0}^\infty Q_n^{(0)}(\vartheta) C_{1,n}^{(1)} \hankel^{(1)}_n(k_1 r).
\end{equation}
Moreover, it is required that the pressure in the innermost fluid domain is bounded~\cite[p. 10]{Fender1972sfa}. Hence, the coefficients $C_{M+1,n}^{(2)}$ must be set to zero as the spherical Bessel function of second kind is unbounded at the origin. The pressure in the innermost fluid is therefore given by (cf.~\cite[p. 10]{Fender1972sfa})
\begin{equation}\label{Eq1:innerFluid}
	p_{M+1}(r,\vartheta) = \sum_{n=0}^\infty Q_n^{(0)}(\vartheta) C_{M+1,n}^{(1)} \besselj_n(k_{M+1} r).
\end{equation}
The total pressure in the $m^{\mathrm{th}}$ fluid domain shall be denoted by
\begin{equation}\label{Eq1:totPressure}
	p_{\mathrm{tot},m} =\begin{cases}
	p_1 + p_{\mathrm{inc}}& m = 1\\
	p_m & \text{otherwise}
	\end{cases}	
\end{equation}
where $p_{\mathrm{inc}}$ is the incident wave.

If the coefficients $A_{m,n}^{(i)}$, $B_{m,n}^{(i)}$ and $C_{m,n}^{(i)}$ can be determined, the solution is fully determined in all domains. Hence, a system of equations will be developed to find these coefficients. Indeed, at the boundaries (at a fixed radius) the series can all be written in terms of the Legendre functions $\legendre_n(\cos\vartheta)$, such that the resulting coefficients can be compared for each $n$. A term in the solution is often referred to as a \textit{mode}, such that the resulting constraints from the boundary conditions form a set of \textit{modal equations}. The terminology comes from the vibration analysis~\cite{Chang1994voa}, where each of these modes represent vibration modes. For example, $u_{\mathrm{r},m,n}$ is referred to be the radial displacement in the $m^{\mathrm{th}}$ solid domain in the $n^{\mathrm{th}}$ mode.

\subsection{Tangential traction conditions}
\Cref{Eq1:traction2} is automatically fulfilled due to the axisymmetric assumption. For the $m^{\mathrm{th}}$ shell, evaluating \Cref{Eq1:traction1} at both the inner and outer radius, yields two equations 
\begin{equation}\label{Eq1:tractionCondInserted}
	\sigma_{\mathrm{r}\upvartheta,m,n}(R_{j,m},\vartheta) = 0, \quad j=0,1.
\end{equation}
As $Q_0^{(1)}(\vartheta)=0$, these equations are automatically satisfied for $n=0$. In addition, since $T_{1,0}^{(i)}(\eta)=0$ and $T_{6,0}^{(i)}(\eta)=0$, the coefficients $B_{m,0}^{(i)}$ are redundant (which is convenient, as two constraints are lost in this case). 

Denote by $\vec{H}_{m,n}^{(1)}$, $m=1,\dots,M$, the eigenfrequency matrix\footnote{As illustrated in~\cite{Chang1994voa}, the matrix $\vec{H}_{m,n}^{(1)}$ represent the modal characteristic equations of the $m^{\mathrm{th}}$ shell. That is, the eigenfrequencies of each shell can be found by solving $\det\vec{H}_{m,n}^{(1)}=0$ in terms of the frequency.}~\cite[p. 17]{Chang1994voa} of the $m^{\mathrm{th}}$ shell
\begin{equation}\label{Eq1:K1n}
	\vec{H}_{m,n}^{(1)} = \begin{bmatrix} S_{5,n}^{(1)}(a_m R_{0,m}) & S_{5,n}^{(2)}(a_m R_{0,m}) & T_{5,n}^{(1)}(b_m R_{0,m}) & T_{5,n}^{(2)}(b_m R_{0,m})\\
	S_{7,n}^{(1)}(a_m R_{0,m}) & S_{7,n}^{(2)}(a_m R_{0,m}) & T_{7,n}^{(1)}(b_m R_{0,m}) & T_{7,n}^{(2)}(b_m R_{0,m})\\
	S_{7,n}^{(1)}(a_m R_{1,m}) & S_{7,n}^{(2)}(a_m R_{1,m}) & T_{7,n}^{(1)}(b_m R_{1,m}) & T_{7,n}^{(2)}(b_m R_{1,m})\\
	S_{5,n}^{(1)}(a_m R_{1,m}) & S_{5,n}^{(2)}(a_m R_{1,m}) & T_{5,n}^{(1)}(b_m R_{1,m}) & T_{5,n}^{(2)}(b_m R_{1,m})\end{bmatrix},
\end{equation}
for $n>0$, and 
\begin{equation}\label{Eq1:K10}
	\vec{H}_{m,0}^{(1)} = \begin{bmatrix}	
	 S_{5,0}^{(1)}(a_m R_{0,m}) & S_{5,0}^{(2)}(a_m R_{0,m})\\
	 S_{5,0}^{(1)}(a_m R_{1,m}) & S_{5,0}^{(2)}(a_m R_{1,m})\end{bmatrix},
\end{equation}
for $n=0$. From \Cref{Eq1:sigma_rt,Eq1:sigma_rr} one observes that the first and the last row of $\vec{H}_{m,n}^{(1)}$ correspond to $\sigma_{\mathrm{rr},m,n}(r)$ at $r=R_{0,m}$ and $r=R_{1,m}$, respectively, and the second and third row (for $n>0$) correspond to $\sigma_{\mathrm{r}\upvartheta,m,n}(r)$ at $r=R_{0,m}$ and $r=R_{1,m}$, respectively. The notation $H_{ij,m,n}^{(1)}$, will be used for the elements of the matrices $\vec{H}_{m,n}^{(1)}$.

For $n>0$, the two conditions in \Cref{Eq1:tractionCondInserted} may be written as
\begin{align}
	H_{21,m,n}^{(1)}A_{m,n}^{(1)} + H_{22,m,n}^{(1)}A_{m,n}^{(2)} + H_{23,m,n}^{(1)}B_{m,n}^{(1)} + H_{24,n}^{(1)}B_{m,n}^{(2)} &= 0\\
	H_{31,m,n}^{(1)}A_{m,n}^{(1)} + H_{32,m,n}^{(1)}A_{m,n}^{(2)} + H_{33,m,n}^{(1)}B_{m,n}^{(1)} + H_{34,m,n}^{(1)}B_{m,n}^{(2)} &= 0.
\end{align}
This gives (for each $n$) $2M$ equations in terms of the $6M$ unknown coefficients $A_{m,n}^{(i)}$, $B_{m,n}^{(i)}$ and $C_{m,n}^{(i)}$, $i=1,2$. Thus, an additional $4M$ equations are needed to determine these coefficients. These equations come from the coupling conditions in \Cref{Eq1:firstBC,Eq1:secondBC} (displacement condition and pressure condition, respectively) which are applied at the outer and inner radius of each shell. The outermost and innermost fluid domains will have to be considered separately.

\subsection{Displacement and pressure condition in intermediate fluid layers}
Consider the $m^{\mathrm{th}}$ fluid domain, with $2\leq m\leq M$, where the pressure field is given by \Cref{Eq1:generalSol}. Inserting \Cref{Eq1:u_r,Eq1:generalSol} into the displacement condition in \Cref{Eq1:firstBC} at $r=R_{1,m-1},R_{0,m}$, yields
\begin{align*}
	&\frac{\rho_{\mathrm{f},m}\omega^2}{R_{j,m-j}}\left[A_{m-j,n}^{(i)} S_{1,n}^{(i)}(a_{m-j} R_{j,m-j}) + B_{m-j,n}^{(i)} T_{1,n}^{(i)}(b_{m-j} R_{j,m-j})\right] \\
	&{\hskip14em\relax} - k_m\left[C_{m,n}^{(1)}\besselj_n'(k_m R_{j,m-j}) + C_{m,n}^{(2)}y_n'(k_m R_{j,m-j})\right] = 0
\end{align*}
which yield the relation
\begin{equation}
	H_{1,m-j,n}^{(4,j)}A_{m-j,n}^{(1)} + H_{2,m-j,n}^{(4,j)}A_{m-j,n}^{(2)} + H_{3,m-j,n}^{(4,j)}B_{m-1,n}^{(1)} + H_{4,m-j,n}^{(4,j)}B_{m-1,n}^{(2)} + H_{i,m,n}^{(3,j)}C_{m,n}^{(i)} = 0,
\end{equation}
for $j=0,1$, where
\begin{align}\label{Eq1:K4}
\begin{split}
	H_{1,m,n}^{(4,j)} &= S_{1,n}^{(1)}(a_m R_{j,m}),\quad H_{2,m,n}^{(4,j)} = S_{1,n}^{(2)}(a_m R_{j,m}),\\ H_{3,m,n}^{(4,j)} &= T_{1,n}^{(1)}(b_m R_{j,m}),\quad H_{4,m,n}^{(4,j)} = T_{1,n}^{(2)}(b_m R_{j,m}),
	\end{split}
\end{align}
and (using \Cref{Eq1:BesselDerivIdentity2} to rewrite the derivative of the Bessel functions)
\begin{align}\label{Eq1:K3}
	H_{i,m,n}^{(3,j)} &= -\frac{1}{\rho_{\mathrm{f},m}\omega^2}\left[nZ_n^{(i)}(\zeta) - \zeta Z_{n+1}^{(i)}(\zeta)\right]\Big\vert_{\zeta=k_mR_{j,m-j}}.
\end{align}

Correspondingly, inserting \Cref{Eq1:sigma_rr,Eq1:innerFluid} into \Cref{Eq1:secondBC} at $r=R_{1,m-1},R_{0,m}$ yields
\begin{equation*}
	\frac{2G_{m-j}}{R_{j,m-j}^2}\left[A_{m-j,n}^{(i)} S_{5,n}^{(i)}(a_{m-j} R_{j,m-j}) + B_{m-j,n}^{(i)} T_{5,n}^{(i)}(b_{m-j} R_{j,m-j})\right] + C_{m,n}^{(i)} Z_n^{(i)}(k_m R_{j,m-j}) = 0
\end{equation*}
which can be rewritten as
\begin{equation}
	H_{11,m-j,n}^{(1)}A_{m-j,n}^{(1)} + H_{12,m-j,n}^{(1)}A_{m-j,n}^{(2)} + H_{13,m-j,n}^{(1)}B_{m-j,n}^{(1)} + H_{14,m-j,n}^{(1)}B_{m-j,n}^{(2)}+ H_{i,m,n}^{(2,j)}C_{m,n}^{(i)} = 0
\end{equation}
where
\begin{equation}\label{Eq1:K2}
	H_{i,m,n}^{(2,j)} = \frac{R_{j,m-j}^2}{2G_{m-j}}Z_n^{(i)}(k_mR_{j,m-j}).
\end{equation}

\subsection{Displacement and pressure condition in the outermost fluid}
It is assumed that the incident wave, $p_{\mathrm{inc}}(\vec{x},\omega)$, and its normal derivative at the outermost solid surface can be written on the form
\begin{align}\label{Eq1:IncidentWaveConds}
\begin{split}
	p_{\mathrm{inc}}\Big\vert_{r=R_{0,1}} &= \sum_{n=0}^\infty F_n^{(1)} \legendre_n(\cos\vartheta),\\
	\pderiv{p_{\mathrm{inc}}}{r}\Big\vert_{r=R_{0,1}} &= \sum_{n=0}^\infty F_n^{(2)} \legendre_n(\cos\vartheta),
	\end{split}
\end{align}
respectively. The coefficients $F_n^{(1)}$ and $F_n^{(2)}$ are discussed in \Cref{Sec1:incidentWave}.

Inserting \Cref{Eq1:u_r,Eq1:outerFluid} into the displacement condition in \Cref{Eq1:firstBC} yields
\begin{align*}
	&\frac{\rho_{\mathrm{f},1}\omega^2}{R_{0,1}}\left[A_{n,1}^{(i)} S_{1,n}^{(i)}(a_1 R_{0,1}) + B_{n,1}^{(i)} T_{1,n}^{(i)}(b_1 R_{0,1})\right]  -k_1 C_{1,n}^{(1)} \deriv{\hankel^{(1)}_n}{\zeta}\Big\vert_{\zeta=k_1R_{0,1}} = F_n^{(2)},
\end{align*}
which yields the relation
\begin{equation}
	H_{1,1,n}^{(4,0)}C_{1,n}^{(1)} + H_{2,1,n}^{(4,0)}C_{1,n}^{(2)} + H_{3,1,n}^{(4,0)}C_{1,n}^{(3)} + H_{4,1,n}^{(4,0)}C_{1,n}^{(4)} + H_{1,1,n}^{(3,0)}C_{1,n}^{(1)} = D_{1,n},
\end{equation}
where $H_{i,1,n}^{(4,0)}$ for $i=1,2,3,4$, are given by \Cref{Eq1:K4} and (using \Cref{Eq1:HankelDerivIdentity2})
\begin{equation}\label{Eq1:K3011}
	H_{1,1,n}^{(3,0)} = -\frac{1}{\rho_{\mathrm{f},1}\omega^2}\left[n\hankel^{(1)}_n(\zeta) - \zeta \hankel^{(1)}_{n+1}(\zeta)\right]\Big\vert_{\zeta=k_1R_{0,1}}
\end{equation}
and
\begin{equation}\label{Eq1:D1}
	D_{1,n} = \frac{R_{0,1}}{\rho_{\mathrm{f},1}\omega^2}F_n^{(2)}.
\end{equation}
Correspondingly, by inserting \Cref{Eq1:sigma_rr,Eq1:outerFluid} into \Cref{Eq1:secondBC} one obtains
\begin{align*}
	&\frac{2G_1}{R_{0,1}^2}\left[C_{n,1}^{(1)} S_{5,n}^{(1)}(a_1 R_{0,1}) + C_{n,1}^{(2)} T_{5,n}^{(1)}(b_1 R_{0,1}) + C_{n,1}^{(3)} S_{5,n}^{(2)}(a_1 R_{0,1}) + C_{n,1}^{(4)} T_{5,n}^{(2)}(b_1 R_{0,1})\right] \\
	&{\hskip27em\relax} + C_{1,n}^{(1)} \hankel^{(1)}_n(k_1R_{0,1})= -F_n^{(1)},
\end{align*}
which yields the relation
\begin{equation}
	H_{1,1,n}^{(1)}C_{1,n}^{(1)} + H_{2,1,n}^{(1)}C_{1,n}^{(2)} + H_{3,1,n}^{(1)}C_{1,n}^{(3)} + H_{4,1,n}^{(1)}C_{1,n}^{(4)} + H_{1,1,n}^{(2,0)}C_{1,n}^{(1)} = D_{2,n},
\end{equation}
where
\begin{equation}\label{Eq1:K2011}
	H_{1,1,n}^{(2,0)} = \frac{R_{0,1}^2}{2G_1}\hankel^{(1)}_n(k_1R_{0,1})
\end{equation}
and
\begin{equation}\label{Eq1:D2}
	D_{2,n} = -\frac{R_{0,1}^2}{2G_1}F_n^{(1)}.
\end{equation}

\subsection{Displacement and pressure condition in the innermost fluid}
For the innermost fluid the pressure field is given by \Cref{Eq1:innerFluid}. Inserting \Cref{Eq1:u_r,Eq1:innerFluid} into the displacement condition in \Cref{Eq1:firstBC} at $r=R_{1,M}$ yields
\begin{equation*}
	\frac{\rho_{\mathrm{f},M+1}\omega^2}{R_{1,M}}\left[A_{M,n}^{(i)} S_{1,n}^{(i)}(a_M R_{1,M}) + B_{M,n}^{(i)} T_{1,n}^{(i)}(b_M R_{1,M})\right] - k_{M+1}C_{M+1,n}^{(1)}\besselj_n'(k_{M+1} R_{1,M}) = 0,
\end{equation*}
which yields the relation
\begin{equation}
	H_{1,M,n}^{(4,1)}A_{M,n}^{(1)} + H_{2,M,n}^{(4,1)}A_{M,n}^{(2)} + H_{3,M,n}^{(4,1)}B_{M,n}^{(1)} + H_{4,M,n}^{(4,1)}B_{M,n}^{(2)} + H_{1,M+1,n}^{(3,1)}C_{M+1,n}^{(1)} = 0,
\end{equation}
where $H_{i,M,n}^{(4,1)}$ for $i=1,2,3,4$, are defined in \Cref{Eq1:K4}, and 
\begin{equation}\label{Eq1:K311Mp1}
	H_{1,M+1,n}^{(3,1)} = -\frac{1}{\rho_{\mathrm{f},M+1}\omega^2}\left[n\besselj_n(\zeta) - \zeta \besselj_{n+1}(\zeta)\right]\Big\vert_{\zeta=k_{M+1}R_{1,M}}.
\end{equation}
Correspondingly, by inserting \Cref{Eq1:sigma_rr,Eq1:innerFluid} into \Cref{Eq1:secondBC} at $r=R_{1,M}$ the following is obtained
\begin{equation*}
	\frac{2G_M}{R_{1,M}^2}\left[A_{M,n}^{(i)} S_{5,n}^{(i)}(a_M R_{1,M}) + B_{M,n}^{(i)} T_{5,n}^{(i)}(b_M R_{1,M})\right] + C_{M+1,n}^{(1)} \besselj_n(k_{M+1} R_{1,M})  = 0,
\end{equation*}
which yields the relation
\begin{equation}
	H_{11,M,n}^{(1)}A_{M,n}^{(1)} + H_{12,M,n}^{(1)}A_{M,n}^{(2)} + H_{13,M,n}^{(1)}B_{M,n}^{(1)} + H_{14,M,n}^{(1)}B_{M,n}^{(2)}+ H_{1,M+1,n}^{(2,1)}C_{M+1,n}^{(1)} = 0,
\end{equation}
where
\begin{equation}\label{Eq1:K211Mp1}
	H_{1,M+1,n}^{(2,1)} = \frac{R_{1,M}^2}{2G_M}\besselj_n(k_{M+1}R_{1,M}).
\end{equation}

\section{Assembling the linear system of equations}
\label{Sec1:assembly}
In the previous section, $6M$ equations for the $6M$ unknowns $A_{m,n}^{(i)}$, $B_{m,n}^{(i)}$ and $C_{m,n}^{(i)}$ for all $n>0$ and $4M$ equations for the $4M$ unknowns for $n = 0$ was established. So far the solution has been presented for $M$ elastic spherical shells with standard displacement and pressure conditions; \textit{the default case with Neumann-to-Neumann conditions}. By some matrix manipulations of the global matrix, one can implement other cases as well, including solid spheres, and single Neumann conditions replacing the Neumann-to-Neumann conditions on the innermost domain.

\subsection{The default case with Neumann-to-Neumann conditions}
For the default case all equations can be collected into one single linear system of equations 
\begin{equation}\label{Eq1:SystemOfEquations}
	\vec{H}_n \vec{C}_n = \vec{D}_n
\end{equation}
where\footnote{Note that the matrix pattern is scaled for the case $n>0$, as $\vec{H}_{m,n}^{(1)}\in\R^{4\times 4}$ and $\vec{H}_{m,n}^{(4,j)}\in\R^{1\times 4}$ for $n>0$, as opposed to $\vec{H}_{m,n}^{(1)}\in\R^{2\times 2}$ and $\vec{H}_{m,n}^{(4,j)}\in\R^{1\times 2}$ when $n=0$ (for $j = 1, 2$).}
\begin{equation*}
	\vec{H}_n = \left[
	\begin{BMAT}(rc){c.c.c.c.c.c.c}{c.c.c.c.c.c.c.c.c}
	\begin{BMAT}(b){c}{c}H_{1,1,n}^{(3,0)}\end{BMAT} & {\hskip2.3em\relax}\vec{H}_{1,n}^{(4,0)}{\hskip2.3em\relax} & {\hskip4em\relax} & {\hskip2.3em\relax}\phantom{\vec{H}_{1,n}^{(4,0)}}{\hskip2.3em\relax}  & {\hskip4em\relax} & {\hskip2.3em\relax}\phantom{\vec{H}_{1,n}^{(4,0)}}{\hskip2.3em\relax}& \\
	\begin{BMAT}(rc){c}{c.ccc}H_{1,1,n}^{(2,0)}\\ \phantom{H_{1,n}^{(2,0)}}\\ \phantom{H_{1,n}^{(2,0)}}\\ \phantom{H_{1,n}^{(2,0)}}\end{BMAT} & \vec{H}_{1,n}^{(1)} & \begin{BMAT}(rc){c}{ccc.c} \phantom{H_{2,n}^{(2,0)}}\\ \phantom{H_{2,n}^{(2,0)}}\\ \phantom{H_{2,n}^{(2,0)}}\\ {\hskip0.5em\relax}\vec{H}_{2,n}^{(2,1)}{\hskip0.5em\relax}\end{BMAT} & & & &\\
	 & \vec{H}_{1,n}^{(4,1)} & \vec{H}_{2,n}^{(3,1)} & & & &\\
	 & & \vec{H}_{2,n}^{(3,0)} & \vec{H}_{2,n}^{(4,0)} & & &\\
	 & & \begin{BMAT}(rc){c}{c.ccc}{\hskip0.5em\relax}\vec{H}_{2,n}^{(2,0)}{\hskip0.5em\relax}\\ \phantom{H_{1,n}^{(2,0)}}\\ \phantom{H_{1,n}^{(2,0)}}\\ \phantom{H_{2,n}^{(2,0)}}\end{BMAT} & \ddots & \begin{BMAT}(rc){c}{ccc.c} \phantom{H_{1,n}^{(2,0)}}\\ \phantom{H_{1,n}^{(2,0)}}\\ \phantom{H_{1,n}^{(2,0)}}\\ {\hskip0.5em\relax}\vec{H}_{M,n}^{(2,1)}{\hskip0.5em\relax}\end{BMAT} & &\\
	 & & & \vec{H}_{M-1,n}^{(4,1)} & \vec{H}_{M,n}^{(3,1)} & &\\
	 & & & & \vec{H}_{M,n}^{(3,0)} & \vec{H}_{M,n}^{(4,0)} &\\
	 & & & & \begin{BMAT}(rc){c}{c.ccc}{\hskip0.5em\relax}\vec{H}_{M,n}^{(2,0)}{\hskip0.5em\relax}\\ \phantom{H_{1,n}^{(2,0)}}\\ \phantom{H_{1,n}^{(2,0)}}\\ \phantom{H_{1,n}^{(2,0)}}\end{BMAT} & \vec{H}_{M,n}^{(1)} & \begin{BMAT}(rc){c}{ccc.c} \phantom{H_{1,n}^{(2,0)}}\\ \phantom{H_{1,n}^{(2,0)}}\\ \phantom{H_{1,n}^{(2,0)}}\\ H_{1,M+1,n}^{(2,1)}\end{BMAT}\\
	 & & & & & \vec{H}_{M,n}^{(4,1)} & H_{1,M+1,n}^{(3,1)}
	\end{BMAT} 
	\right] 
\end{equation*}
and
\begin{equation*} 
	\vec{C}_n = \begin{bmatrix}
		C_{1,n}^{(1)}\\
		\vec{A}_{1,n}\\
		\vec{B}_{1,n}\\
		\vec{C}_{2,n}\\
		\vdots\\
		\vec{A}_{M-1,n}\\
		\vec{B}_{M-1,n}\\
		\vec{C}_{M,n}\\
		\vec{A}_{M,n}\\
		\vec{B}_{M,n}\\
		C_{M+1,n}^{(1)}
	\end{bmatrix}\quad
	\vec{A}_{m,n} = \begin{bmatrix}
		A_{m,n}^{(1)}\\
		A_{m,n}^{(2)}\\
	\end{bmatrix}\quad
	\vec{B}_{m,n} = \begin{bmatrix}
		B_{m,n}^{(1)}\\
		B_{m,n}^{(2)}\\
	\end{bmatrix}\quad
	\vec{C}_{m,n} = \begin{bmatrix}
		C_{m,n}^{(1)}\\
		C_{m,n}^{(2)}\\
	\end{bmatrix}\quad
	\vec{D}_n = \begin{bmatrix}
		D_{1,n}\\
		D_{2,n}\\
		0\\
		0\\
		\vdots\\
		0
	\end{bmatrix}.
\end{equation*} 
The submatrices $\vec{H}_{m,n}^{(1)}$ has entries given in \Cref{Eq1:K1n} and \Cref{Eq1:K10}. The submatrices 
\begin{equation*}
	\vec{H}_{m,n}^{(2,j)} = \begin{bmatrix}
		H_{1,m,n}^{(2,j)} & H_{2,m,n}^{(2,j)}
	\end{bmatrix}
\end{equation*}
has entries given in \Cref{Eq1:K2}. The submatrices
\begin{equation*}
	\vec{H}_{m,n}^{(3,j)} = \begin{bmatrix}
		H_{1,m,n}^{(3,j)} & H_{2,m,n}^{(3,j)}
	\end{bmatrix}
\end{equation*}
has entries given in \Cref{Eq1:K3}. The submatrices
\begin{equation*}
	\vec{H}_{m,n}^{(4,j)} = \begin{bmatrix}
		H_{1,m,n}^{(4,j)} & H_{2,m,n}^{(4,j)} & H_{3,m,n}^{(4,j)} & H_{4,m,n}^{(4,j)}
	\end{bmatrix}
\end{equation*}
for $n>0$, and 
\begin{equation*}
	\vec{H}_{m,n}^{(4,j)} = \begin{bmatrix}
		H_{1,m,n}^{(4,j)} & H_{2,m,n}^{(4,j)}
	\end{bmatrix}
\end{equation*}
for $n=0$, has entries given in \Cref{Eq1:K4}. The entries $H_{1,1,n}^{(3,0)}$, $H_{1,1,n}^{(2,0)}$, $H_{1,M+1,n}^{(3,1)}$ and $H_{1,M+1,n}^{(2,1)}$, are given in \Cref{Eq1:K3011,Eq1:K2011,Eq1:K311Mp1,Eq1:K211Mp1}, respectively. The entries $D_{1,n}$ and $D_{2,n}$, are given in \Cref{Eq1:D1,Eq1:D2}, respectively.

\subsection{Alternative boundary conditions}
\label{Subsec1:alternativeBoundaryConditions}
By removing the last five (three) rows and columns of $\vec{H}_n$ for $n>0$ ($n=0$), the Neumann-to-Neumann boundary condition (NNBC) is replaced by a single Neumann condition\footnote{That is, the normal velocity component of the fluid at the surface is zero, such that $u_{\mathrm{r}}=0$ in \Cref{Eq1:firstBC}. This is often referred to as a sound-hard boundary condition, SHBC.}
\begin{equation}
	\pderiv{p_{\mathrm{tot},M}}{r} = 0
\end{equation}
at the innermost solid domain. This Neumann boundary condition may be replaced by other boundary conditions like the Robin boundary condition (impedance boundary condition) by corresponding manipulation of the matrix $\vec{H}_n$. By removing the last row and column of the matrix $\vec{H}_n$ a Neumann condition ($\sigma_{\mathrm{rr}}=0$) is obtained on the inside of the innermost shell\footnote{This is often referred to as a sound-soft boundary condition, SSBC.}.  Moreover, one can model scattering on solid spheres\footnote{This type of boundary conditions is named elastic sphere boundary conditions, ESBC.} (such that the innermost domain is no longer fluid, but solid) by removing the three last rows (corresponding to the boundary conditions at $R_{1,M}$) and three columns (corresponding to the coefficients $A_{M,n}^{(2)}$, $B_{M,n}^{(2)}$ and $C_{M+1,n}^{(1)}$) of the matrix $\vec{H}_n$ (and corresponding entries of $\vec{C}_n$ and $\vec{D}_n$). The reason for not using the coefficients $A_{M,n}^{(2)}$ and $B_{M,n}^{(2)}$ is that the corresponding spherical Bessel functions of second kind are unbounded at the origin, such that these coefficients must be set to zero. The displacement of the inner solid sphere is then given by
\begin{equation}
	\vec{u}_M = u_{\mathrm{r},M} \vec{e}_{\mathrm{r}} + u_{\upvartheta,M} \vec{e}_\upvartheta
\end{equation}
where
\begin{align}
	u_{\mathrm{r},M}(r,\vartheta) &= \frac{1}{r}\sum_{n=0}^\infty Q_n^{(0)}(\vartheta)\left[A_{M,n}^{(1)}S_{1,n}^{(1)}(a_M r)+B_{M,n}^{(1)}T_{1,n}^{(1)}(b_M r)\right]\\
	u_{\upvartheta,M}(r,\vartheta) &= \frac{1}{r}\sum_{n=0}^\infty Q_n^{(1)}(\vartheta)\left[A_{M,n}^{(1)}S_{2,n}^{(1)}(a_M r)+B_{M,n}^{(1)}T_{2,n}^{(1)}(b_M r)\right].
\end{align} 
It should be noted that the solution is well defined also at the origin due to the formulas in \Cref{Eq1:BesselLimits,Eq1:BesselLimits2}. In fact, on can show that (using \Cref{Eq1:SphericalToXfun})
\begin{equation*}
	\lim_{r\to 0}\vec{u}_M(r,\vartheta) = \frac13 \left[a_MA_{M,1}^{(1)}-2b_MB_{M,1}^{(1)}\right]\vec{e}_3,
\end{equation*}
and (using \Cref{Eq1:SphericalToXJacobian})
\begin{align*}
	\lim_{r\to 0}\pderiv{u_{1,M}}{x_1}(r,\vartheta) &= \frac{G_M}{9K_M}\left(4a_M^2-3b_M^2\right)A_{M,0}^{(1)} - \frac{1}{15}\left(a_M^2 A_{M,2}^{(1)} - 3b_M^2B_{M,2}^{(1)}\right)\\
	\lim_{r\to 0}\pderiv{u_{2,M}}{x_2}(r,\vartheta) &= \frac{G_M}{9K_M}\left(4a_M^2-3b_M^2\right)A_{M,0}^{(1)} - \frac{1}{15}\left(a_M^2 A_{M,2}^{(1)} - 3b_M^2B_{M,2}^{(1)}\right)\\
	\lim_{r\to 0}\pderiv{u_{3,M}}{x_3}(r,\vartheta) &= \frac{G_M}{9K_M}\left(4a_M^2-3b_M^2\right)A_{M,0}^{(1)} + \frac{2}{15}\left(a_M^2 A_{M,2}^{(1)} - 3b_M^2B_{M,2}^{(1)}\right)\\
	\lim_{r\to 0}\pderiv{u_{i,M}}{x_j}(r,\vartheta) &= 0\quad\text{for}\quad i\neq j
\end{align*}
where $u_{i,M}$ is the $i^{\mathrm{th}}$ Cartesian component of $\vec{u}$. The stress field can then be computed in the origin as (using \Cref{Eq1:StressTransformFromCartToSpherical})
\begin{align*}
	\lim_{r\to 0}\sigma_{11,M}(r,\vartheta) &= \frac{G_M}{15}\left[5\left(4a_M^2-3b_M^2\right)A_{M,0}^{(1)} - 2a_M^2A_{M,2}^{(1)} +6b_M^2B_{M,2}^{(1)}\right]\\
	\lim_{r\to 0}\sigma_{22,M}(r,\vartheta) &= \frac{G_M}{15}\left[5\left(4a_M^2-3b_M^2\right)A_{M,0}^{(1)} - 2a_M^2A_{M,2}^{(1)} +6b_M^2B_{M,2}^{(1)}\right]\\
	\lim_{r\to 0}\sigma_{33,M}(r,\vartheta) &= \frac{G_M}{15}\left[5\left(4a_M^2-3b_M^2\right)A_{M,0}^{(1)} + 4a_M^2A_{M,2}^{(1)} -12b_M^2B_{M,2}^{(1)}\right]\\
	\lim_{r\to 0}\sigma_{23,M}(r,\vartheta) &= 0\\
	\lim_{r\to 0}\sigma_{13,M}(r,\vartheta) &= 0\\
	\lim_{r\to 0}\sigma_{12,M}(r,\vartheta) &= 0
\end{align*}
where $\sigma_{ij,M}$ is the stress field in the solid sphere in Cartesian coordinates. If the innermost domain is a fluid, then
\begin{align*}
	\lim_{r\to 0}p_{M+1}(r,\vartheta) &= C_{M+1,0}^{(1)}\\
	\lim_{r\to 0}\nabla p_{M+1}(r,\vartheta) &= \frac{k_{M+1}}{3}C_{M+1,1}^{(1)} \vec{e}_3\\
	\lim_{r\to 0}\nabla^2 p_{M+1}(r,\vartheta) &= -k_{M+1}^2 C_{M+1,0}^{(1)}.
\end{align*}
Finally, note that one can model connected fluid or solid layers by manipulating the the matrix $\vec{H}_n$ to match the pressure and displacement condition between these domains. An example of such an application is air bubbles in water~\cite{Fender1972sfa}.
\newpage

\subsection{Summary of solution formulas}
\begin{figure}
	\centering
	\includegraphics[scale=1]{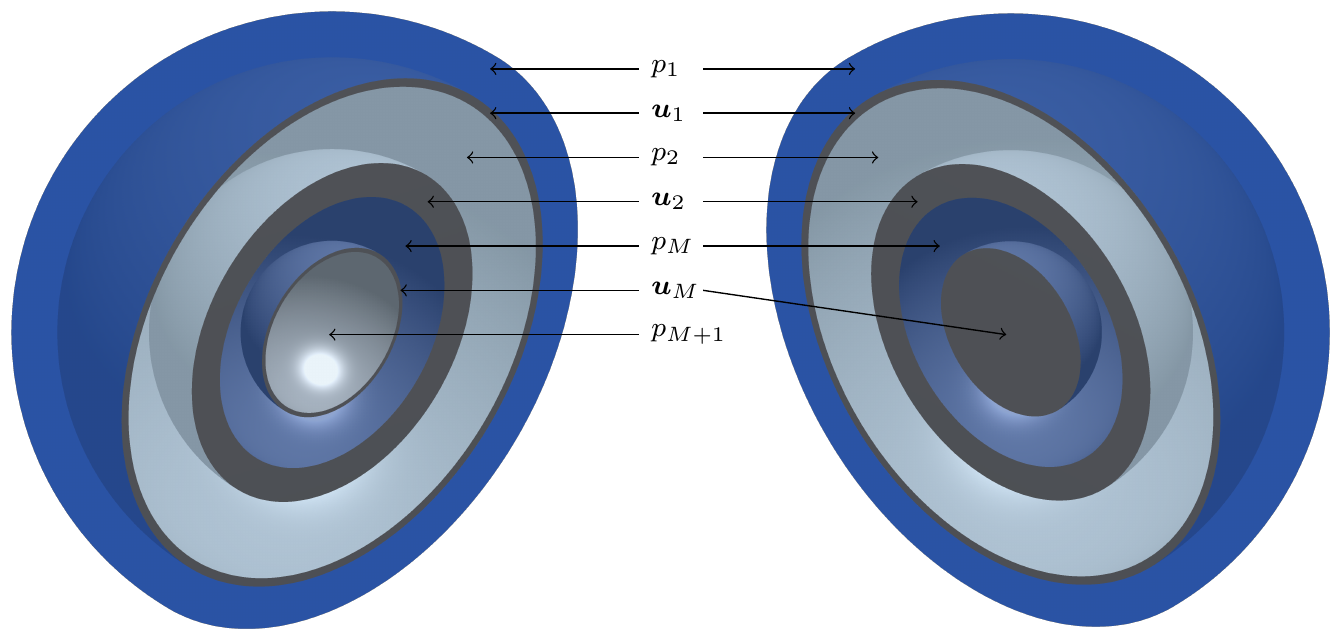}
	\caption{Illustration (clip view) of a model (to the left) with 3 steel shells and a model (to the right) with 2 steel shells surrounding a solid steel sphere, illustrating the distribution of functions (in the case $M=3$). The model to the left models a fluid as the innermost domain, while the model to the right models a solid domain as the innermost domain (note that the expression $\vec{u}_M$ is slightly altered in this case).}
	\label{Fig1:function_distribution}
\end{figure}

In this sub section, the final expressions has been summarized (see \Cref{Fig1:function_distribution}).
Recall that $\hankel_n^{(i)}$, $Z_{j,n}^{(i)}$, $S_{j,n}^{(i)}$ and $T_{j,n}^{(i)}$ are all derived from spherical Bessel functions ($\besselj_n$ and $\bessely_n$), while $Q_n^{(i)}$ are derived from Legendre functions. All coefficients ($A_{m,n}^{(i)}$, $B_{m,n}^{(i)}$ and $C_{m,n}^{(i)}$) are found by solving the linear system of equations in \Cref{Eq1:SystemOfEquations}. The scattered pressure field in the outermost (unbounded) fluid domain, the $m^{\mathrm{th}}$ fluid layer (for $2\leq m \leq M$), and the innermost fluid domain (if present), are given by
\begin{align}\label{Eq1:Summary_p1}
p_1(r,\vartheta) &= \sum_{n=0}^\infty Q_n^{(0)}(\vartheta) C_{1,n}^{(1)} \hankel^{(1)}_n(k_1 r)\\\label{Eq1:Summary_pm}
p_m(r,\vartheta) &= \sum_{n=0}^\infty Q_n^{(0)}(\vartheta)C_{m,n}^{(i)} Z_n^{(i)}(k_m r)\\
	p_{M+1}(r,\vartheta) &= \sum_{n=0}^\infty Q_n^{(0)}(\vartheta) C_{M+1,n}^{(1)} \besselj_n(k_{M+1} r),\label{Eq1:Summary_pM1}
\end{align}
respectively. The displacement field in the $m^{\mathrm{th}}$ solid domain is given by
\begin{equation}
	\vec{u}_m = u_{\mathrm{r},m} \vec{e}_{\mathrm{r}} + u_{\upvartheta,m} \vec{e}_\upvartheta\label{Eq1:Summary_u}
\end{equation}
where
\begin{align}
	u_{\mathrm{r},m}(r,\vartheta) &= \frac{1}{r}\sum_{n=0}^\infty Q_n^{(0)}(\vartheta)\left[A_{m,n}^{(i)}S_{1,n}^{(i)}(a_m r)+B_{m,n}^{(i)}T_{1,n}^{(i)}(b_m r)\right]\\
	u_{\upvartheta,m}(r,\vartheta) &= \frac{1}{r}\sum_{n=0}^\infty Q_n^{(1)}(\vartheta)\left[A_{m,n}^{(i)}S_{2,n}^{(i)}(a_m r)+B_{m,n}^{(i)}T_{2,n}^{(i)}(b_m r)\right].
\end{align} 
If the inner domain is a solid domain, the terms involving $S_{1,n}^{(2)}$ and $T_{1,n}^{(2)}$ in $\vec{u}_M$, are not present.
\section{Computational aspects}
\label{Sec1:computationalAspects}
Several computational issues arise when implementing the exact solution (which has been implemented in \href{mathworks.com}{MATLAB}). The source code can be downloaded from GitHub \href{https://github.com/Zetison/e3Dss}{here}. In this section, a discussion of some of these issues will be presented.

\subsection{Matrix manipulations}
Note that the only complex valued matrix entries of $\vec{H}_n$ are the first two entries in the first column. So instead of using a complex matrix solution routine to solve the system, one can exploit this fact to solving a real valued linear system of equations with two right hand sides. Refer to Fender~\cite[pp. 18-20]{Fender1972sfa} for details. Moreover, Fender shows that some further matrix manipulation may reduce the overall computational time by 30\% (when doing a frequency sweep). By using the same ideas, the size of $\vec{H}_n$ can be reduced from $6M$ to $4M$.

\begin{figure}
	\centering
	\begin{subfigure}[t]{0.48\textwidth}
		\centering
		\includegraphics[scale=1]{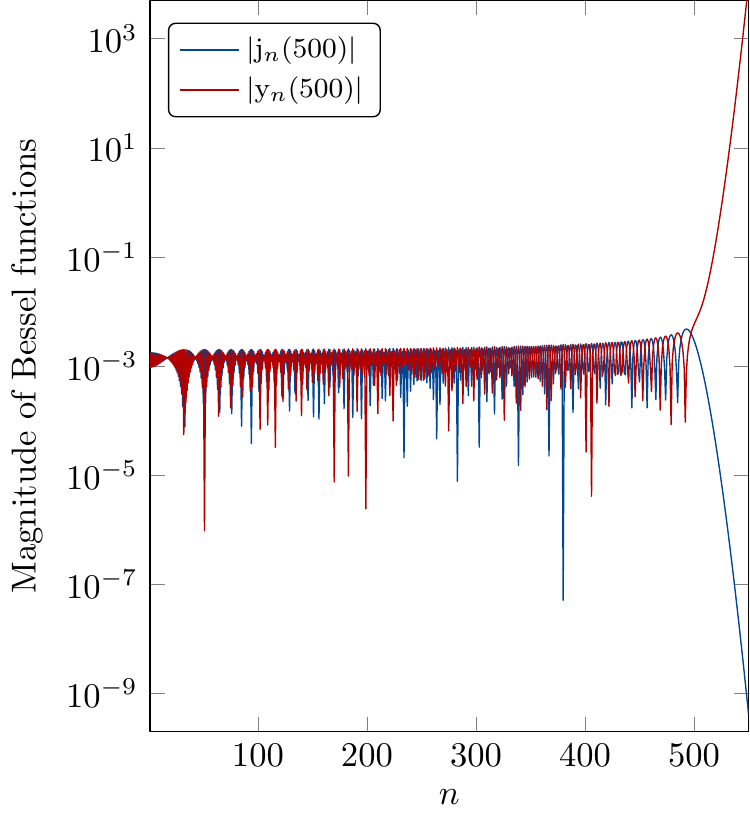}
		\caption{Magnitude of Bessel functions plotted for $n\in[0,550]$.}
	\end{subfigure}
	~
	\begin{subfigure}[t]{0.48\textwidth}
		\centering
		\includegraphics[scale=1]{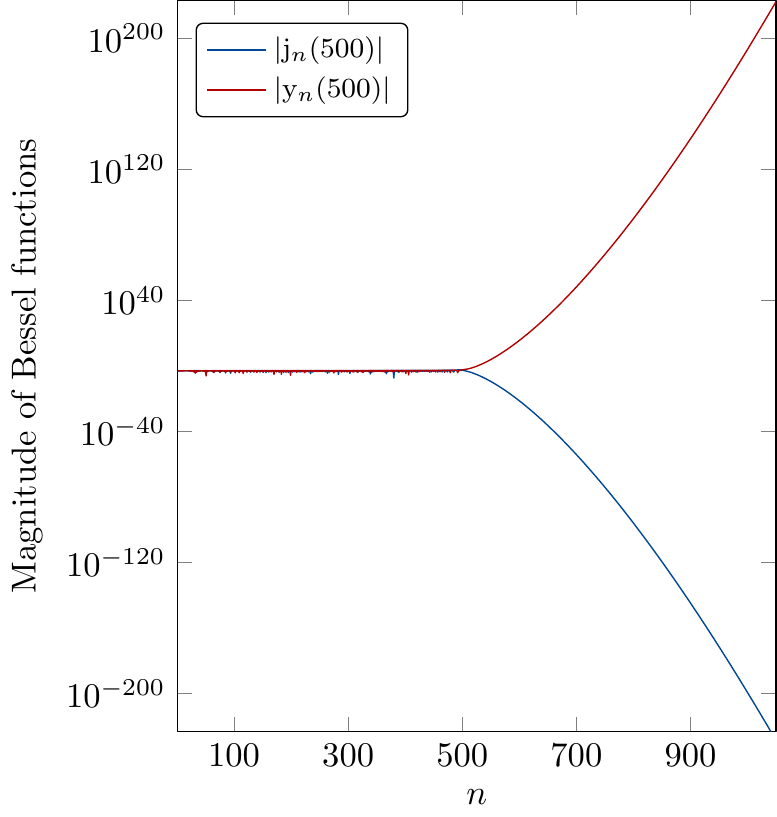}
		\caption{Magnitude of Bessel functions plotted for $n\in[0,1050]$.}
	\end{subfigure}
	\caption{Illustration of the asymptotic behavior, of the spherical Bessel functions of first ($\besselj_n(x)$) and second ($\bessely_n(x)$) kind, as a function of $n$, for a fixed argument $x=500$.}
	\label{Fig1:besselPlotLargeN}
\end{figure}

\begin{figure}
	\centering
	\begin{subfigure}{\textwidth}
		\centering
		\includegraphics[scale=1]{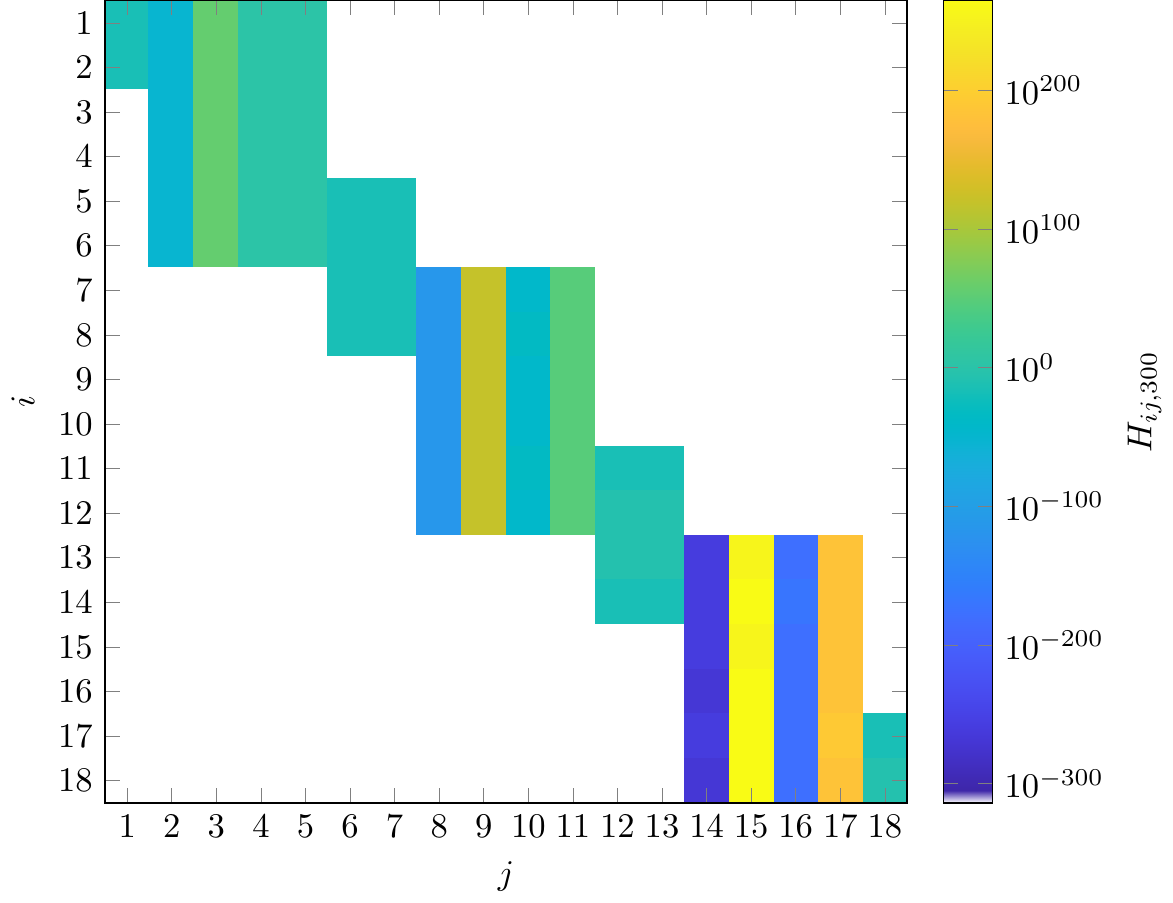}
		\caption{Before preconditioning.}
		\label{Fig1:Spy2K}
	\end{subfigure}
	\par\bigskip
	\begin{subfigure}{\textwidth}
		\centering
		\includegraphics[scale=1]{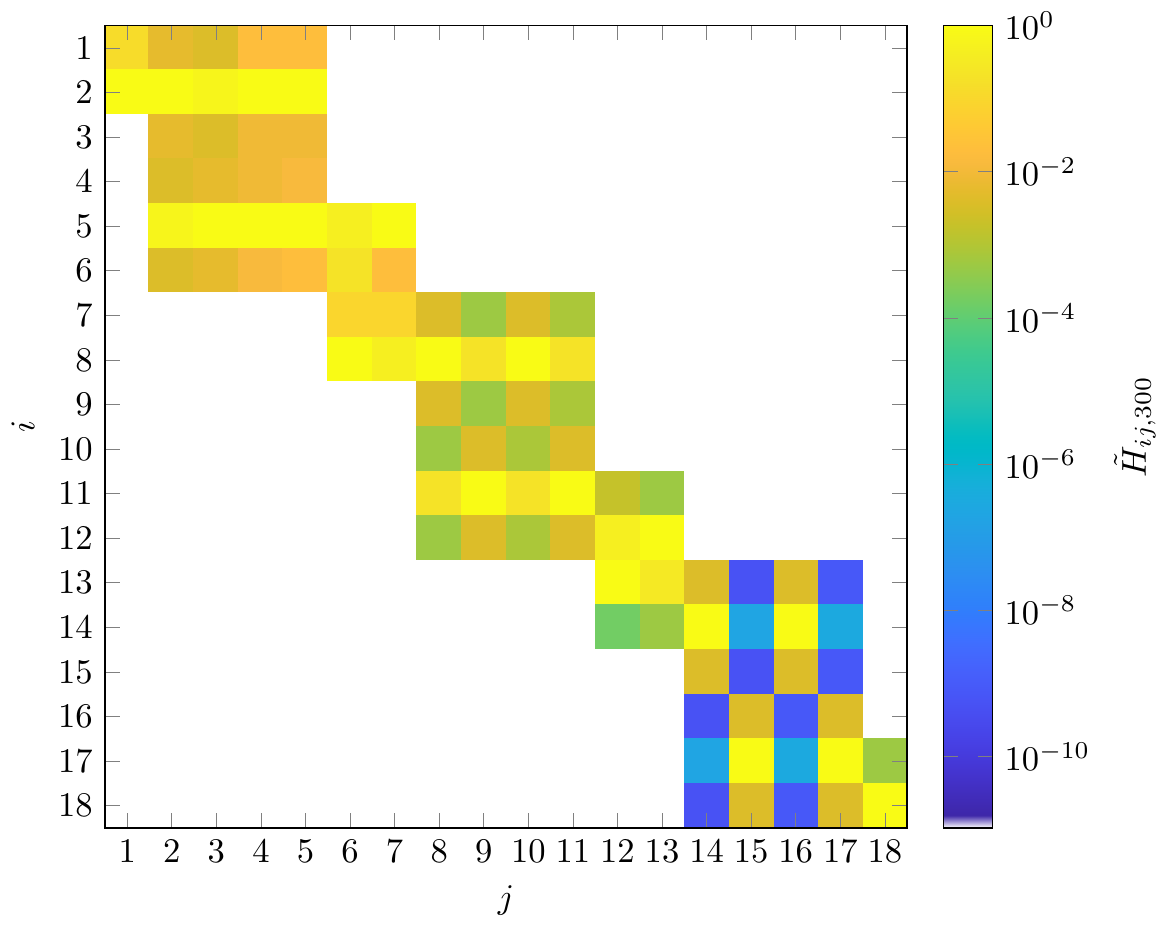}
		\caption{After preconditioning.}
		\label{Fig1:Spy2K2}
	\end{subfigure}
	\caption{\textbf{Matrix manipulations}: Plot of the magnitude of the matrix entries of $\vec{H}_{300}$ and $\tilde{\vec{H}}_{300}$.}
	\label{Fig1:SpyK}
\end{figure}

Note that for $n>0$, column number $2l$, $l=1, 2, \dots, 3M$, of $\vec{H}_n$ contains entries which are linear combinations of $\besselj_n$ and $\besselj_{n+1}$ (and no spherical Bessel functions of second kind), while column number $2l-1$, $l=1, 2, \dots, 3M$, of $\vec{H}_n$ contains entries which are linear combinations of $\bessely_n$ and $\bessely_{n+1}$. So since
\begin{equation}
	\lim_{n\to\infty} |\besselj_n(\zeta)| = 0\quad\text{and}\quad \lim_{n\to\infty} |\bessely_n(\zeta)| = \infty,
\end{equation}
(which is illustrated in \Cref{Fig1:besselPlotLargeN})
the matrix $\vec{H}_n$ becomes poorly scaled for large $n$. This issue can be solved by scaling the matrix with a (diagonal) preconditioning matrix $\vec{P}_n$ where the diagonal entries are given by the maximal modulus of the corresponding column vectors of $\vec{H}_n$. Defining the vector $\tilde{\vec{C}}_n = \vec{P}_n\vec{C}_n$ and solving the system $\tilde{\vec{H}}_n\tilde{\vec{C}}_n = \vec{D}_n$ with $\tilde{\vec{H}}_n = \vec{H}_n \vec{P}_n^{-1}$, the solution is obtained by $\vec{C}_n = \vec{P}_n^{-1}\tilde{\vec{C}}_n$. In \Cref{Fig1:Spy2K,Fig1:Spy2K2} the magnitude of the entries in $\vec{H}_n$ is visualized before and after preconditioning, respectively.
This example is the matrix $\vec{H}_{300}$ of the S135 benchmark problem (described in \Cref{Subsec1:benchmarkProblem}) at $f=\SI{30}{kHz}$. The condition number was improved from $\operatorname{cond}(\vec{H}_{300}) \approx 7.4\cdot 10^{278}$ to $\operatorname{cond}(\tilde{\vec{H}}_{300}) \approx 9.4\cdot 10^4$.

\subsection{Series evaluation}
\label{Subsec1:seriesEval}
As the series involves summation over infinitely many terms, the series needs to be truncated at some number $n=N_\varepsilon$. Denote by $p_1^{(N)}$, the truncated sum for the scattered pressure in the outer domain (\Cref{Eq1:Summary_pm}), 
\begin{align}\label{Eq1:truncated_p1}
p_1^{(N)}(r,\vartheta) &= \sum_{n=0}^N Q_n^{(0)}(\vartheta) C_{1,n}^{(1)} \hankel^{(1)}_n(k_1 r),
\end{align}
and correspondingly for the other fields in \Cref{Eq1:Summary_pm,Eq1:Summary_pM1,Eq1:Summary_u}.
In~\cite[pp. 32-35]{Ihlenburg1998fea}, Ihlenburg discusses such a value based on the decay of Bessel functions, in which he suggests to use $N\approx 2kr$. In this work, however, the summation is terminated whenever the magnitude of term $n=N_\varepsilon$ divided by the magnitude of the partial sum (based on the first $N_\varepsilon+1$ terms) is less than some prescribed tolerance $\varepsilon$. Typically machine epsilon is used for this number, i.e. $\varepsilon \approx 2.220446\cdot 10^{-16}$. As with the suggestion of Ihlenburg, the number of terms, $N_\varepsilon$, grows linearly with the frequency. The computational complexity of the problem is thus $\bigoh(\omega)$. 

The solution is often needed at several points, or frequencies (or a combination of both). In this case, one should compute the solutions at all points at once, such that the calls to the implemented Bessel functions are minimized.

\subsection{Round-off errors}
\label{Subsec1:RoundoffErrors}
\begin{figure}
	\centering
	\begin{subfigure}[t]{0.48\textwidth}
		\includegraphics[scale=0.95]{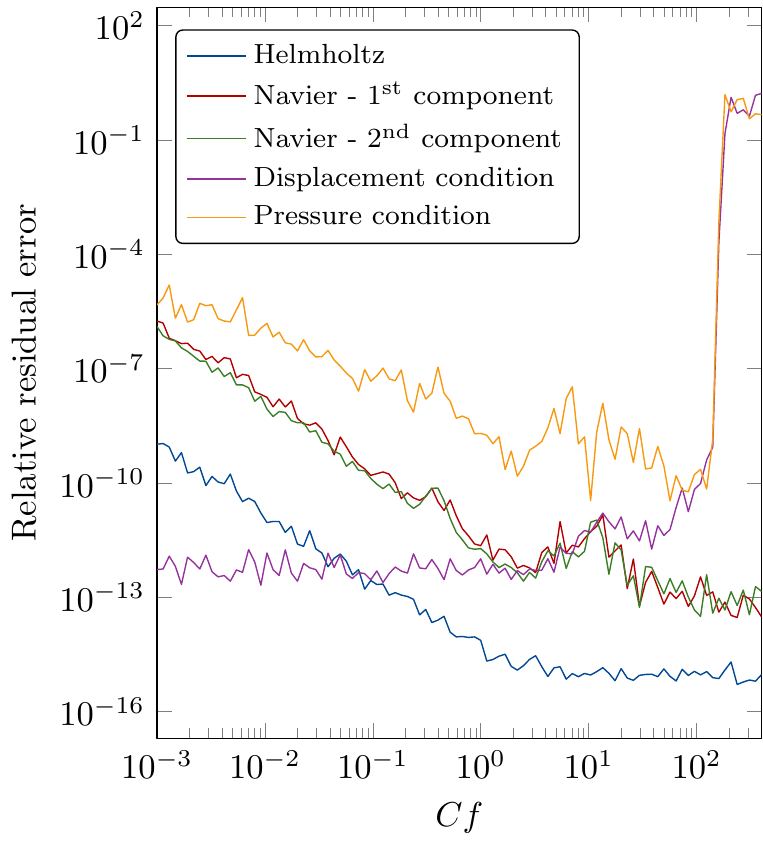}
		\caption{Double precision, $\varepsilon \approx 10^{-16}$}
	\end{subfigure}
	~
	\begin{subfigure}[t]{0.48\textwidth}
		\includegraphics[scale=0.95]{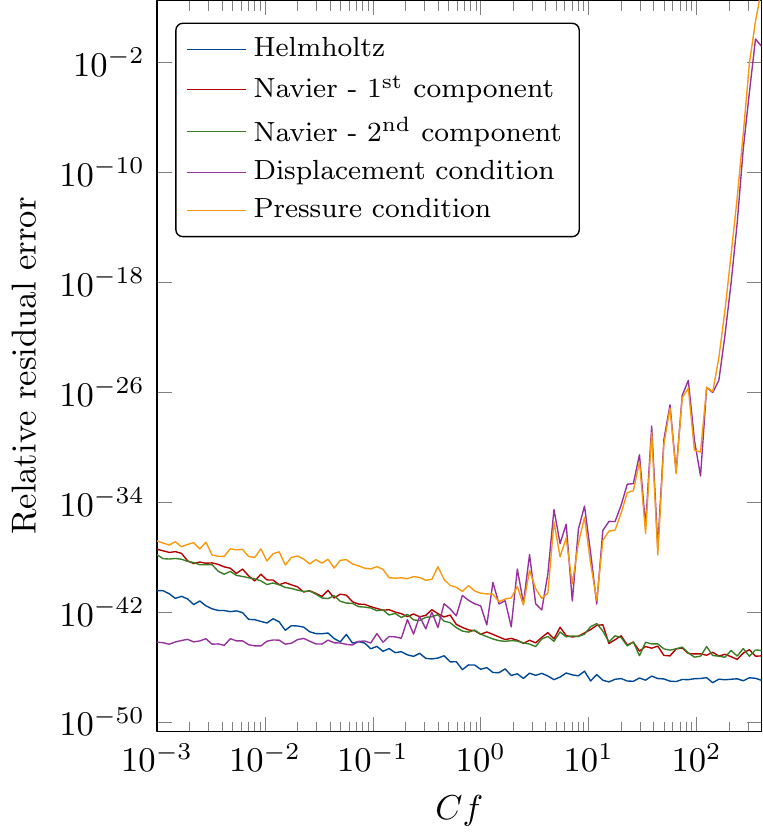}
		\caption{Symbolic precision $\varepsilon \approx 10^{-40}$}
	\end{subfigure}
	\caption{\textbf{Round-off errors}: Residual errors for the governing equations and boundary conditions. The use of symbolic precision in MATLAB illustrate that the errors are due to round-off errors. The relative residual formulas for the Helmholtz equation, the first and second components of the Navier equation, the displacement condition and the pressure conditions are given by \Cref{Eq1:residualHelmhotlz,Eq1:residualNavier1,Eq1:residualNavier2,Eq1:residualDisplacement,Eq1:residualPressure}, respectively.}
	\label{Fig1:errorsS123_ASI-6NN}
\end{figure}

Although the products $A_{m,n}^{(2)}S_{j,n}^{(2)}(\zeta)$, $B_{m,n}^{(2)}T_{j,n}^{(2)}(\zeta)$ and $C_{m,n}^{(2)}Z_n^{(2)}(\eta)$ all goes to zero as $n\to\infty$, the functions $S_{j,n}^{(2)}(a_m r)$, $T_{j,n}^{(2)}(\zeta)$ and $Z_n^{(2)}(\eta)$ does not. In fact, these functions become unbounded when $n\to\infty$ because they are all superposition of Bessel functions of second kind with this property. So since the floating point number has an upper bound\footnote{For double precision this is typically $V_{\mathrm{max}} \approx 1.797693134862316\cdot 10^{308}$.}, there is a limit to the number of terms that can be used. A naive solution to this problem is to try higher precision, which can easily be done with MATLAB symbolic class. This however, increases the computational time drastically. 
In \Cref{Fig1:errorsS123_ASI-6NN} several round-off phenomena which typically arises are illustrated.
The specific example used here is the S135 benchmark problem with SSBC (described in \Cref{Subsec1:benchmarkProblem}). The incident wave, $p_{\mathrm{inc}}$, is a plane wave traveling in the direction given by $\vartheta=\ang{60}$ and $\varphi=\ang{240}$ (see \Cref{Sec1:resultsDisc}). An uniform (relative to the spherical coordinate system) set of sample points are distributed in all domains\footnote{It is placed 32 points in each domain except for the inner domain with 25 points. The distribution of point in the radial direction in the exterior domain is limited to the interval $[R_{0,1}, 2R_{0,1}]$.} where the residual error in the Helmholtz equation (\Cref{Eq1:helmholtz}), the $1^{\mathrm{st}}$ and $2^{\mathrm{nd}}$ component of Navier's equation in spherical coordinates (\Cref{Eq1:navierSphericalSimplified1,Eq1:navierSphericalSimplified2}, respectively), the displacement condition (\Cref{Eq1:firstBC}) and the pressure condition (\Cref{Eq1:secondBC}), is measured. By using the infinity norm, $\|\cdot\|_\infty$, for each residual, and dividing by the magnitude of the terms involved, the relative residual error is obtained. The maximal relative residual in all domains can then be calculated. In particular, these relative residual errors are given by
\begin{align}
	&\max_{1\leq m\leq M+1}\frac{\left\|\left(\nabla^2 + k_m^2\right)p_m\right\|_\infty}{\left\|k_m^2p_m\right\|_\infty}\label{Eq1:residualHelmhotlz}\\
	&\max_{1\leq m\leq M}\frac{\left\|\pderiv{\sigma_{\mathrm{rr},m}}{r} + \frac{1}{r}\pderiv{\sigma_{\mathrm{r}\upvartheta,m}}{\vartheta} + \frac{1}{r}\left(2\sigma_{\mathrm{r}\mathrm{r},m} - \sigma_{\upvartheta\upvartheta,m} - \sigma_{\upvarphi\upvarphi,m} + \sigma_{\mathrm{r}\upvartheta,m}\cot\vartheta\right) + \omega^2\rho_{\mathrm{s},m}u_{\mathrm{r},m}\right\|_\infty}{\left\|\omega^2\rho_{\mathrm{s},m}u_{\mathrm{r},m}\right\|_\infty}\label{Eq1:residualNavier1}\\
	&\max_{1\leq m\leq M}\frac{\left\|\pderiv{\sigma_{\mathrm{r}\upvartheta,m}}{r} + \frac{1}{r}\pderiv{\sigma_{\upvartheta\upvartheta,m}}{\vartheta} + \frac{1}{r}\left[(\sigma_{\upvartheta\upvartheta,m} - \sigma_{\upvarphi\upvarphi,m})\cot\vartheta + 3\sigma_{\mathrm{r}\upvartheta,m} \right] +\omega^2\rho_{\mathrm{s},m}u_{\upvartheta,m}\right\|_\infty}{\left\|\omega^2\rho_{\mathrm{s},m}u_{\upvartheta,m}\right\|_\infty}\label{Eq1:residualNavier2}\\
	&\max_{1\leq m\leq M}\frac{\left\|\rho_{\mathrm{f}} \omega^2 u_{\mathrm{r}} - \pderiv{p_{\mathrm{tot}}}{r}\right\|_\infty}{\left\|\pderiv{p_{\mathrm{tot}}}{r}\right\|_\infty}\label{Eq1:residualDisplacement}\\
	&\max_{1\leq m\leq M}\frac{\left\|\sigma_{\mathrm{rr}} + p_{\mathrm{tot}}\right\|_\infty}{\left\|p_{\mathrm{tot}}\right\|_\infty}.\label{Eq1:residualPressure}
\end{align}
By comparing these error results for both double precision and symbolic precision in MATLAB, one can conclude that the errors indeed originates from round-off errors. When using double precision, the summation is ended whenever $|\bessely_n(\eta)| > 10^{290}$, such that invalid solutions is obtained for sufficiently large $n$. Since one needs to have enough terms for the solution to converge, and at the same time have to avoid computing $\bessely_n(\eta)$ for low $\eta$ (and large $n$), the following bound on the frequency based on experimental data is suggested
\begin{equation}\label{Eq1:Upsilon}
	f \lesssim \frac{100}{C}  \quad\text{where}\quad C =\left(\frac{R_{0,1}}{c_{\mathrm{f},1}}\right)^{\frac{3}{2}}\frac{1}{\sqrt{\Upsilon}},\quad\Upsilon = \min\left\{\min_{1\leq m\leq M}\frac{R_{1,m}}{\max\{c_{\mathrm{s,1},m},c_{\mathrm{s,2},m}\}},\min_{1\leq m\leq M}\frac{R_{0,m}}{c_{\mathrm{f},m}}\right\},
\end{equation}
where $c_{\mathrm{s,1},M}$ and $c_{\mathrm{s,2},M}$ is the transverse and longitudinal wave velocity for the $M^{\mathrm{th}}$ spherical shell, respectively. The constant $\Upsilon$ corresponds to the lowest argument $\eta$ used for the Bessel functions of second kind. An addendum will be given to yield more numerical evidence for this bound. In particular, plots similar to the ones in \Cref{Fig1:errorsS123_ASI-6NN} will be presented for all benchmarks and corresponding boundary conditions in \Cref{Subsec1:benchmarkProblem}. However, it would certainly be possible to construct models in which this bound is not valid.

One can also observe significant round-off errors for very low frequencies which is again due to the evaluation of the spherical Bessel functions of the second kind with the property
\begin{equation}
	\lim_{\zeta\to 0} |\bessely_n(\zeta)| = \infty.
\end{equation}
Finally, observe that problems for higher frequencies also occur when using the symbolic class in MATLAB (even though the summation is not terminated prematurely), which calls for a more mathematically sound way of solving this issue. To avoid evaluating the Bessel functions directly, one could include a scaling such that one need to evaluate products of the form $\besselj_n(\xi)\bessely_n(\eta)$, $\besselj_{n+1}(\xi)\bessely_n(\eta)$, $\besselj_n(\xi)\bessely_{n+1}(\eta)$ and $\besselj_{n+1}(\xi)\bessely_{n+1}(\eta)$ (which will be $0\cdot\infty$ type products). One would then probably need to use relations like~\cite[\href{http://functions.wolfram.com/03.21.26.0047.01}{03.21.26.0047.01} and \href{http://functions.wolfram.com/03.21.26.0049.01}{03.21.26.0049.01}]{WolframResearch2016m}
\begin{align}
\besselj_n(\sqrt{z})\bessely_n(\sqrt{z}) &=-\frac{\sqrt{\PI}}{2}G_{1,3}^{2,0}\left(z\Bigg\vert\begin{matrix}
	0\\
	-\frac{1}{2}, n, -n-1
	\end{matrix}\right)\\
	\besselj_{n+1}(\sqrt{z})\bessely_n(\sqrt{z}) &=\frac{\sqrt{\PI}}{2}G_{2,4}^{2,1}\left(z\Bigg\vert\begin{matrix}
	0,-\frac{1}{2}\\
	0, n+\frac{1}{2},-1,-n-\frac{3}{2}
	\end{matrix}\right)
\end{align}
where $G$ is the Meijer G-function. This investigation is left as future work.

\section{Numerical examples} 
\label{Sec1:resultsDisc}
To give further evidence for the correctness of the implemented code, comparison to existing benchmark solutions by Chang~\cite{Chang1994soa}, Ihlenburg~\cite{Ihlenburg1998fea} and Fender~\cite{Fender1972sfa}, will be presented. A final benchmark problem in the time-domain will be added.

It is customary to present results in the \textit{far-field}. For the scattered pressure $p_1$, it is defined by
\begin{equation}
	p_0(\hat{\vec{x}},\omega) =  r \euler^{-\imag k_1 r}p_1(\vec{x},\omega),\quad r = |\vec{x}| \to \infty,
\end{equation}
with $\hat{\vec{x}} = \vec{x}/|\vec{x}|$. As a side note, using \Cref{Eq1:sphericalHankelLimit}, the far-field pattern of the scattered pressure in \Cref{Eq1:outerFluid}, is given by (in the axisymmetric case)
\begin{equation}
	p_0 = \frac{1}{k_1}\sum_{n=0}^\infty \imag^{-n-1} Q_n^{(0)}(\vartheta) C_{1,n}^{(1)}
\end{equation}
which yields a very efficient way of computing the far-field pattern.

From the far-field pattern, the \textit{target strength}, $\TS$, can be computed. It is defined by
\begin{equation}\label{Eq1:TS}
	\TS = 20\log_{10}\left(\frac{|p_0(\hat{\vec{x}},\omega)|}{|P_{\mathrm{inc}}(\omega)|}\right)
\end{equation}
where $P_{\mathrm{inc}}$ is the amplitude of the incident wave at the geometric center of the scatterer (i.e. the origin). Note that $\TS$ is independent of $P_{\mathrm{inc}}$.

The directional vector, $\vec{d}_{\mathrm{s}}$, in spherical coordinates, is given by
\begin{equation}
	\vec{d}_{\mathrm{s}} = -\begin{bmatrix}
		\sin\vartheta_{\mathrm{s}}\cos\varphi_{\mathrm{s}}\\
		\sin\vartheta_{\mathrm{s}}\sin\varphi_{\mathrm{s}}\\
		\cos\vartheta_{\mathrm{s}}		
	\end{bmatrix}.
\end{equation}
If the source of the incident wave is located at
\begin{equation}\label{Eq1:x_s}
	\vec{x}_{\mathrm{s}} = -r_{\mathrm{s}}\vec{d}_{\mathrm{s}},
\end{equation}
the far-field pattern of an incident wave from the point source
\begin{equation}
	p_{\mathrm{inc}}(\vec{x},\omega) = P_{\mathrm{inc}}(\omega)\frac{\euler^{\imag k_1 |\vec{x}_{\mathrm{s}}-\vec{x}|}}{|\vec{x}_{\mathrm{s}}-\vec{x}|},
\end{equation}
is actually a plane wave
\begin{equation}
	\lim_{r_{\mathrm{s}}\to\infty} r_{\mathrm{s}}\euler^{-\imag k_1 r_{\mathrm{s}}}p_{\mathrm{inc}}(\vec{x},\omega) = P_{\mathrm{inc}}(\omega)\euler^{\imag k_1\vec{d}_{\mathrm{s}}\cdot\vec{x}}.
\end{equation}
Unless stated otherwise, plane waves will be used for the incident wave. Note that the direction of plane waves and location of far-field points is often expressed in the \textit{aspect angle}, $\alpha=\varphi$, and the \textit{elevation angle}, $\beta=\ang{90}-\vartheta$.

\subsection{Chang benchmark problem} 
\begin{figure}
	\centering
	\begin{subfigure}[t]{\textwidth}
		\centering
		\includegraphics{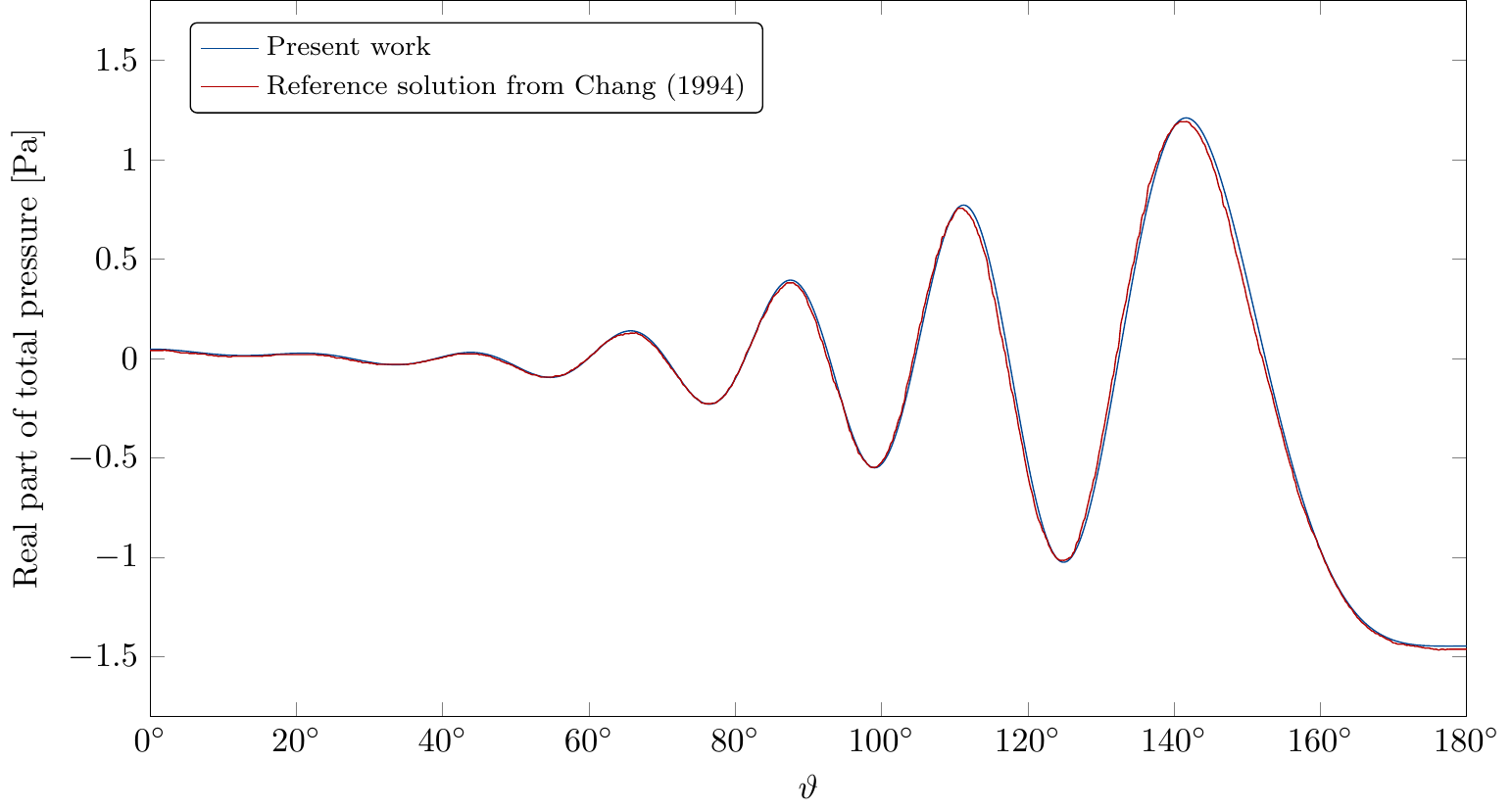}
		\caption{Wave number $k_1=\SI{15}{m^{-1}}$ and series truncation at $N_\varepsilon = 46$.}
		\label{Fig1:Chang1}
	\end{subfigure}
	\par\bigskip
	\begin{subfigure}[t]{\textwidth}
		\centering
		\includegraphics{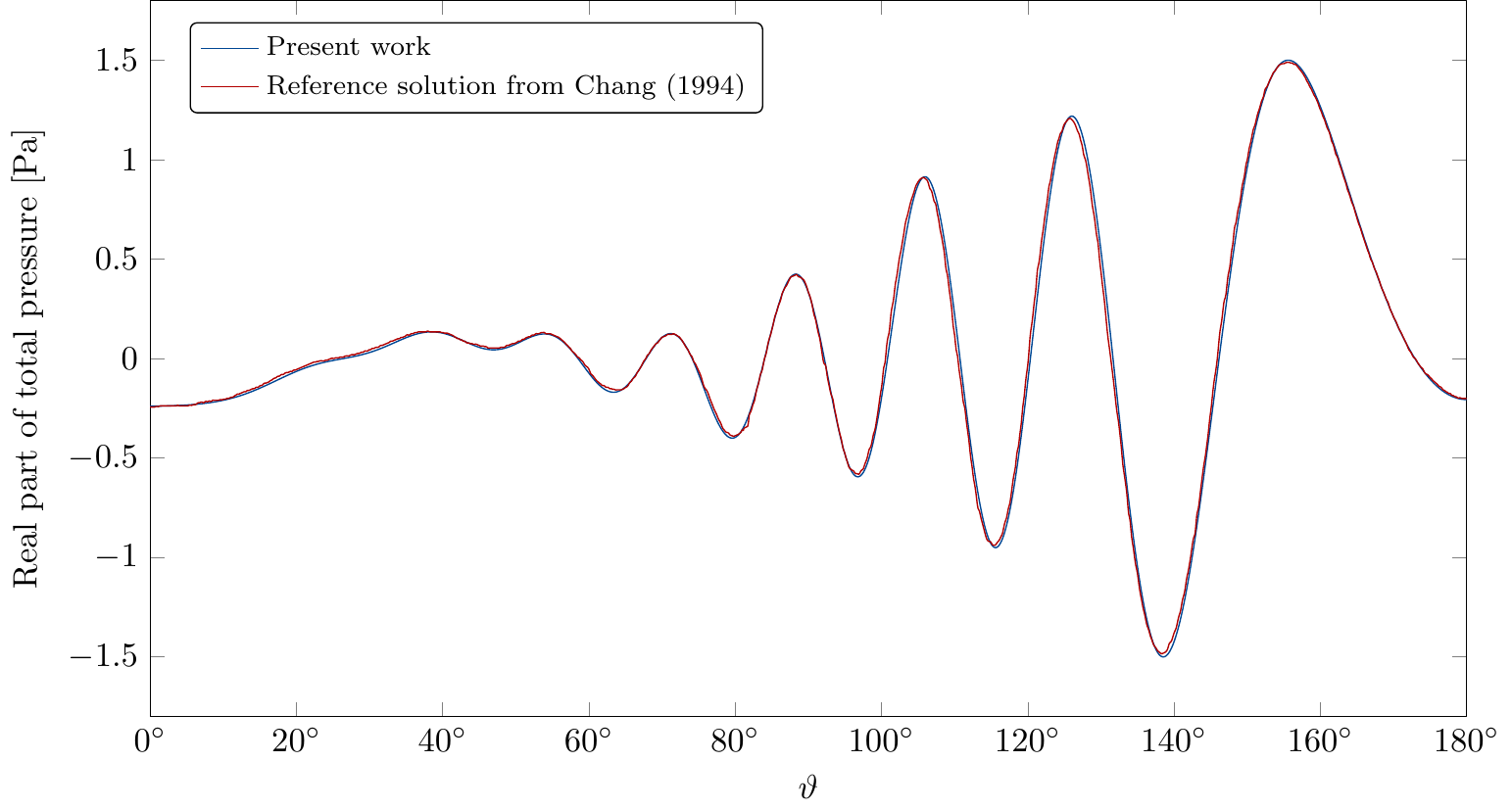}
		\caption{Wave number $k_1=\SI{20}{m^{-1}}$ and series truncation at $N_\varepsilon = 54$.}
		\label{Fig1:Chang2}
	\end{subfigure}
	\caption{\textbf{Chang benchmark problem}: Predicted total pressure as a function of the polar angle $\vartheta$.}
\end{figure}

Chang~\cite{Chang1994soa} considers a single spherical shell, with a single homogeneous Neumann condition (sound-soft boundary conditions, SSBC) on the inside of the shell, scattering an incident plane wave (with amplitude $P_{\mathrm{inc}}=\SI{1}{Pa}$). Chang sends the incident plane wave along the positive $x_3$-axis, and uses the parameters in \Cref{Tab1:Chang}.
\begin{table}
	\centering
	\caption{\textbf{Chang parameters:} Parameters for the examples in figure 16 and figure 17 in \cite{Chang1994soa}.}
	\label{Tab1:Chang}
	\begin{tabular}{l l}
		\toprule
		Parameter & Description\\
		\midrule
		$E = \SI{2.0e11}{Pa}$ & Young's modulus\\
		$\nu = 0.3$ & Poisson's ratio\\
		$\rho_{\mathrm{s}} = \SI{7800}{kg.m^{-3}}$ & Density of solid\\
		$\rho_{\mathrm{f}} = \SI{1000}{kg.m^{-3}}$ & Density of water\\
		$c_{\mathrm{f},1} = \SI{1460}{m.s^{-1}}$ & Speed of sound in fluid\\
		$R_{0,1} = \SI{1.005}{m}$ & Outer radius of spherical shell\\
		$R_{1,1} = \SI{0.995}{m}$ & Inner radius of spherical shell\\
		\bottomrule
	\end{tabular}
\end{table}
Moreover, the total pressure (\Cref{Eq1:totPressure}) is measured at the surface. In \Cref{Fig1:Chang1,Fig1:Chang2} the results are found with $k=\SI{15}{m^{-1}}$ and $k=\SI{20}{m^{-1}}$, respectively\footnote{The discrepancies probably comes from the fact that the data set is collected by the software \href{https://automeris.io/WebPlotDigitizer/}{WebPlotDigitizer} where a digital scan of figure 16 and figure 17~\cite[pp. 32-33]{Chang1994soa} has been made.}. In both cases, the shadow region of the scatterer, $\vartheta\in [0,\ang{90}]$, is clearly visible (with total pressure significantly lower than $P_{\mathrm{inc}}$).

\begin{figure}
	\centering
	\includegraphics{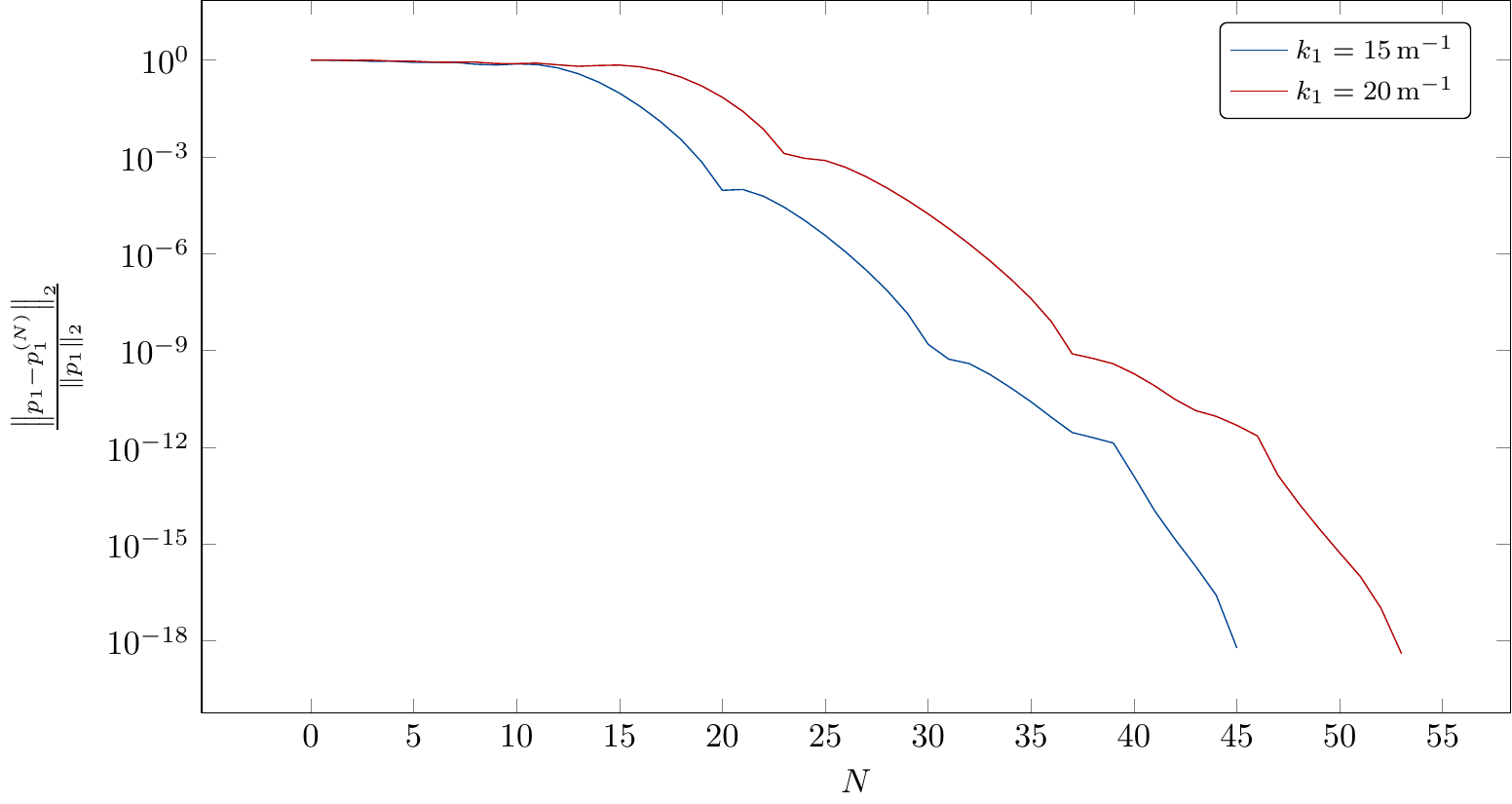}
	\caption{\textbf{Chang benchmark problem}: Relative error (with 2000 sample points uniformly placed in the $\vartheta$-direction) of the truncated series in \Cref{Eq1:truncated_p1} as a function of $N$.}
	\label{Fig1:ChangErrors}
\end{figure}

A simple convergence analysis is shown in \Cref{Fig1:ChangErrors} where the error in the truncated series in \Cref{Eq1:truncated_p1} is plotted. As discussed in \Cref{Subsec1:seriesEval} the convergence is delayed by the increased frequency from $k_1=\SI{15}{m^{-1}}$ to $k_1=\SI{20}{m^{-1}}$. To obtain machine epsilon precision (double precision) $N_\varepsilon=45$ and $N_\varepsilon=53$ is needed for these frequencies, respectively.

\subsection{Ihlenburg benchmark problem} 
Ihlenburg~\cite{Ihlenburg1998fea} considers a single spherical shell with a single homogeneous Neumann condition (sound-soft boundary conditions, SSBC) on the inside of the shell, scattering an incident plane wave. Building upon this example, the corresponding rigid scattering (sound-hard boundary conditions, SHBC) case and scattering with fluid fill will be presented (Neumann-Neumann boundary conditions, NNBC). The parameters in \Cref{Tab1:IhlenburgParameters} are here used. Frequency sweeps of the target strength (in \Cref{Eq1:TS}) are plotted in \Cref{Fig1:Fender1,Fig1:Fender2} at the polar angles $\vartheta=\ang{180}$ and $\vartheta=\ang{0}$, respectively.
\begin{table}
	\centering
	\caption{\textbf{Ihlenburg parameters:} Parameters for the Ihlenburg benchmark problem.}
	\label{Tab1:IhlenburgParameters}
	\begin{tabular}{l l}
		\toprule
		Parameter & Description\\
		\midrule
		$E = \SI{2.07e11}{Pa}$ & Young's modulus\\
		$\nu = 0.3$ & Poisson's ratio\\
		$\rho_{\mathrm{s}} = \SI{7669}{kg.m^{-3}}$ & Density of solid\\
		$\rho_{\mathrm{f}} = \SI{1000}{kg.m^{-3}}$ & Density of water\\
		$c_{\mathrm{f}} = \SI{1524}{m.s^{-1}}$ & Speed of sound in fluid\\
		$R_{0,1}=\SI{5.075}{m}$ & Outer radius\\
		$R_{1,1}=\SI{4.925}{m}$ & Inner radius\\
		\bottomrule
	\end{tabular}
\end{table}

\begin{figure}
	\centering
	\begin{subfigure}[t]{\textwidth}
		\centering
		\includegraphics{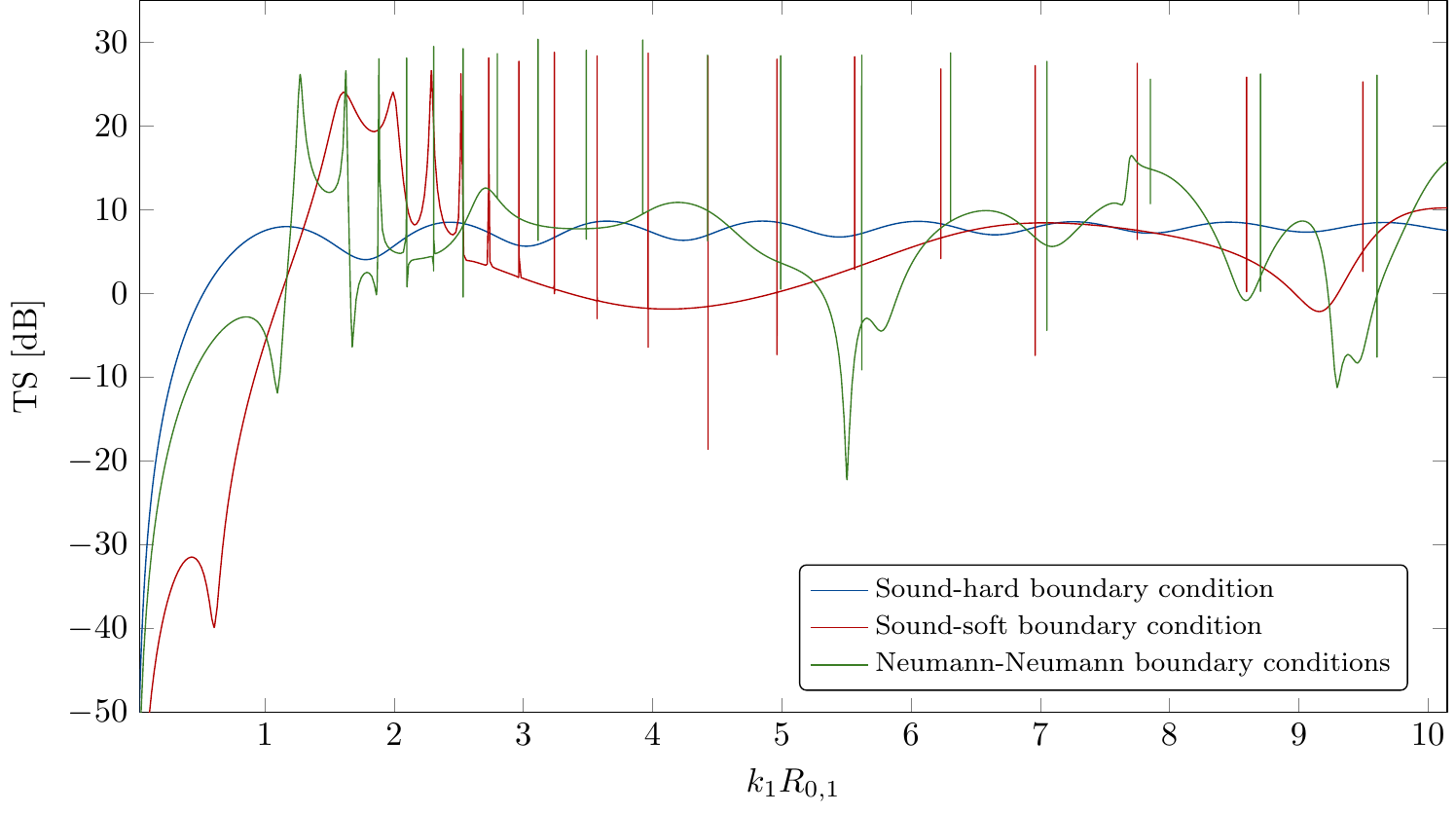}
		\caption{Measured at $\vartheta = \ang{180}$.}
		\label{Fig1:Ihlenburg1}
	\end{subfigure}
	\par\bigskip
	\begin{subfigure}[t]{\textwidth}
		\centering
		\includegraphics{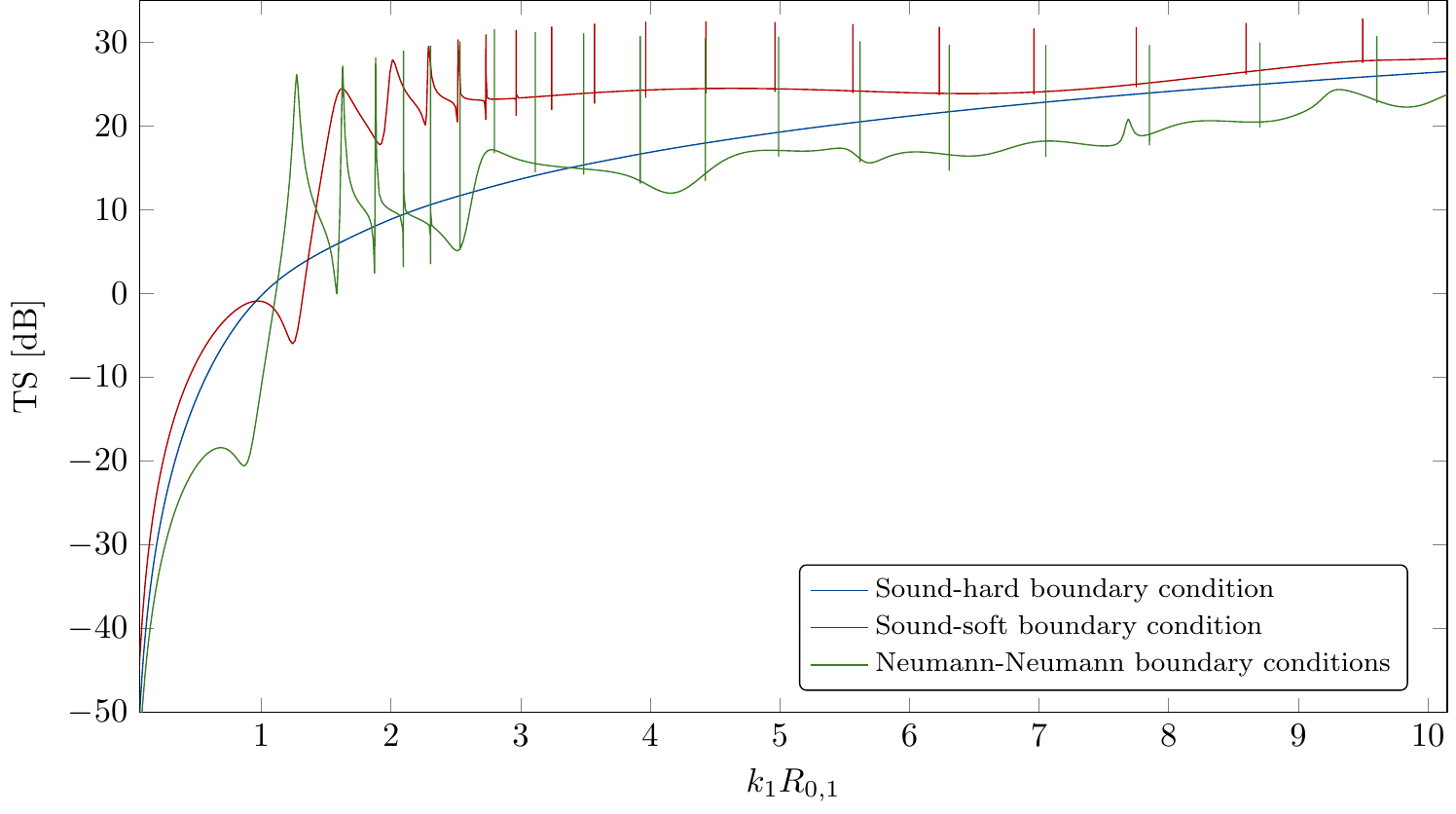}
		\caption{Measured at $\vartheta = \ang{0}$.}
		\label{Fig1:Ihlenburg2}
	\end{subfigure}
	\caption{\textbf{Ihlenburg benchmark problem}: Plots of the target strength, $\TS$. The backscattered pressure will correspond to $\vartheta=\ang{180}$, which is also the specific case considered by Ihlenburg \cite[p. 192]{Ihlenburg1998fea} (note that Ihlenburg plots the far field instead of the target strength).}
\end{figure}
\begin{figure}
	\centering
	\includegraphics{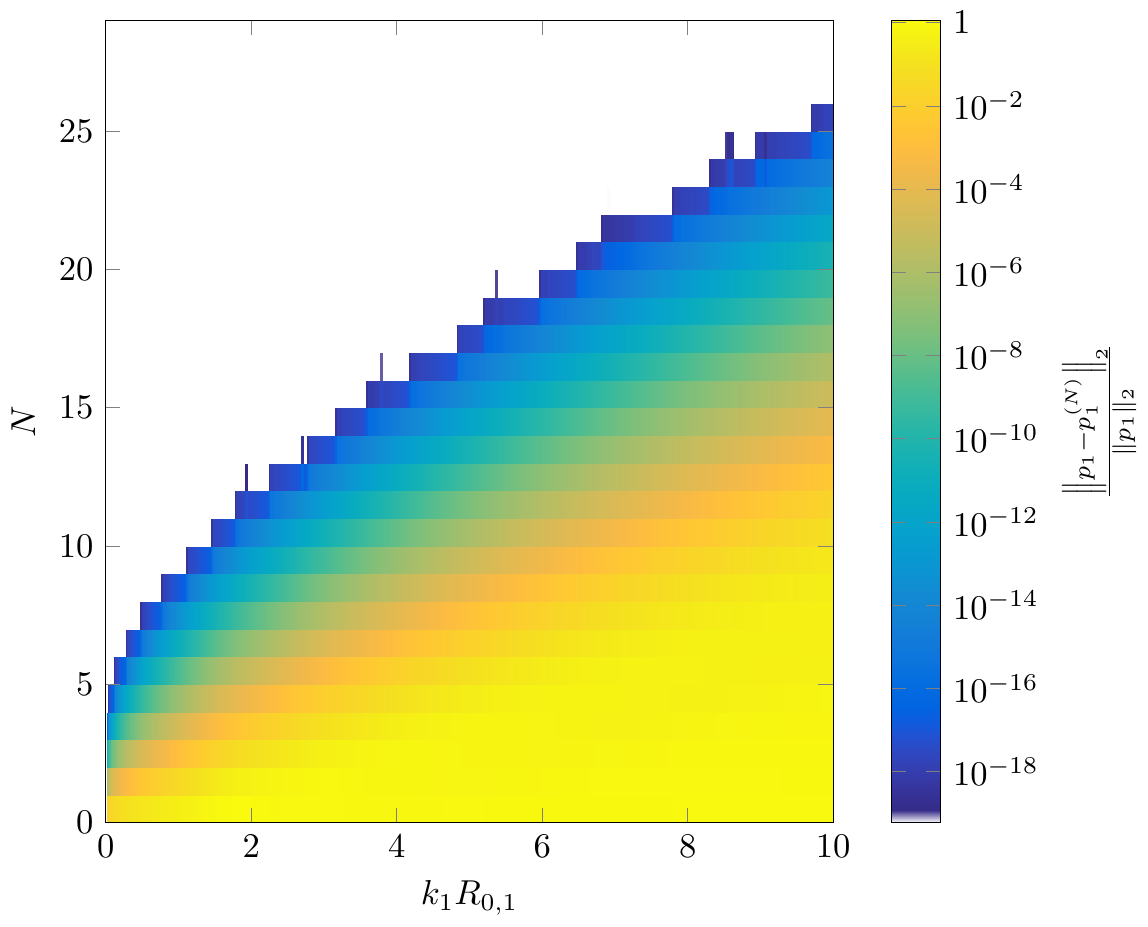}
		\caption{\textbf{Ihlenburg benchmark problem - the sound-hard case}: Relative error in the $l_2$-norm (with two sample points at $\vartheta = \ang{0}$ and $\vartheta = \ang{180}$) of the truncated series in \Cref{Eq1:truncated_p1} as a function of $N$. The ``exact'' solution, $p_1$, is obtained as described in \Cref{Subsec1:seriesEval}.}
	\label{Fig1:IhlenburgErrors1}
\end{figure}
\begin{figure}
	\centering
	\includegraphics{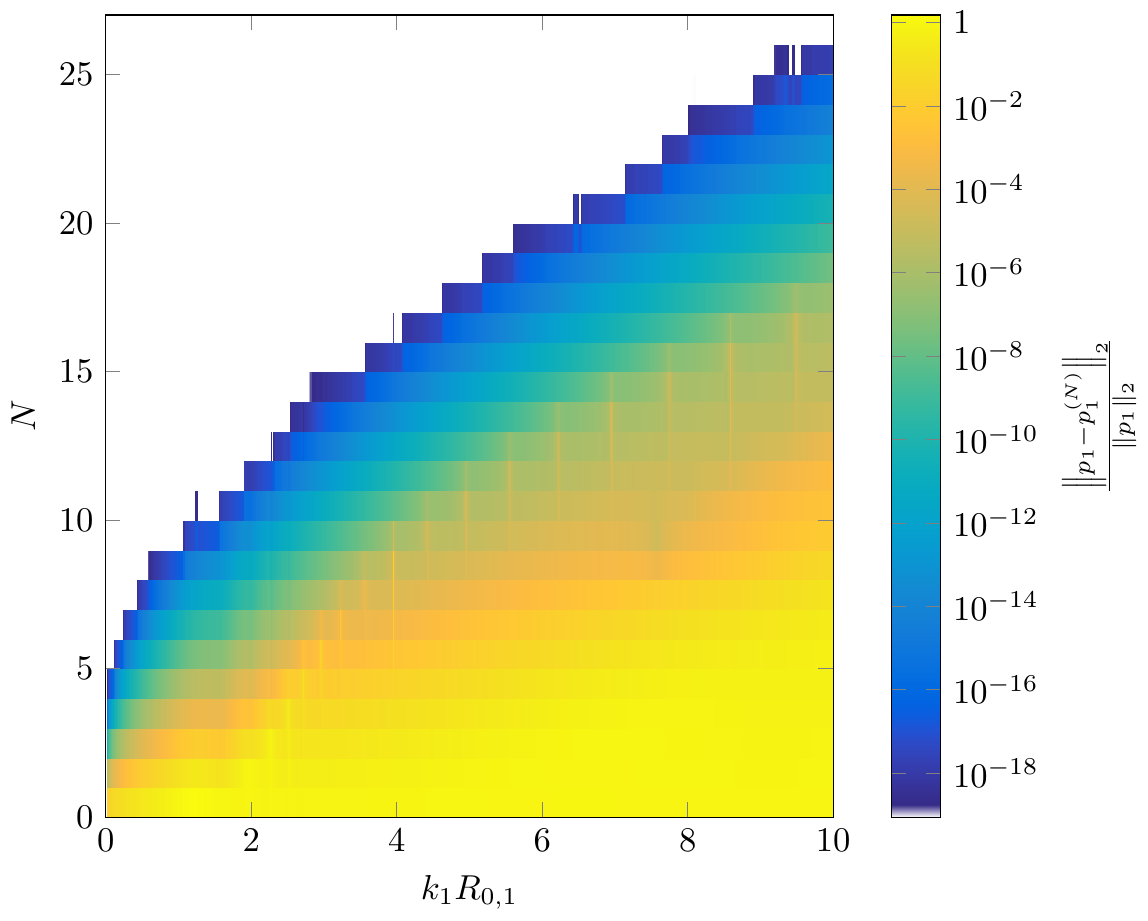}
		\caption{\textbf{Ihlenburg benchmark problem - the sound-soft case}: Relative error in the $l_2$-norm (with two sample points at $\vartheta = \ang{0}$ and $\vartheta = \ang{180}$) of the truncated series in \Cref{Eq1:truncated_p1} as a function of $N$. The ``exact'' solution, $p_1$, is obtained as described in \Cref{Subsec1:seriesEval}.}
	\label{Fig1:IhlenburgErrors2}
\end{figure}
\begin{figure}
	\centering
	\includegraphics{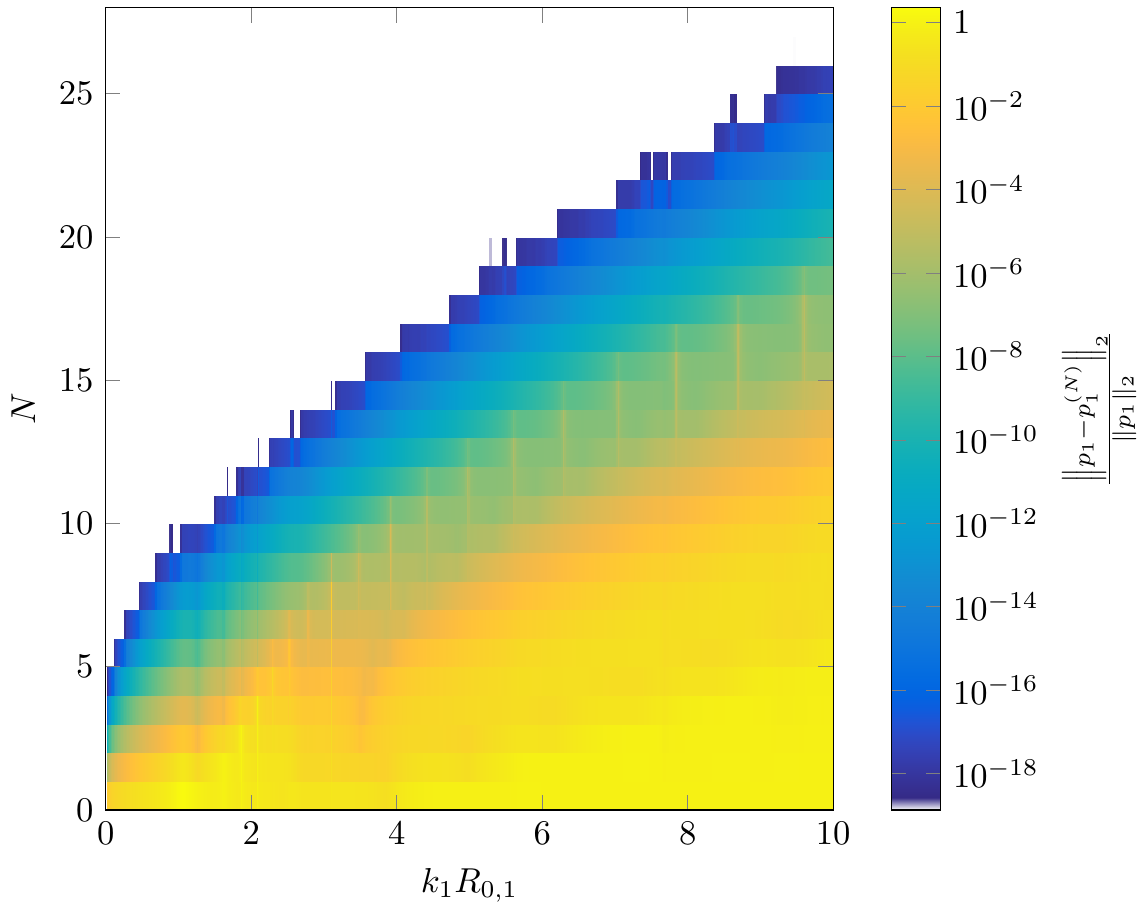}
		\caption{\textbf{Ihlenburg benchmark problem - the Neumann-Neumann case}: Relative error in the $l_2$-norm (with two sample points at $\vartheta = \ang{0}$ and $\vartheta = \ang{180}$) of the truncated series in \Cref{Eq1:truncated_p1} as a function of $N$. The ``exact'' solution, $p_1$, is obtained as described in \Cref{Subsec1:seriesEval}.}
	\label{Fig1:IhlenburgErrors3}
\end{figure}

Convergence plots for the three different cases are plotted in \Cref{Fig1:IhlenburgErrors1,Fig1:IhlenburgErrors2,Fig1:IhlenburgErrors3}, respectively. The linear computational complexity discussed in \Cref{Subsec1:seriesEval} is revealed. Moreover, by comparing the SHBC case in \Cref{Fig1:IhlenburgErrors1} to the SSBC and NNBC cases in \Cref{Fig1:IhlenburgErrors2,Fig1:IhlenburgErrors3}, it is clear that the eigenmodes requires more terms (larger $N$) in order to achieve better than 1\% error precision (this is in particular the case for eigenmodes at higher frequencies). However, the eigenmodes has no need of more terms in order to reach machine epsilon precision. So in the case of elastic scattering, the series termination strategy described in \Cref{Subsec1:seriesEval} is more rigorous than termination of the series at a given $N$ linearly depending on the frequency.

\subsection{Fender benchmark problem} 
\begin{figure}
	\centering
	\begin{subfigure}[t]{\textwidth}
		\centering
	\includegraphics{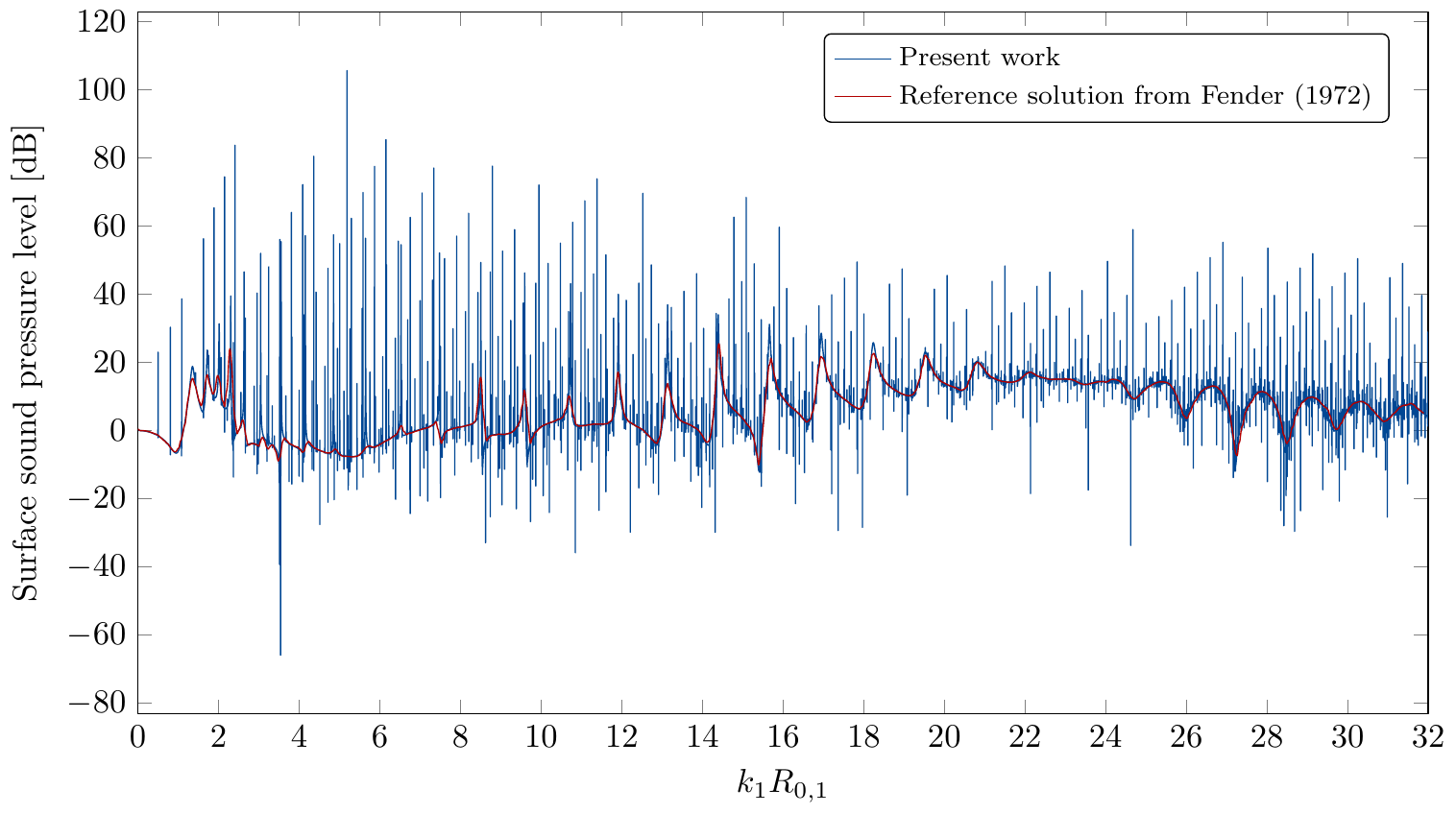}
		\caption{Measured at $\vartheta = \ang{0}$.}
		\label{Fig1:Fender1}
	\end{subfigure}
	\par\bigskip
	\begin{subfigure}[t]{\textwidth}
		\centering
	\includegraphics{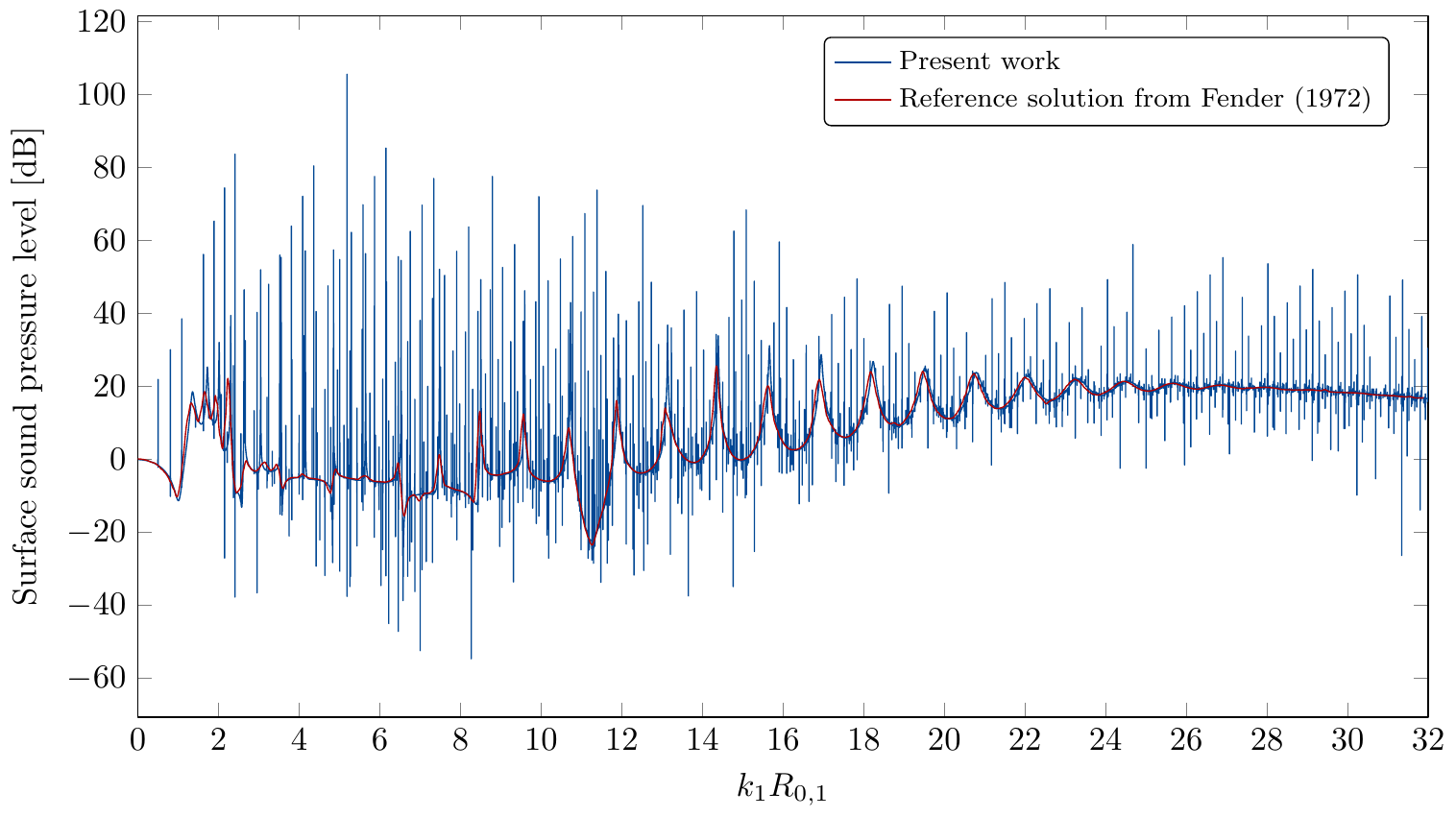}
		\caption{Measured at $\vartheta = \ang{180}$.}
		\label{Fig1:Fender2}
	\end{subfigure}
	\caption{\textbf{Fender benchmark problem}: Predicted total pressure as a function of $k_1R_{0,1}$ at the surface of the shell.}
\end{figure}

Fender~\cite{Fender1972sfa} consider a single air filled spherical shell scattering an incident plane wave (with amplitude $P_{\mathrm{inc}}=\SI{1}{Pa}$). The parameters in \Cref{Tab1:Fender} are here used,
\begin{table}
	\centering
	\caption{\textbf{Fender parameters:} Parameters for the examples in figure 2 and figure 3 in \cite{Fender1972sfa}.}
	\label{Tab1:Fender}
	\begin{tabular}{l l}
		\toprule
		Parameter & Description\\
		\midrule
		$c_{\mathrm{s},1} = \SI{6412}{m.s^{-1}}$ & Longitudinal wave velocity\\
		$c_{\mathrm{s},2} = \SI{3043}{m.s^{-1}}$ & Transverse wave velocity\\
		$\rho_{\mathrm{s},1} = \SI{2700}{kg.m^{-3}}$ & Density of solid\\
		$\rho_{\mathrm{f},1} = \SI{1026}{kg.m^{-3}}$ & Density of outer fluid (water)\\
		$\rho_{\mathrm{f},2} = \SI{1.21}{kg.m^{-3}}$ & Density of inner fluid (air)\\
		$c_{\mathrm{f},1} = \SI{1500}{m.s^{-1}}$ & Speed of sound in water\\
		$c_{\mathrm{f},2} = \SI{343}{m.s^{-1}}$ & Speed of sound in air\\
		$R_{0,1} = \SI{1}{m}$ & Outer radius of spherical shell\\
		$R_{1,1} = \SI{0.95}{m}$ & Inner radius of spherical shell\\
		\bottomrule
	\end{tabular}
\end{table}
where the following conversion formulas is of convenience
\begin{equation}
	E = \rho_s c_{\mathrm{s},2}^2\frac{3c_{\mathrm{s},1}^2-4c_{\mathrm{s},2}^2}{c_{\mathrm{s},1}^2-c_{\mathrm{s},2}^2}\quad\text{and}\quad
	\nu = \frac12 \frac{c_{\mathrm{s},1}^2-2c_{\mathrm{s},2}^2}{c_{\mathrm{s},1}^2-c_{\mathrm{s},2}^2}.
\end{equation}
Fender also sends the incident plane wave along the $x_3$-axis, but in negative direction. The frequency sweep results of the total pressure (in \Cref{Eq1:totPressure}) are measured at the surface. In \Cref{Fig1:Fender1,Fig1:Fender2} the results are found at polar angles $\vartheta=\ang{0}$ and $\vartheta=\ang{180}$, respectively\footnote{The discrepancies again probably comes from the fact that the data set is collected by the software \href{https://automeris.io/WebPlotDigitizer/}{WebPlotDigitizer} where a digital scan of Figure 2 and Figure 3~\cite[pp. 30-31]{Fender1972sfa} has been made. Moreover, the spectrum has been sampled rather closely, revealing small (less significant) eigenmodes not shown by Fender.}.

\begin{figure}
	\centering
	\includegraphics{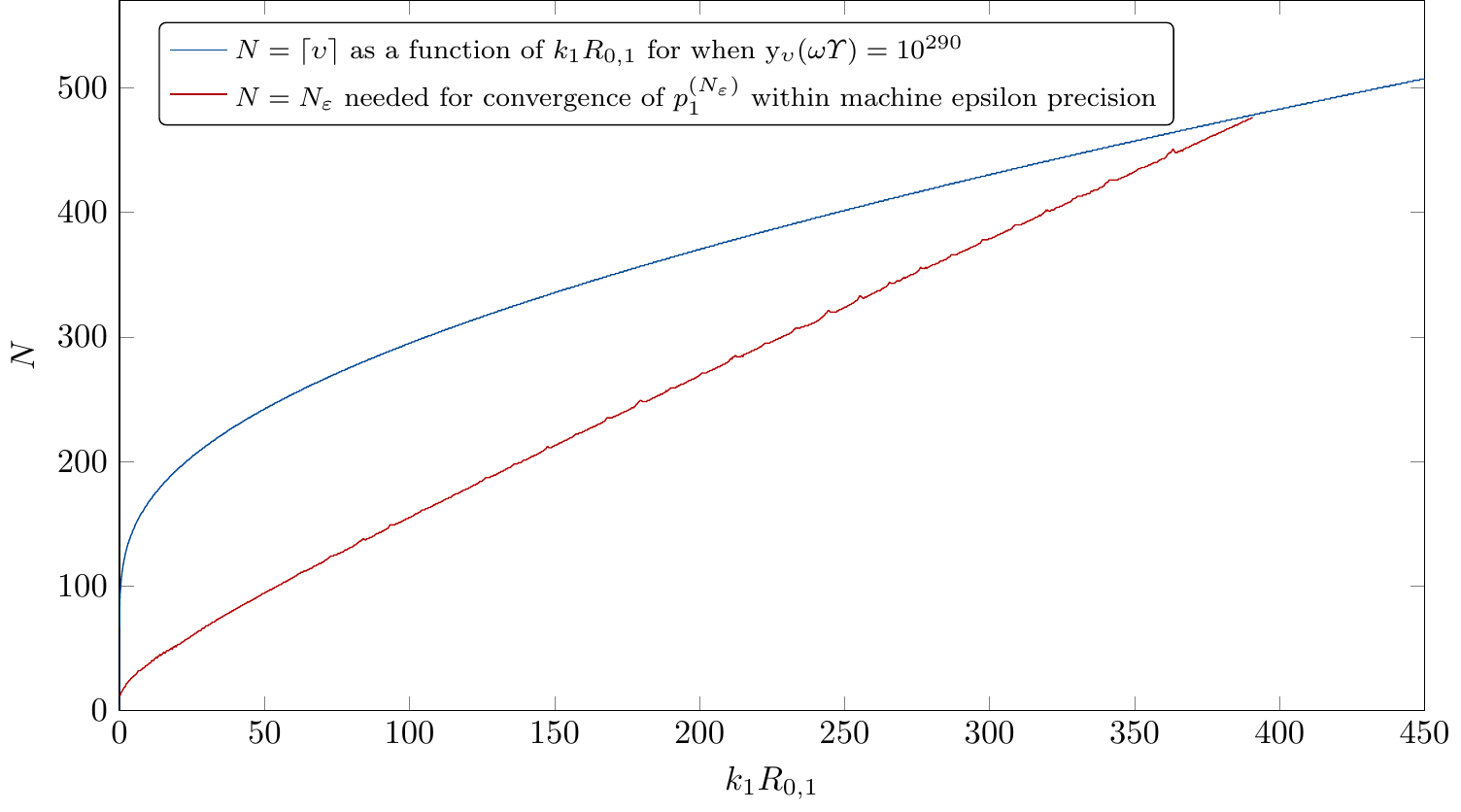}
		\caption{\textbf{Fender benchmark problem}: The intersection of these to graphs marks the largest frequency for which the algorithms presented in this work will give satisfactory results. Here, $\Upsilon$ is given by \Cref{Eq1:Upsilon} and the truncated series $p_1^{(N_\varepsilon)}$ is given by \Cref{Eq1:truncated_p1}.}
	\label{Fig1:FenderConvergence}
\end{figure}

In \Cref{Fig1:FenderConvergence}, another convergence study is illustrated. The Fender benchmark problem was run with increasing frequency until a Bessel function was evaluated to be above $10^{290}$ (the termination criterion as described in \Cref{Subsec1:RoundoffErrors}). Due to the linear behavior of $N$ as a function of $\omega$ needed for convergence (computational complexity) and the concave behavior of the smallest number $N$ such that $|\bessely_N(\omega\Upsilon)|>10^{290}$ (where $\Upsilon$ is given by \Cref{Eq1:Upsilon}), prematurely termination of the series is inevitable for large enough frequencies.

\subsection{Benchmark problems}
\label{Subsec1:benchmarkProblem}
\begin{figure}
	\centering
	\begin{subfigure}{0.3\textwidth}
		\centering
		\includegraphics[width=0.8\textwidth]{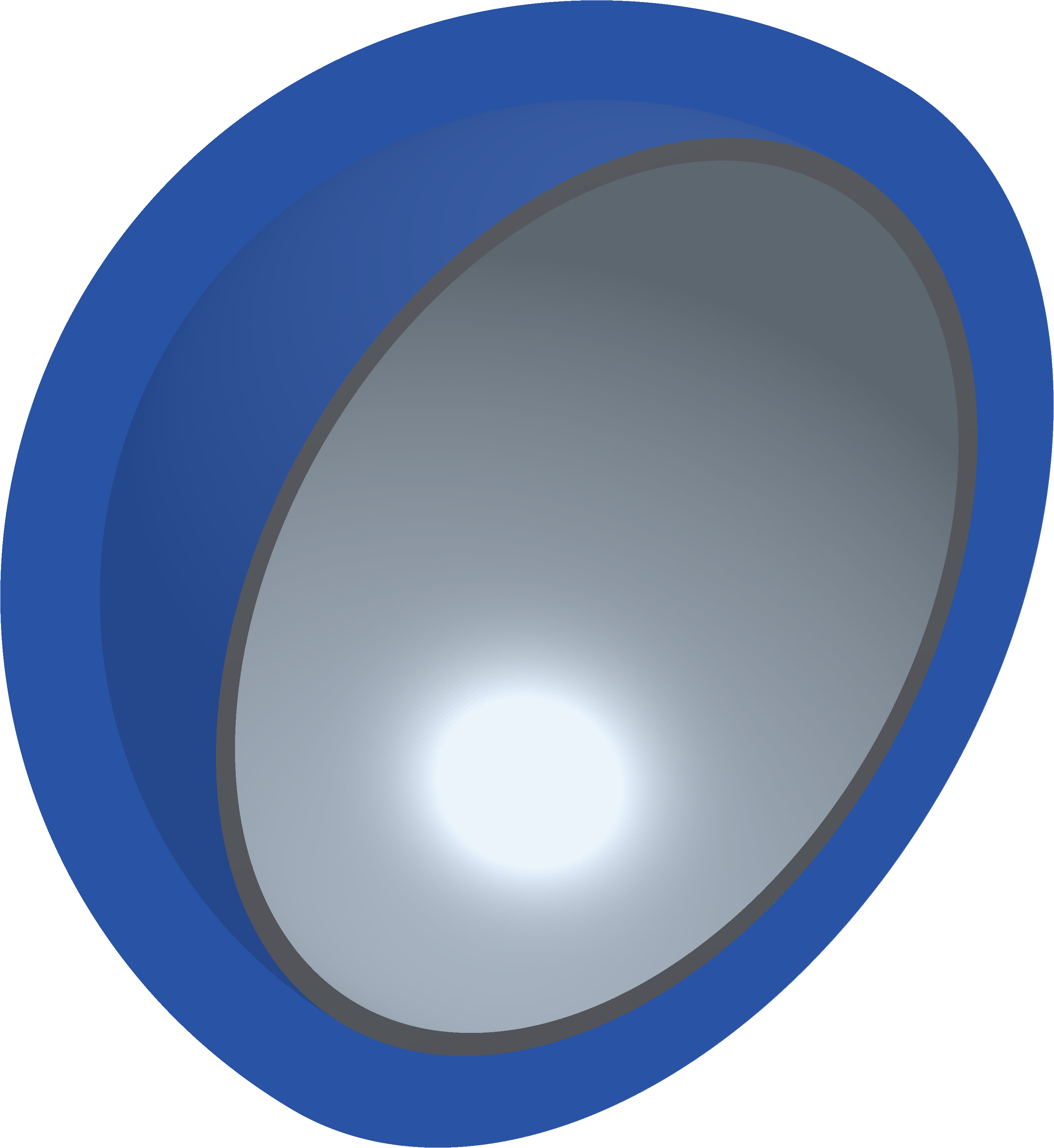}
		\caption{S1}
    \end{subfigure}
	~
	\begin{subfigure}{0.3\textwidth}
		\centering
		\includegraphics[width=0.8\textwidth]{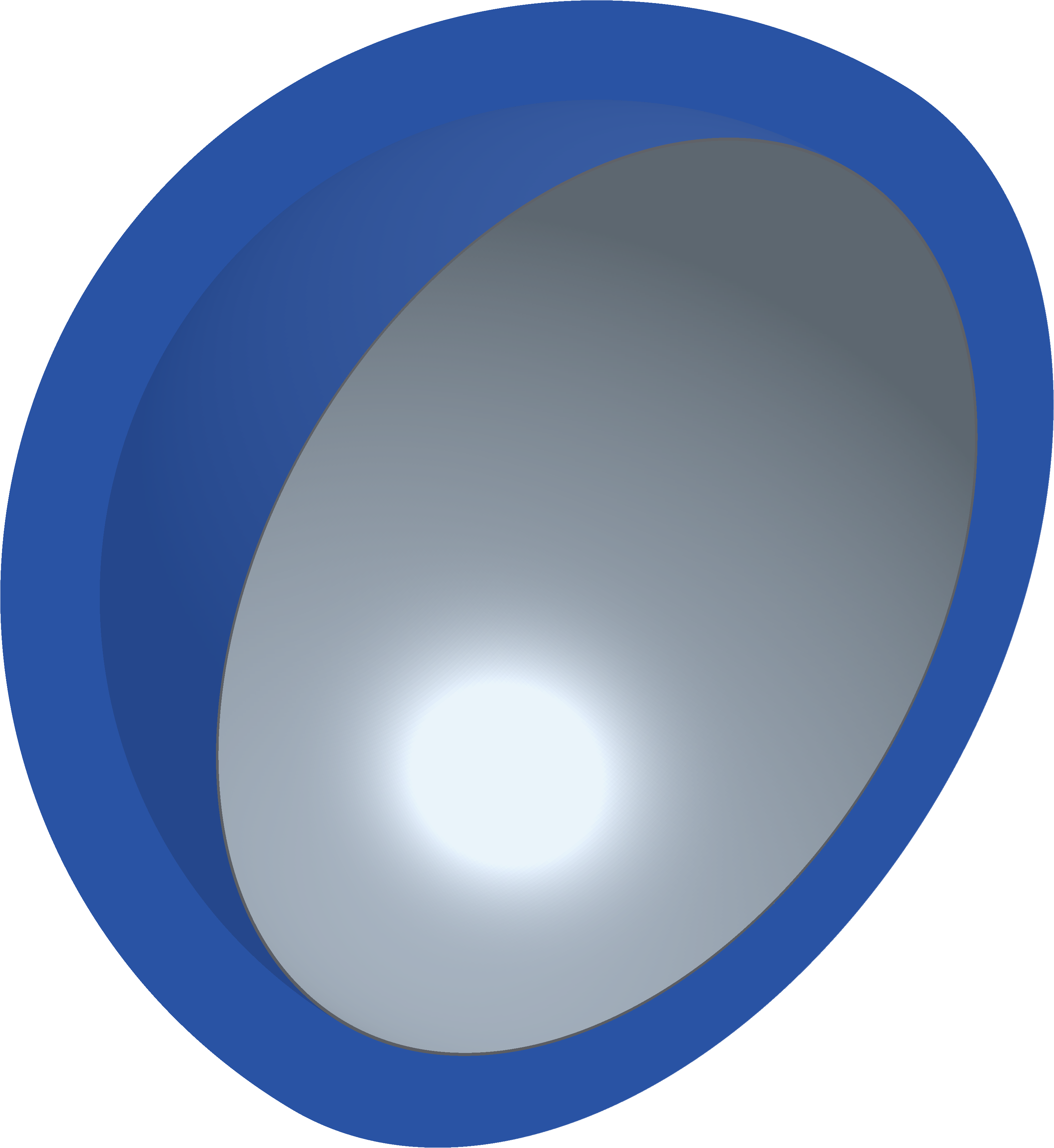}
		\caption{S3}
    \end{subfigure}
	~
	\begin{subfigure}{0.3\textwidth}
		\centering
		\includegraphics[width=0.8\textwidth]{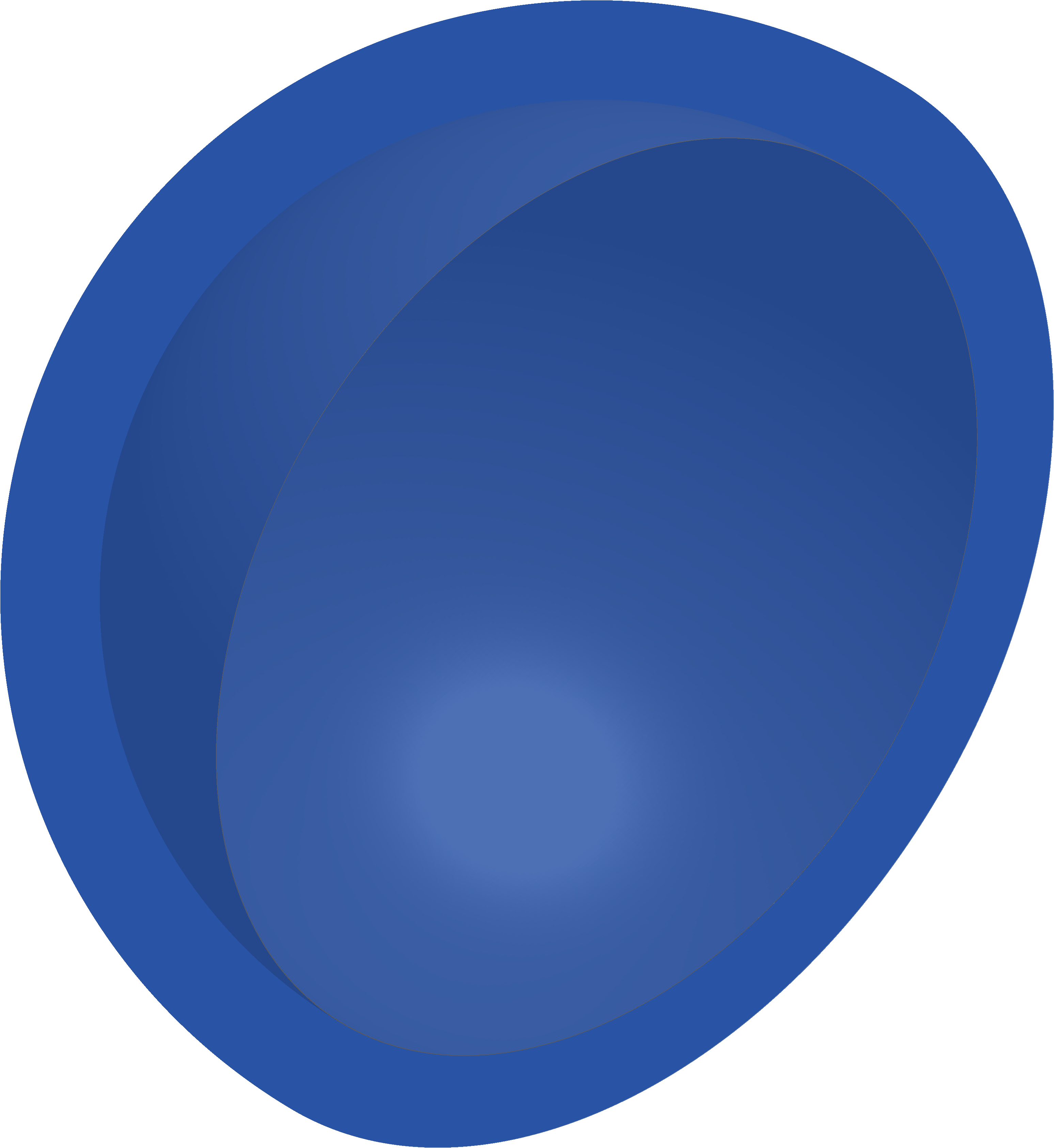}
		\caption{S5}
    \end{subfigure}
	\par\bigskip
	\begin{subfigure}{0.3\textwidth}
		\centering
		\includegraphics[width=0.8\textwidth]{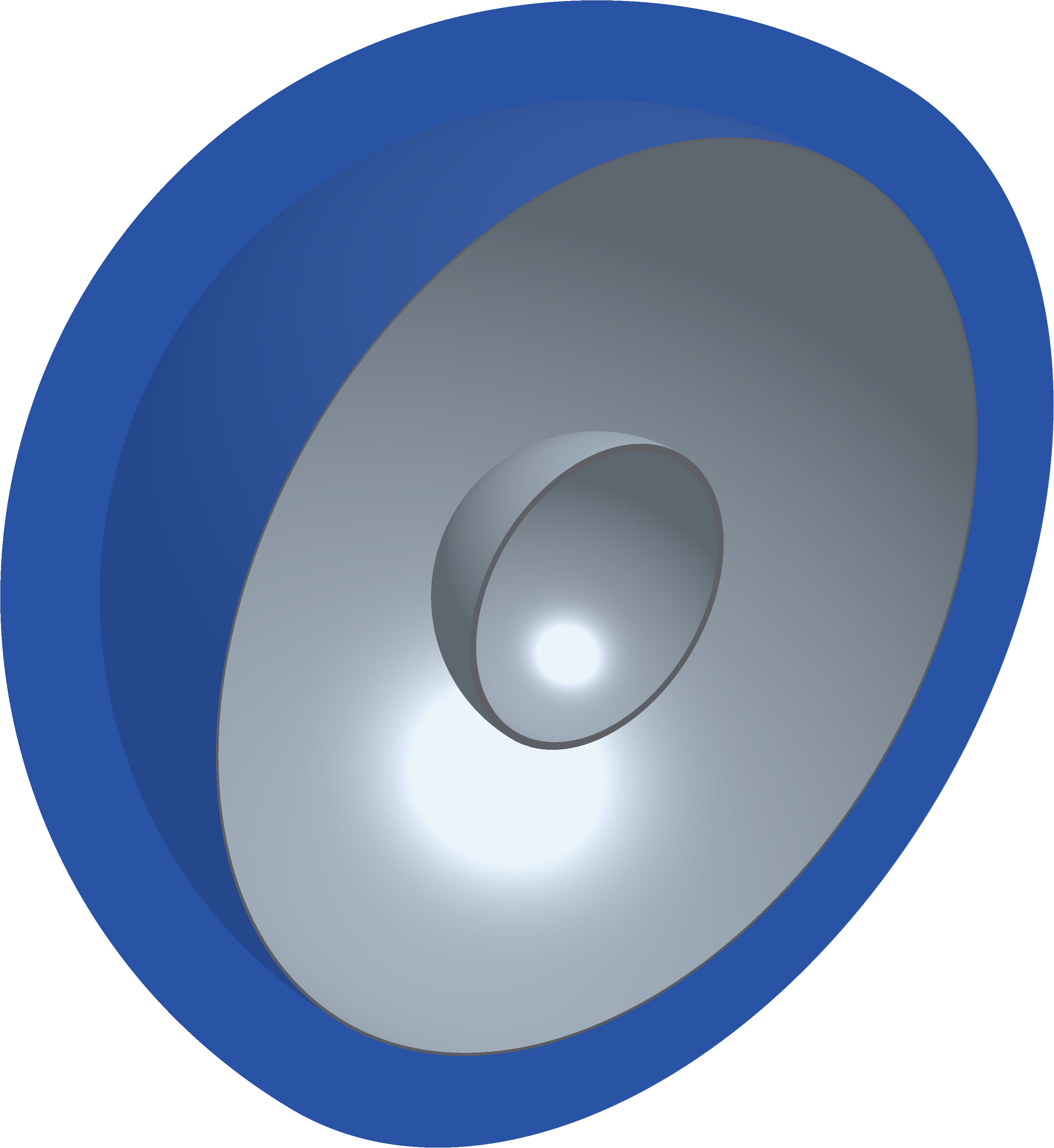}
		\caption{S13}
    \end{subfigure}
	~
	\begin{subfigure}{0.3\textwidth}
		\centering
		\includegraphics[width=0.8\textwidth]{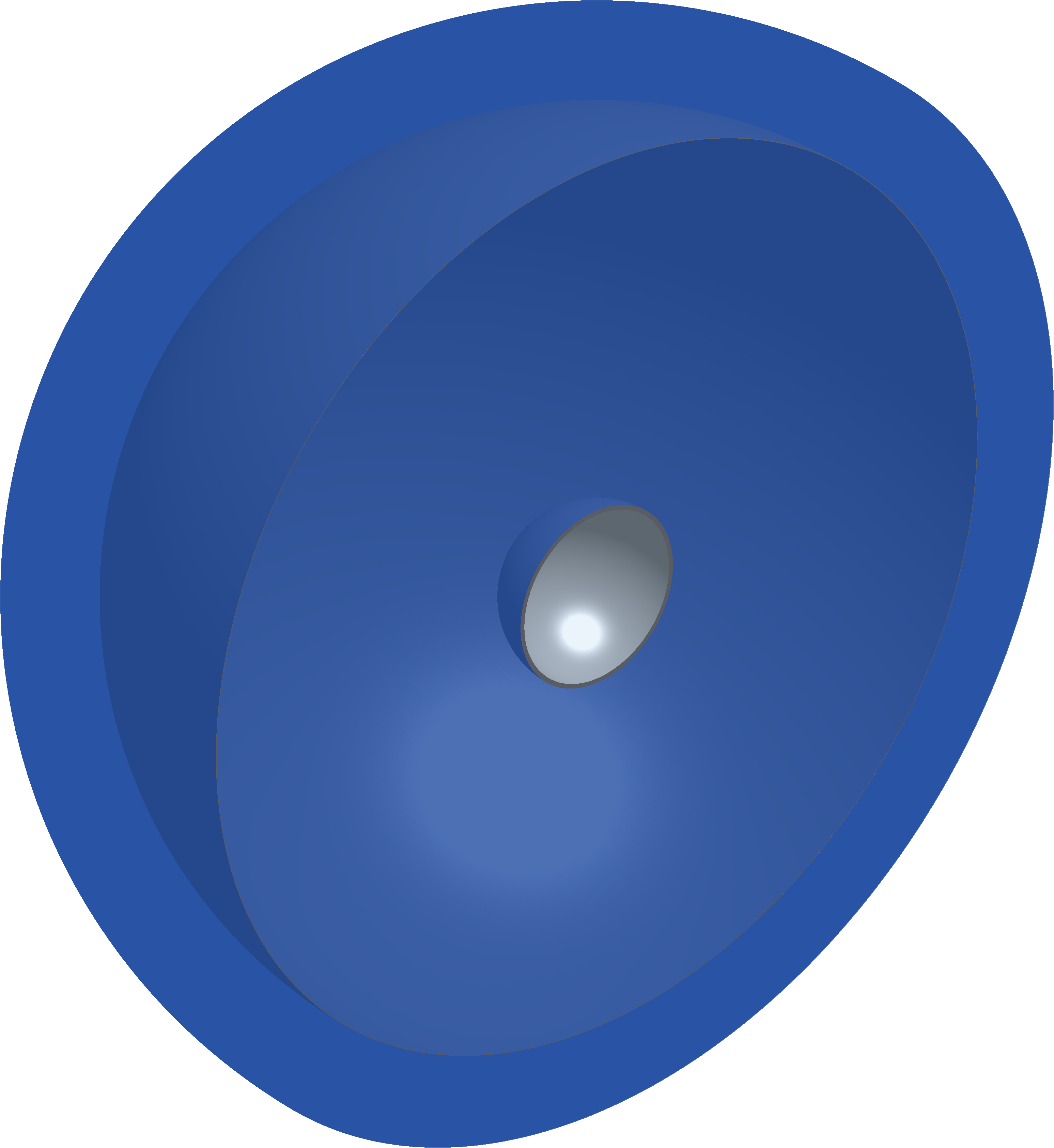}
		\caption{S15}
    \end{subfigure}
	~
	\begin{subfigure}{0.3\textwidth}
		\centering
		\includegraphics[width=0.8\textwidth]{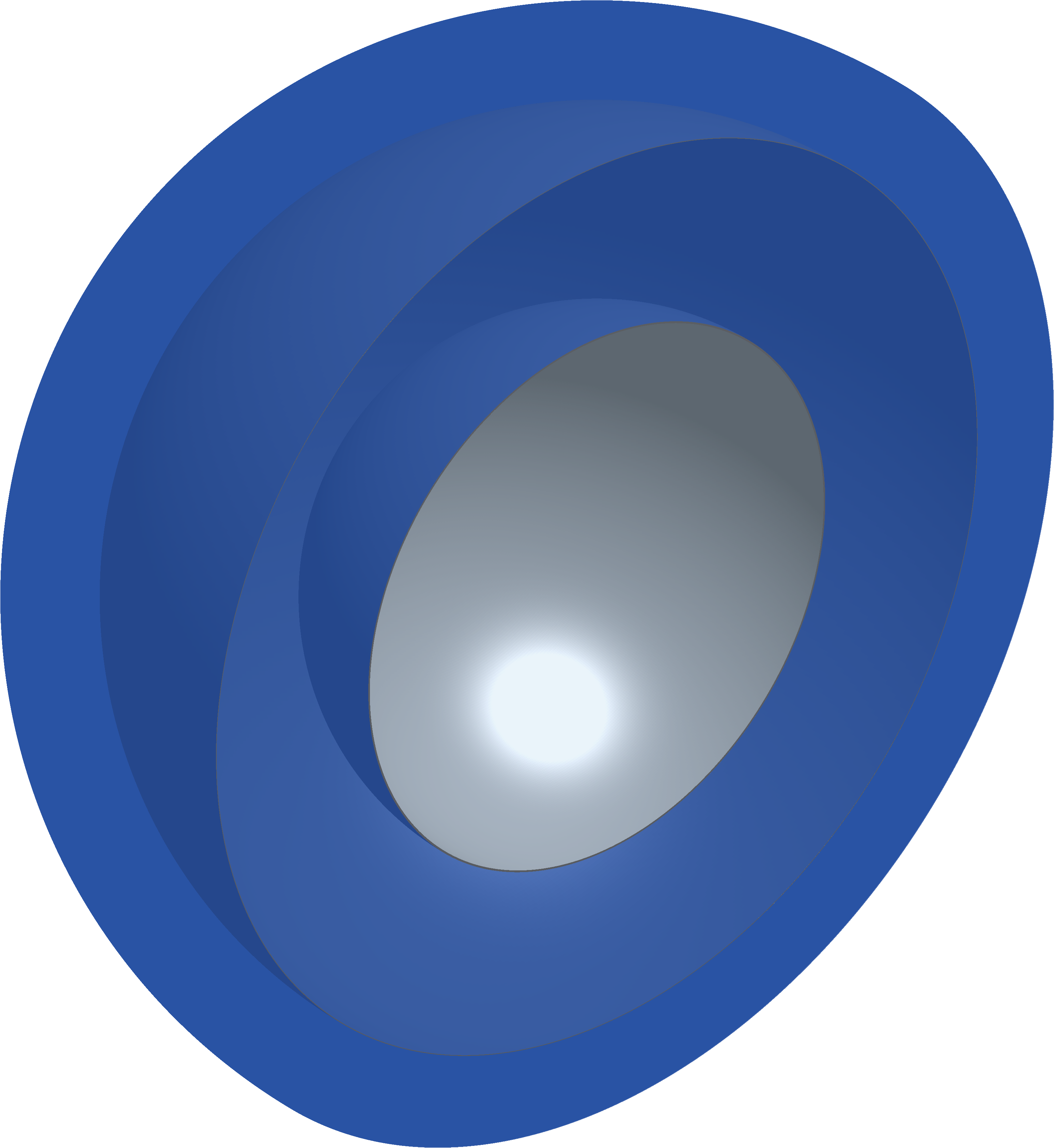}
		\caption{S35}
    \end{subfigure}
	\par\bigskip
	\begin{subfigure}{0.3\textwidth}
		\centering
		\includegraphics[width=0.8\textwidth]{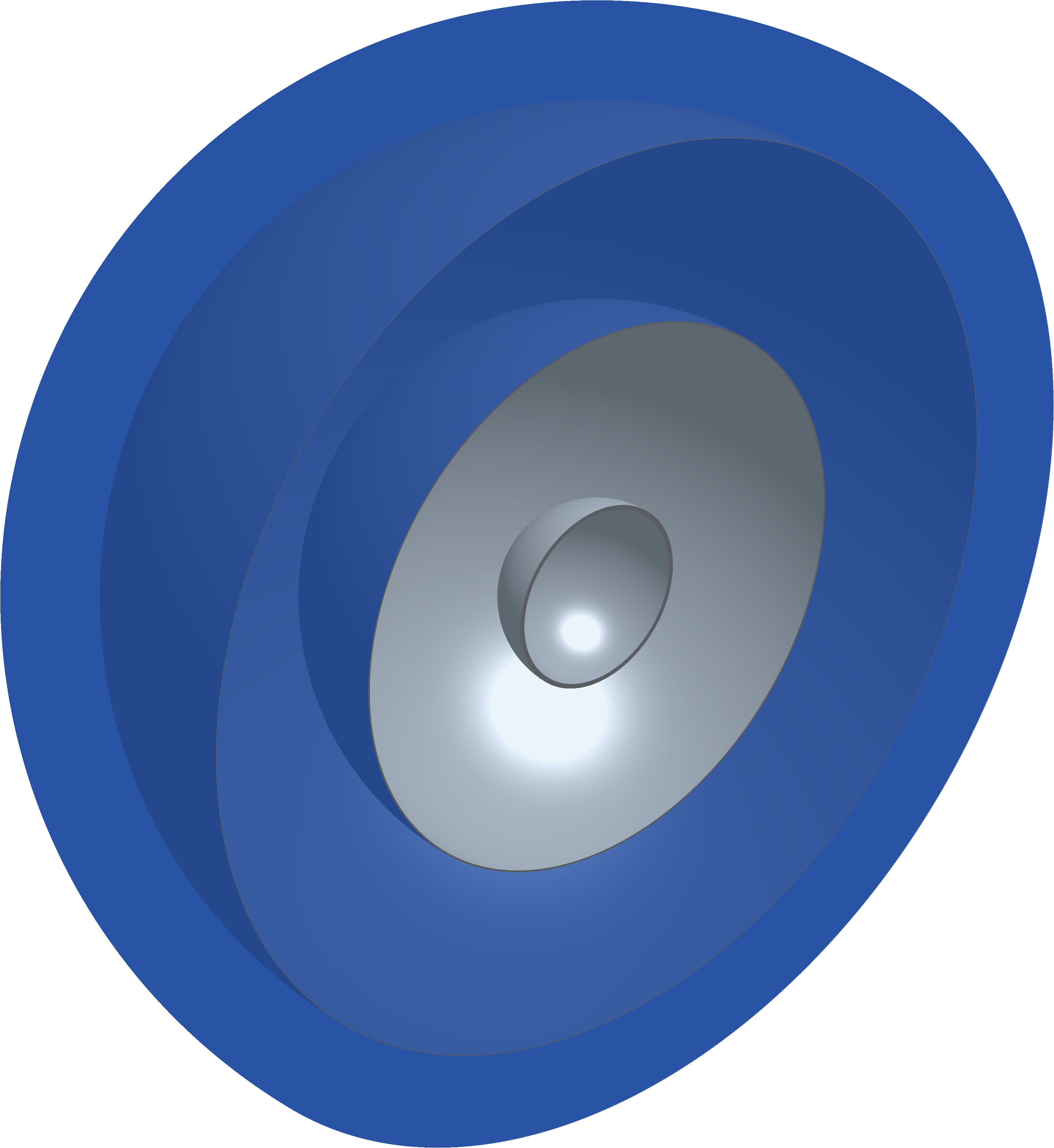}
		\caption{S135}
    \end{subfigure}
	~
	\begin{subfigure}{0.3\textwidth}
		\centering
		\includegraphics[width=0.8\textwidth]{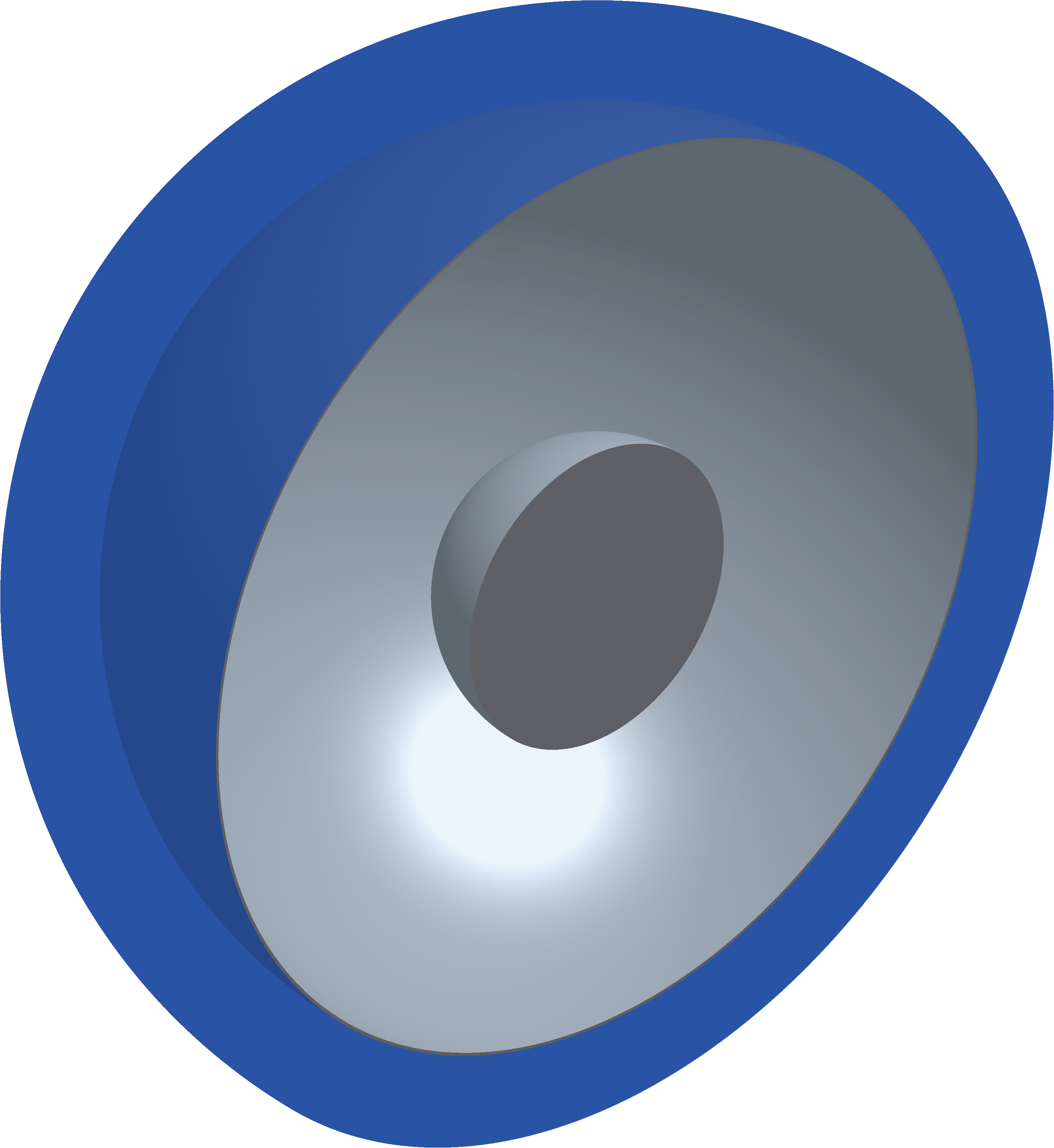}
		\caption{S13 with ESBC}
		\label{Fig1:S12_ASI_3NN}
    \end{subfigure}
	~
	\begin{subfigure}{0.3\textwidth}
		\centering
		\includegraphics[width=0.8\textwidth]{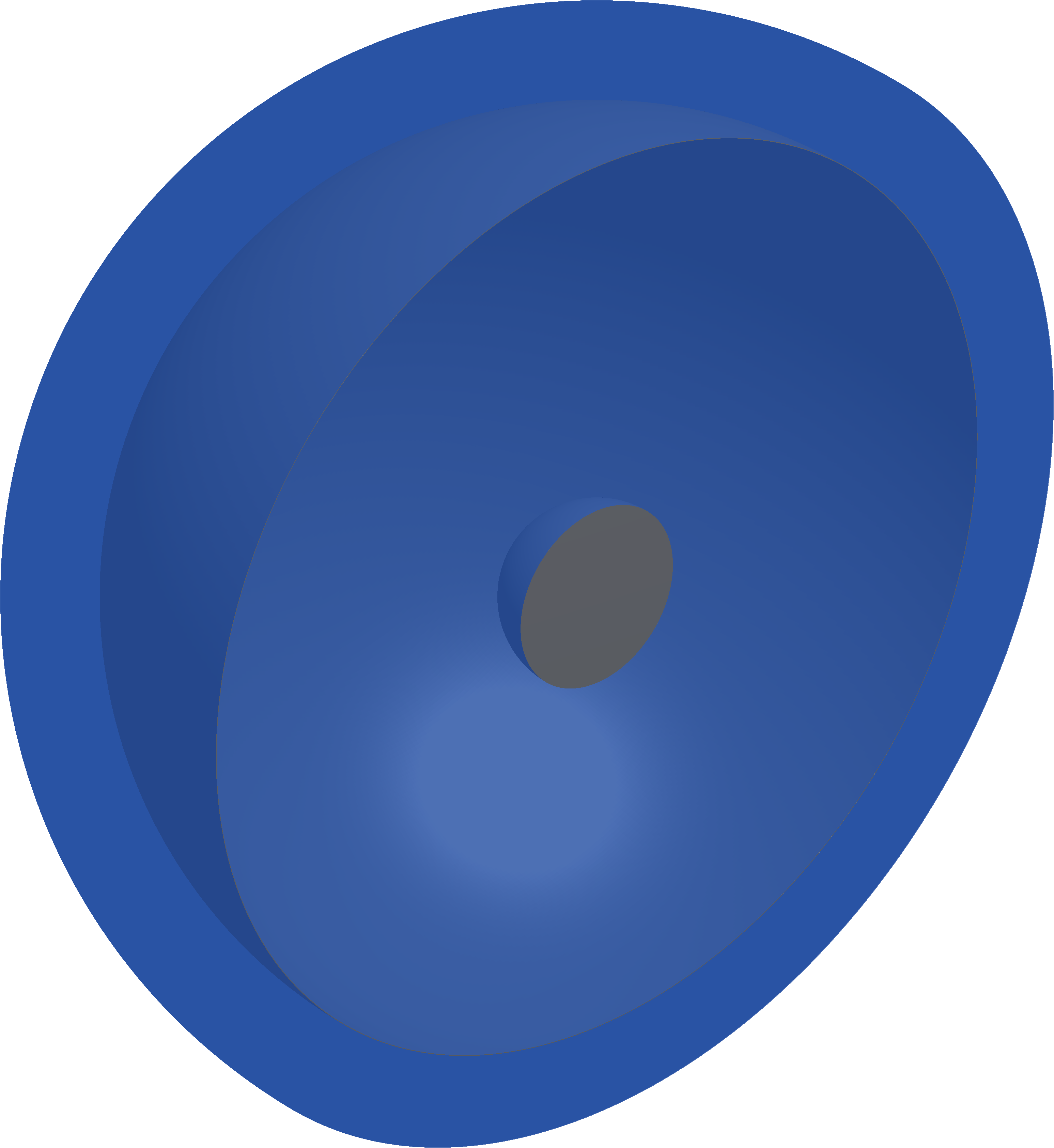}
		\caption{S15 with ESBC}
		\label{Fig1:S23_ASI_3NN}
    \end{subfigure}
    \caption{\textbf{Benchmark problems}: The first row (S1, S2, S3) represent the default set of benchmarks from which the others are built (clip view). The model S5 and S3 is, respectively, 5 and 3 times the size of S1 (the figures are thus not to scale). S13 is a combination of S1 and S2, S13 is a combination of S1 and S3, and S23 is a combination of S2 and S3. S123 is a combination of S1, S2 and S3. The final two figures are derived models with a solid sphere as the innermost domain.}
	\label{Fig1:BenchmarksProblems}
\end{figure}

Let S1, S3 and S5 be benchmark models of spherical shells characterized by the outer radius $R_{0,1}$ and the inner radius $R_{1,1}$ of the shell. The shells are filled with the given fluid (\Cref{Tab1:S1S3S5}) and embedded in water.
\begin{table}
	\centering%
	\caption{\textbf{Benchmark problems:} Parameters for S1, S3 and S5.}
	\label{Tab1:S1S3S5}
	\begin{tabular}{l l l l}
		\toprule
		 & S1 & S3 & S5 \\
		\midrule
		Outer radius, $R_{0,1}$ & $\SI{1}{m}$ & $\SI{3}{m}$ & $\SI{5}{m}$\\
		Inner radius, $R_{1,1}$ & $\SI{0.95}{m}$ & $\SI{2.98}{m}$ & $\SI{4.992}{m}$\\
		Fluid fill & air & air & water\\
		\bottomrule
	\end{tabular}
\end{table}
The remaining parameters are given in \Cref{Tab1:sphericalShellParameters}. 
\begin{table}
	\centering
	\caption{\textbf{Benchmark problems:} Common parameters for the benchmark problems.}
	\label{Tab1:sphericalShellParameters}
	\begin{tabular}{l l}
		\toprule
		Parameter & Description\\
		\midrule
		$E = \SI{2.10e11}{Pa}$ & Young's modulus\\
		$\nu = 0.3$ & Poisson's ratio\\
		$\rho_{\mathrm{s}} = \SI{7850}{kg.m^{-3}}$ & Density of solid\\
		$\rho_{\mathrm{f,water}} = \SI{1000}{kg.m^{-3}}$ & Density of water\\
		$\rho_{\mathrm{f,air}} = \SI{1.2}{kg.m^{-3}}$ & Density of air\\
		$c_{\mathrm{f,water}} = \SI{1500}{m.s^{-1}}$ & Speed of sound in water\\
		$c_{\mathrm{f,air}} = \SI{340}{m.s^{-1}}$ & Speed of sound in air\\
		\bottomrule
	\end{tabular}
\end{table}
These models can be combined into a new set of benchmark problems: S13 (S1 inside S3 with air in between), S15 (S1 inside S5 with water in between), S35 (S3 inside S5 with water in between) and S135 (S1 inside S3 inside S5 with air in between S1 and S3 and water in between S3 and S5). These benchmark problems are illustrated in \Cref{Fig1:BenchmarksProblems}.

By default all benchmarks has acoustic structure interaction (ASI) on all of the interfaces between fluid and solid domains (with Neumann-to-Neumann boundary conditions, NNBC, in \Cref{Eq1:firstBC,Eq1:secondBC}). As described in \Cref{Subsec1:alternativeBoundaryConditions}, the boundary conditions on the innermost shell may be replaced by other boundary conditions like the sound-soft (SSBC) and sound-hard boundary condition (SHBC). The elastic sphere boundary condition (ESBC) results from filling the innermost shell with the given elastic material. 

\begin{figure}
	\centering
	\begin{subfigure}[t]{0.48\textwidth}
		\centering
		\includegraphics{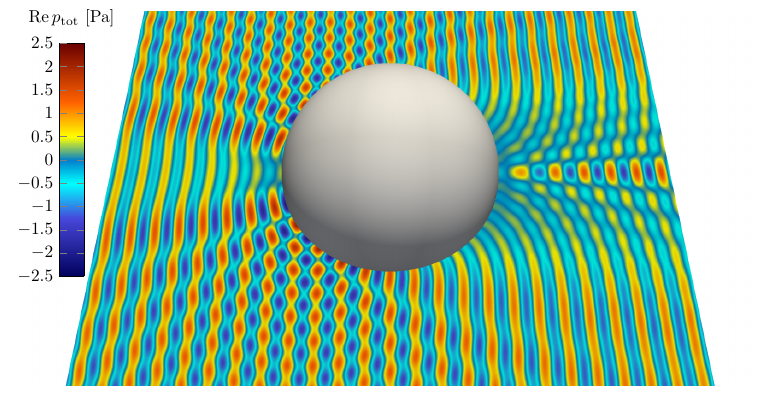}
		\caption{\textbf{S5 with SHBC}: Plot of the real part of $p_{\mathrm{tot}}$.}
		\label{Fig1:S5_SHBC_real}
	\end{subfigure}
	~
	\begin{subfigure}[t]{0.48\textwidth}
		\centering
		\includegraphics{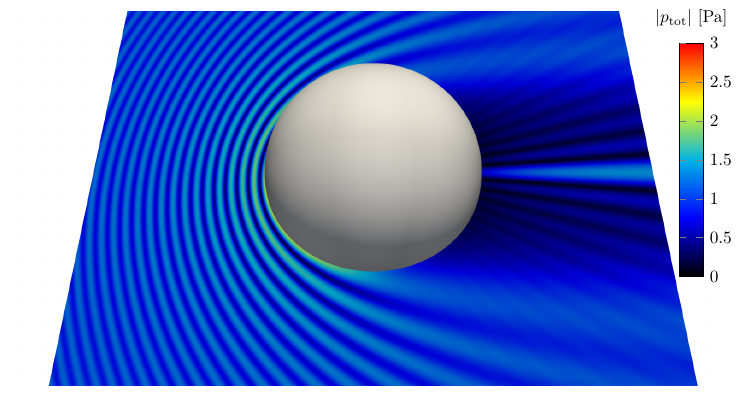}
		\caption{\textbf{S5 with SHBC}: Plot of the modulus of $p_{\mathrm{tot}}$.}
		\label{Fig1:S5_SHBC_abs}
	\end{subfigure}
	\par\bigskip
	\begin{subfigure}[t]{0.48\textwidth}
		\centering
		\includegraphics{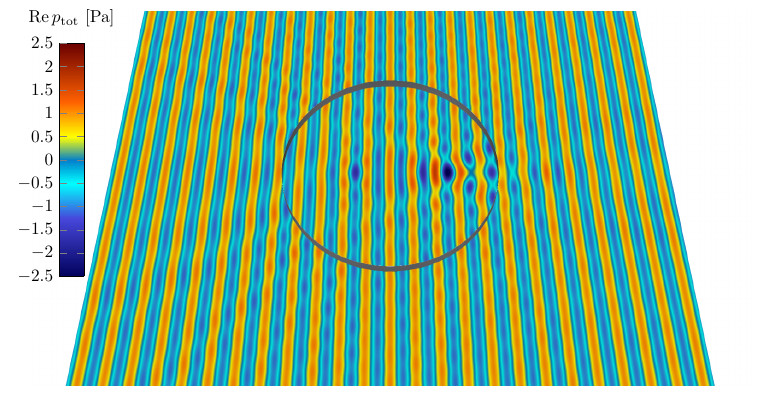}
		\caption{\textbf{S5 with NNBC}: Plot of the real part of $p_{\mathrm{tot}}$.}
		\label{Fig1:S5_NNBC_real}
	\end{subfigure}
	~
	\begin{subfigure}[t]{0.48\textwidth}
		\centering
		\includegraphics{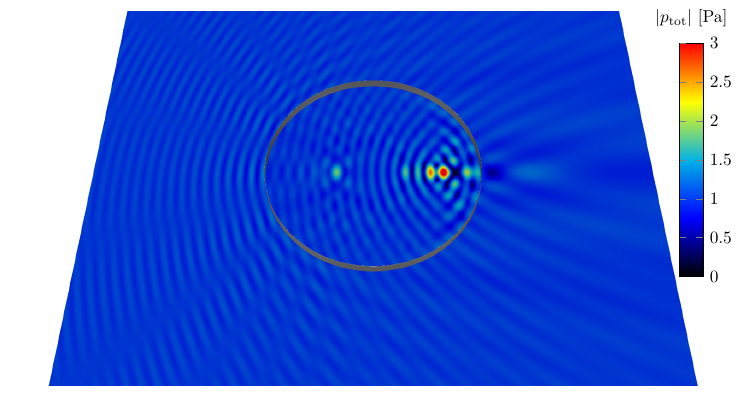}
		\caption{\textbf{S5 with NNBC}: Plot of the modulus of $p_{\mathrm{tot}}$.}
		\label{Fig1:S5_NNBC_abs}
	\end{subfigure}
	\par\bigskip
	\begin{subfigure}[t]{0.48\textwidth}
		\centering
		\includegraphics{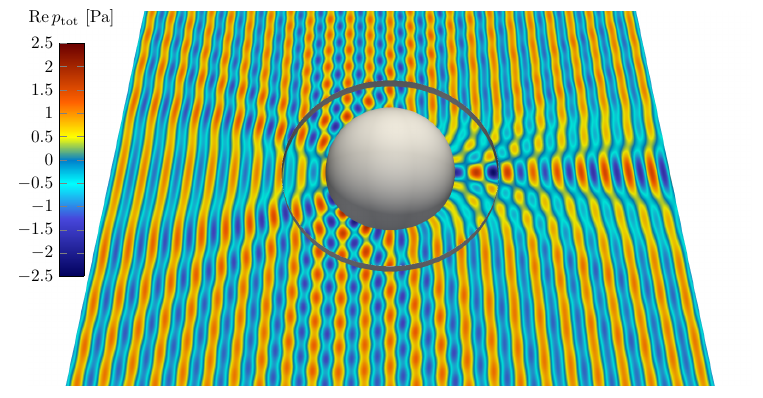}
		\caption{\textbf{S35 with SHBC}: Plot of the real part of $p_{\mathrm{tot}}$.}
		\label{Fig1:S35_SHBC_real}
	\end{subfigure}
	~
	\begin{subfigure}[t]{0.48\textwidth}
		\centering
		\includegraphics{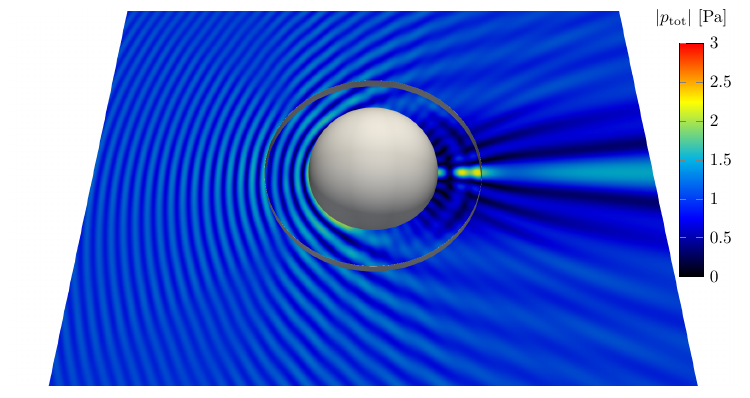}
		\caption{\textbf{S35 with SHBC}: Plot of the modulus of $p_{\mathrm{tot}}$.}
		\label{Fig1:S35_SHBC_abs}
	\end{subfigure}
	\par\bigskip
	\begin{subfigure}[t]{0.48\textwidth}
		\centering
		\includegraphics{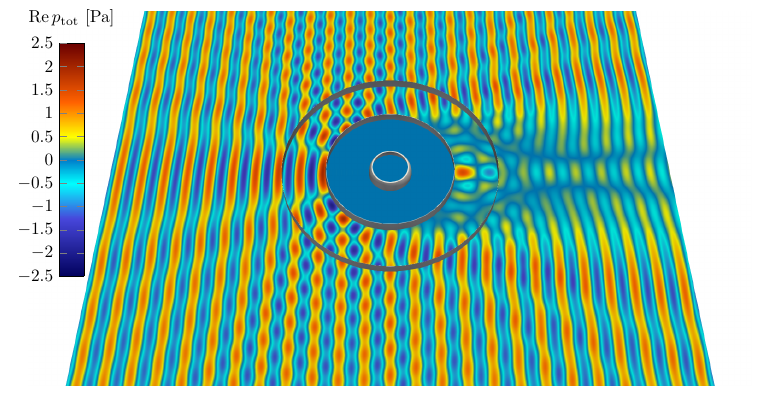}
		\caption{\textbf{S135 with NNBC}: Plot of the real part of $p_{\mathrm{tot}}$.}
		\label{Fig1:S35_NNBC_real}
	\end{subfigure}
	~
	\begin{subfigure}[t]{0.48\textwidth}
		\centering
		\includegraphics{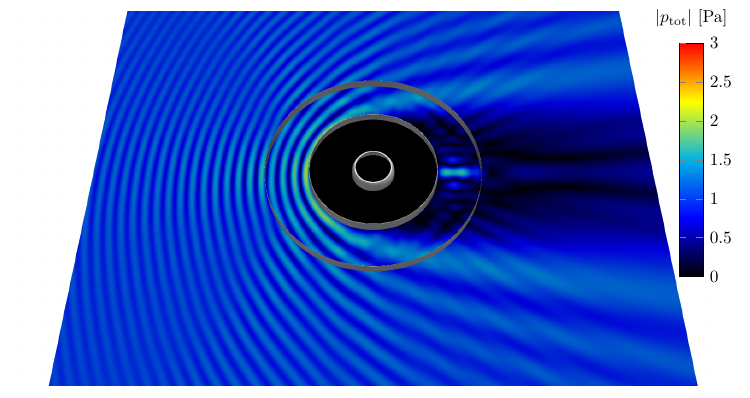}
		\caption{\textbf{S135 with NNBC}: Plot of the modulus of $p_{\mathrm{tot}}$.}
		\label{Fig1:S35_NNBC_abs}
	\end{subfigure}
	\caption{\textbf{Benchmark problems}: Plots of the near-field of some benchmark problems. The shells are cut open whenever a domain inside the shell is present. The visualization was done in \href{http://www.paraview.org/}{Paraview}.}
	\label{Fig1:Benchmarks_NearField}
\end{figure}

In \Cref{Fig1:Benchmarks_NearField}, the near field is plotted for some of these benchmarks. In \Cref{Fig1:S5_SHBC_abs} the classical interference pattern emerging behind the rigid scatterer S5 (with SHBC) can be observed. In contrast, the corresponding case with NNBC in \Cref{Fig1:S5_NNBC_real,Fig1:S5_NNBC_abs} gives a different picture entirely because most of the energy simply passes straight through the thin spherical shell. The example is expanded further in \Cref{Fig1:S35_SHBC_real,Fig1:S35_SHBC_abs,Fig1:S35_NNBC_real,Fig1:S35_NNBC_abs}. For the latter case, the energy transmitted is greatly reduced due to air filled fluid inside the second shell. The S135 benchmark was visually identical to this benchmark due to this fact (that is, it is hard to reveal objects inside air filled domains). However, it is clear that sound-hard boundary conditions are not a good approximation of NNBC in this case. The more natural approximation would be to use SSBC. Indeed, the SSBC approximate the innermost fluid with $p_{M+1}=0$, which is clearly a good approximation in this case.

\subsection{Benchmark problems in the time domain}
\begin{figure}
	\centering
	\begin{subfigure}[t]{0.48\textwidth}
		\centering
		\includegraphics{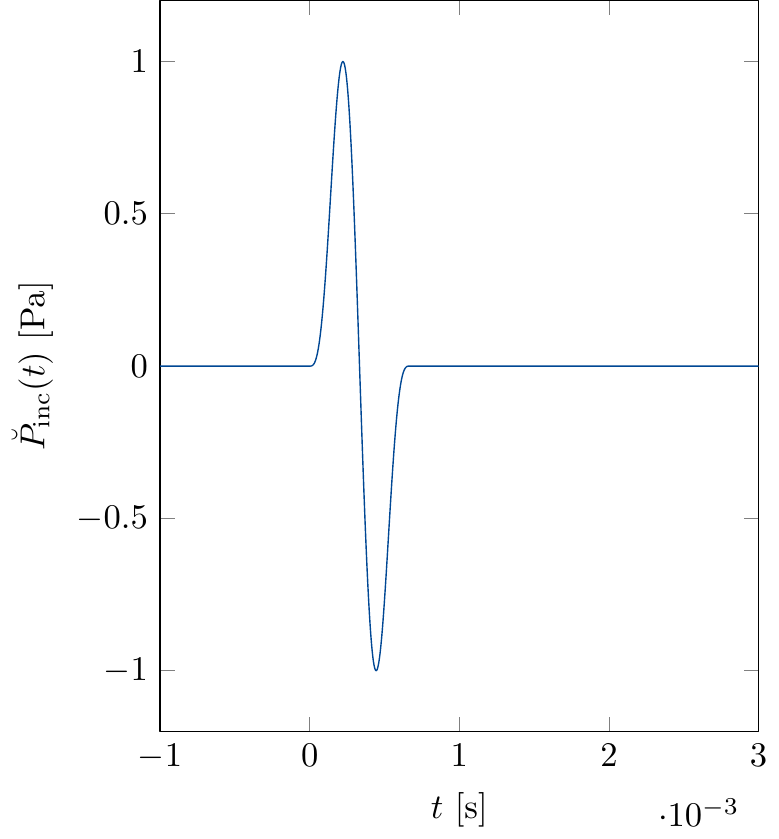}
		\caption{Wavelet in time domain.}
		\label{Fig1:Pb_inc}
	\end{subfigure}
	~
	\begin{subfigure}[t]{0.48\textwidth}
		\centering
		\includegraphics{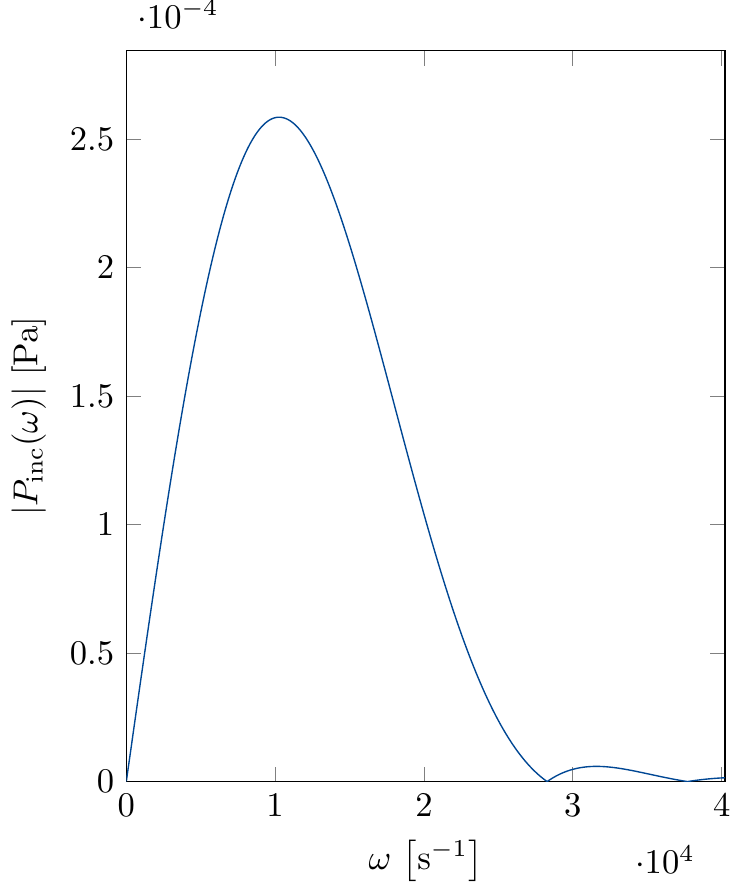}
		\caption{Wavelet in frequency domain.}
		\label{Fig1:P_inc}
	\end{subfigure}
	\caption{\textbf{Benchmark problems in the time domain}: Wavelet $\breve{P}_{\mathrm{inc}}(t)$ and corresponding frequency spectrum $|P_{\mathrm{inc}}(\omega)|$. The wavelet has compact support on the interval $[0,1/f_{\mathrm{c}}]$, where $f_{\mathrm{c}}=\SI{1.5e3}{Hz}$. The frequency spectrum is plotted for positive frequencies to the end of the bandwidth, $f=B/2=\SI{6.4e3}{Hz}$ (with $B=\check{N}/T$).}
\end{figure}

Finally, the application of the work in the time domain will be presented. In particular, consider scattering by a single wavelet given by (from~\cite[p. 633]{Jensen2011coa})
\begin{equation}\label{Eq1:Pb_inc}
	\breve{P}_{\mathrm{inc}}(t) = \begin{cases} \frac{4}{3\sqrt{3}} \left[\sin(\omega_{\mathrm{c}} t)-\frac12 \sin\left(2\omega_{\mathrm{c}} t\right)\right] & 0 < t < \frac{1}{f_{\mathrm{c}}}\\
	0 & \text{otherwise},\end{cases}
\end{equation}
with $\omega_{\mathrm{c}} = 2\PI f_{\mathrm{c}}$ and $k_{\mathrm{c}} = \omega_{\mathrm{c}}/c_{\mathrm{f},1}$, and where $f_{\mathrm{c}}$ is the center frequency  (\Cref{Fig1:Pb_inc}). The corresponding frequency spectrum (using the definition of the Fourier transform in \Cref{Eq1:Psi}), is given by (\Cref{Fig1:P_inc})
\begin{equation}
	P_{\mathrm{inc}}(\omega) = \left(\fourier\breve{P}_{\mathrm{inc}}\right)(\omega) = \begin{cases}
		\frac{4}{3\sqrt{3}}\frac{\imag\PI}{\omega}\euler^{\imag\PI\omega/\omega_{\mathrm{c}}} & \omega \in\{\pm\omega_{\mathrm{c}},\pm 2\omega_{\mathrm{c}}\}\\
		\frac{4}{\sqrt{3}}\frac{\omega_{\mathrm{c}}^3}{\left(\omega^2-\omega_{\mathrm{c}}^2\right)\left(\omega^2-4\omega_{\mathrm{c}}^2\right)}\left(1-\euler^{-2\PI\imag\omega/\omega_{\mathrm{c}}}\right) & \text{otherwise.}
		\end{cases}
\end{equation}
A plane wave with this wavelet in the time-domain then takes the form
\begin{equation}\label{Eq1:PlaneWaveTimeDomain}
	\breve{p}_{\mathrm{inc}}(\vec{x},t) = \breve{P}_{\mathrm{inc}}\left(t-\frac{x_3}{c_{\mathrm{f},1}}\right),
\end{equation}
with corresponding field in the frequency domain given by
\begin{equation}
	p_{\mathrm{inc}}(\vec{x},\omega)=\left(\fourier\breve{p}_{\mathrm{inc}}(\vec{x},\cdot)\right)(\omega) = P_{\mathrm{inc}}(\omega)\euler^{\imag k_1 x_3}.
\end{equation}
An alternative to plane waves, is waves due to point sources. Using the wavelet in \Cref{Eq1:Pb_inc}, these waves are given by
\begin{equation}
	\breve{p}_{\mathrm{inc}}(\vec{x},t) = \breve{P}_{\mathrm{inc}}\left(t-\frac{|\vec{x}_{\mathrm{s}}-\vec{x}|}{c_{\mathrm{f},1}}\right)\frac{r_{\mathrm{s}}}{|\vec{x}_{\mathrm{s}}-\vec{x}|} \quad\text{and}\quad p_{\mathrm{inc}}(\vec{x},\omega) = P_{\mathrm{inc}}(\omega)\frac{r_{\mathrm{s}}}{|\vec{x}_{\mathrm{s}}-\vec{x}|}\euler^{\imag k_1 |\vec{x}_{\mathrm{s}}-\vec{x}|}\label{Eq1:P_incOmega}
\end{equation}
where $\vec{x}_{\mathrm{s}}$ is location of the point source given in \Cref{Eq1:x_s} (at a finite distance $r_{\mathrm{s}}=|\vec{x}_{\mathrm{s}}|$). 

As $\Psi(\vec{x},\omega)$ cannot be computed for infinitely many frequencies, an approximation of the time dependent fields in \Cref{Eq1:Psit} by~\cite[p. 614]{Jensen2011coa} can be used
\begin{equation}
	\breve{\Psi}(\vec{x},t_m) \approx \frac{2}{T}\Re\left\{\sum_{n=1}^{\check{N}/2-1} \Psi(\vec{x},\omega_n)\euler^{-2\PI\imag n m/\check{N}}\right\}
\end{equation}
where
\begin{equation}
	t_m = m\Diff t, \quad \Diff t = \frac{T}{\check{N}}, \quad \omega_n = n\Diff\omega, \quad \Diff\omega = \frac{2\PI}{T}. 
\end{equation}
Note that the contribution from the static case ($n=0$) has not been included as the incident wave, $p_{\mathrm{inc}}(\vec{x},0) = 0$, results in the trivial solution $\Psi(\vec{x},0)=0$. The Fourier series approximation results in periodic time-dependent fields, with period $T$, sampled in the interval $[0,T]$ with $\check{N}$ time steps. The parameter $\check{N}$ also quantifies the number of terms in the Fourier series approximation, such that it also controls the error (aliasing). By choosing $\check{N}$ to be powers of two, the approximation can be very efficiently evaluated by the fast Fourier transformation. 

\begin{figure}
	\centering
	\includegraphics{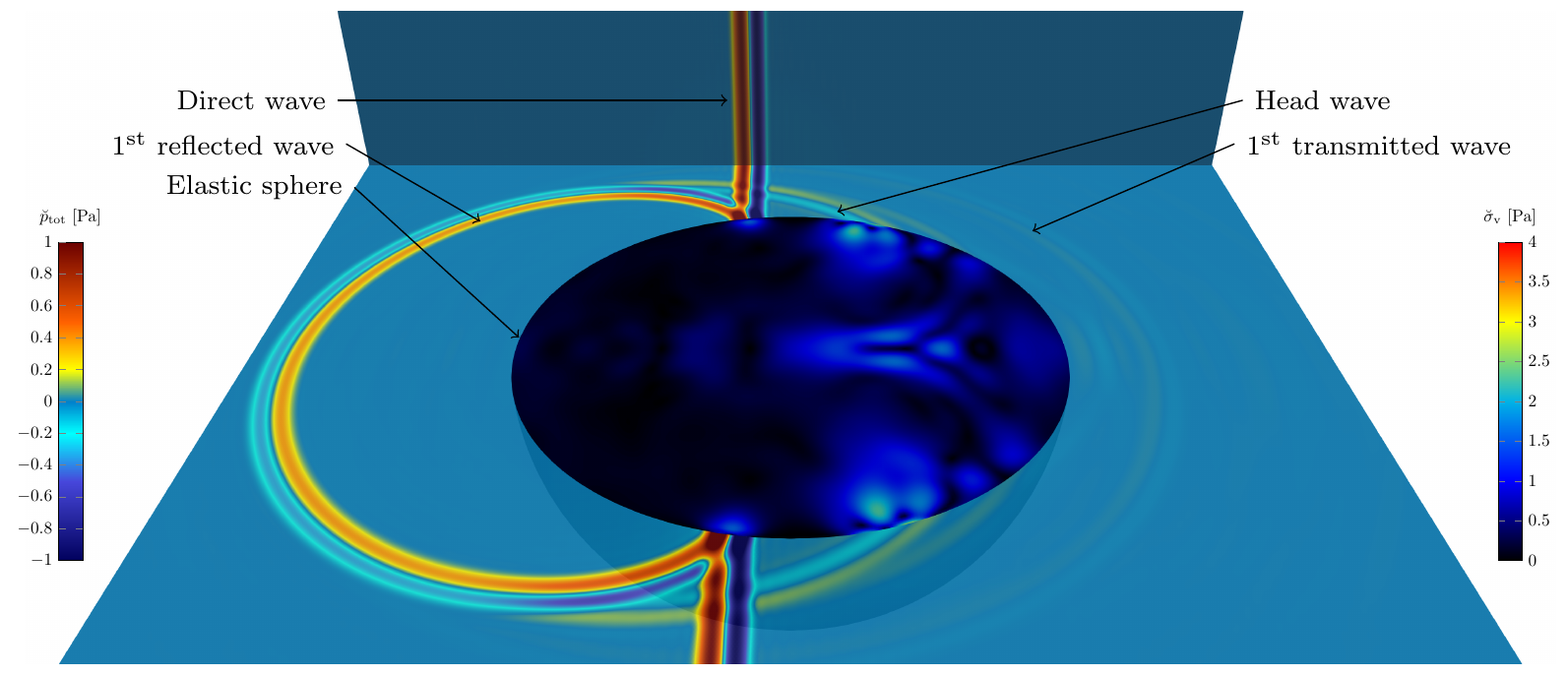}
    \caption{\textbf{Benchmark problems in the time domain}: Visualization of a wavelet (from a far-field point) in the time domain which transmits (the $1^{\mathrm{st}}$ transmitted wave) and reflects (the $1^{\mathrm{st}}$ reflected wave) an elastic sphere (which is cut open for visualization purposes). The $2^{\mathrm{nd}}$ transmitted wave through the elastic sphere takes a lead of the direct wave as the wave speed in the elastic material is larger than that of the fluid. Aliasing is also visible. Two transparent planes have been inserted to visualize the total pressure field. The von Mises stress, $\breve{\sigma}_{\mathrm{v}}$, is visualized in the solid. The visualization was done in \href{http://www.paraview.org/}{Paraview}.}
	\label{Fig1:S5_ESBC_planeWave}
\end{figure}

In \Cref{Fig1:S5_ESBC_planeWave} an example based on the S5 benchmark problem with ESBC is illustrated; An elastic sphere (with parameters given in \Cref{Tab1:sphericalShellParameters}) is impinged by the incident wave in \Cref{Eq1:PlaneWaveTimeDomain}. In this example, the following parameters has been used: $f_{\mathrm{c}}=\SI{1.5}{kHz}$, $\check{N}=2^{10}=1024$ and $T=120/f_{\mathrm{c}}$. In \Cref{Fig1:S5_ESBC_planeWave}, the total pressure is plotted in the fluid, and the von Mises stress given by
\begin{equation}
	\sigma_{\mathrm{v}} =  \sqrt{\frac{(\sigma_{11} - \sigma_{22})^2 + (\sigma_{22} - \sigma_{33})^2 + (\sigma_{11} - \sigma_{33})^2 + 6(\sigma_{23}^2 + \sigma_{13}^2 + \sigma_{12}^2)}{2}} 
\end{equation}
is plotted in the solid domain.

For the benchmark problems, the longitudinal speed and transverse speed in the solid is $c_{\mathrm{s},1}\approx \SI{6001}{ms^{-1}}$ and $c_{\mathrm{s},2}\approx \SI{3208}{ms^{-1}}$, respectively. So since, $c_{\mathrm{s},1}\approx 4c_{\mathrm{f}}$ and $c_{\mathrm{s},2}\approx 2c_{\mathrm{f}}$, the waves traveling through the elastic sphere with the longitudinal wave speed, will transmit through the solid 4 times as fast as the waves in the surrounding fluid (this wave correspond to the $1^{\mathrm{st}}$ transmitted wave in \Cref{Fig1:S5_ESBC_planeWave}). Correspondingly for the waves traveling with the transverse wave speed. This can indeed be observed as well, but the amplitude of the transverse wave traveling at the speed $c_{\mathrm{s},1}$ is only about $2\%$ of the amplitude of the incident wave $P_{\mathrm{inc}}$, and is thus barely visible. Much more energy is transmitted through the wave with the transverse wave speed (the amplitude is approximately $14\%$ of $P_{\mathrm{inc}}$).

Consider finally the S15 benchmark problem with ESBC; A thin shell surrounding a solid sphere is impinged by the incident wave in \Cref{Eq1:P_incOmega} (that is, the incident wave originates from a point source in the near field). The point source is located at a radius of $r_{\mathrm{s}}=\SI{10}{m}$ away from the center of the scatterer (the origin). All other parameters remains the same as in the previous example. 

\begin{figure}
	\centering
	\includegraphics{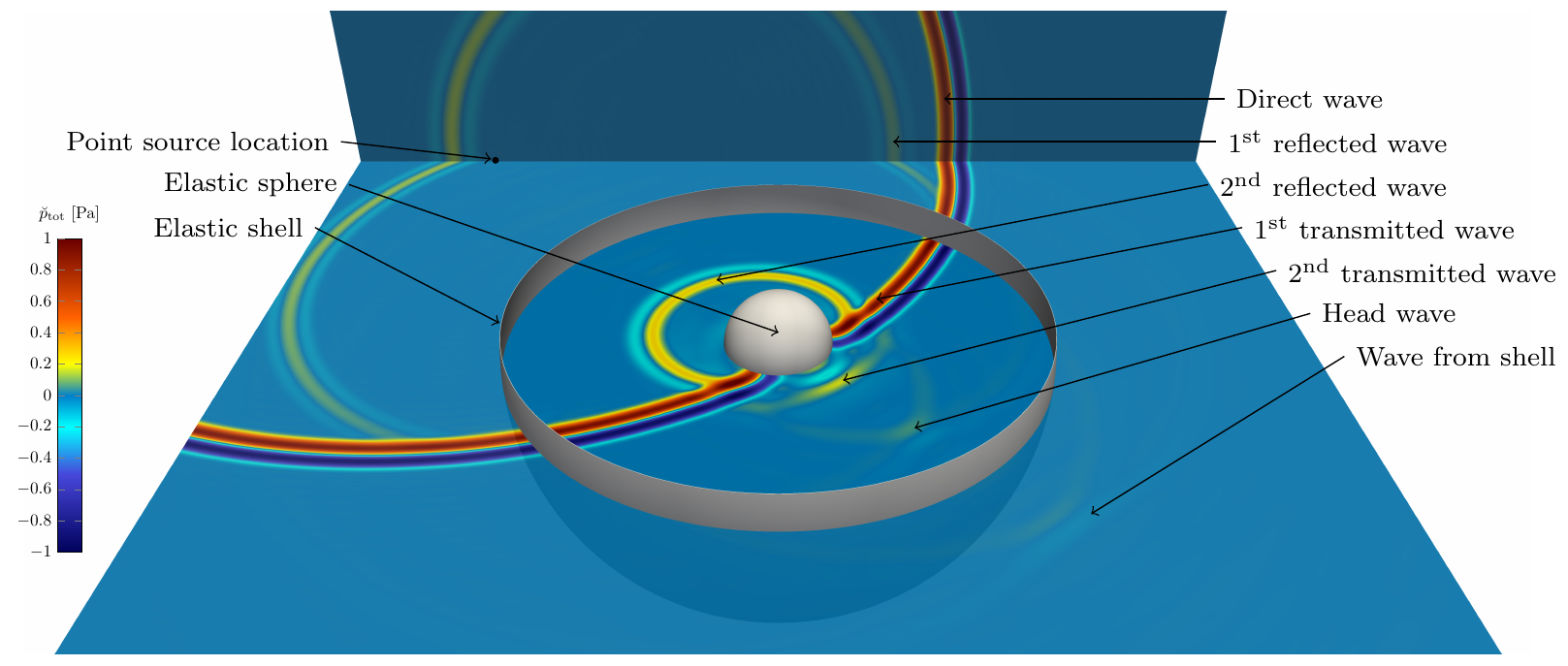}
    \caption{\textbf{Benchmark problems in the time domain}: Visualization of a wavelet (from a point source) in the time domain which transmits (the $1^{\mathrm{st}}$ transmitted wave) and reflects (the $1^{\mathrm{st}}$ reflected wave) the outermost thin shell (which is cut open for visualization purposes), and is then scattered (the $2^{\mathrm{nd}}$ reflected wave) by the innermost solid sphere. The $2^{\mathrm{nd}}$ transmitted wave through the elastic sphere takes a lead of the direct wave as the wave speed in the elastic material is larger than that of the fluid. Aliasing is also visible. Two transparent planes have been inserted to visualize the total pressure field. The displacement field is here not visualized. The visualization was done in \href{http://www.paraview.org/}{Paraview}.}
	\label{Fig1:S15_ESBC_pointCharge}
\end{figure}

In \Cref{Fig1:S15_ESBC_pointCharge} the effects from previous example can again be observed. In addition, waves traveling in the thin shell are reflected backwards through a head wave in the intermediate fluid. In addition, the corresponding waves are transmitted through the shell denoted as ``Wave from shell''.

It should be pointed out that aliasing (pollution of the solution from the previous incident waves due to periodicity) is present, although not visible in these plots. The aliasing can be decreased further by increasing the period $T$. To preserve the size of the \textit{bandwidth} $B=\check{N}/T$, $\check{N}$ must be correspondingly increased. 
\section{Conclusions}
\label{Sec1:conclusions}
An exact solution to 3D scattering problems on spherical symmetric scatterers has been presented. From a computational point of view, the solution is exact in the sense that round-off errors are the only source of errors. However, these round-off errors play a crucial role for higher frequencies (and also for very low frequencies) when implementing the solution naively. In any case, the computational complexity of the solution is $\bigoh(\omega)$.

A set of benchmark problems have been presented for future references. Results have been presented for some of these benchmarks in both the far-field (frequency domain) and the time domain (near-field). The exact solution presents a vast set of parameters for large ranges, which makes it a good reference solution, as many numerical phenomena can occur for different combinations of these parameters.

%

\section*{Acknowledgements}
This work was supported by the Department of Mathematical Sciences at the Norwegian University of Science and Technology and by the Norwegian Defence Research Establishment. The authors wish to thank Trond Kvamsdal who assisted in the proof-reading of the manuscript.

\appendix
\section{The spherical coordinate system}
\label{Sec1:sphericalCoordinates}
\begin{figure}
	\centering
	\includegraphics[scale=0.9]{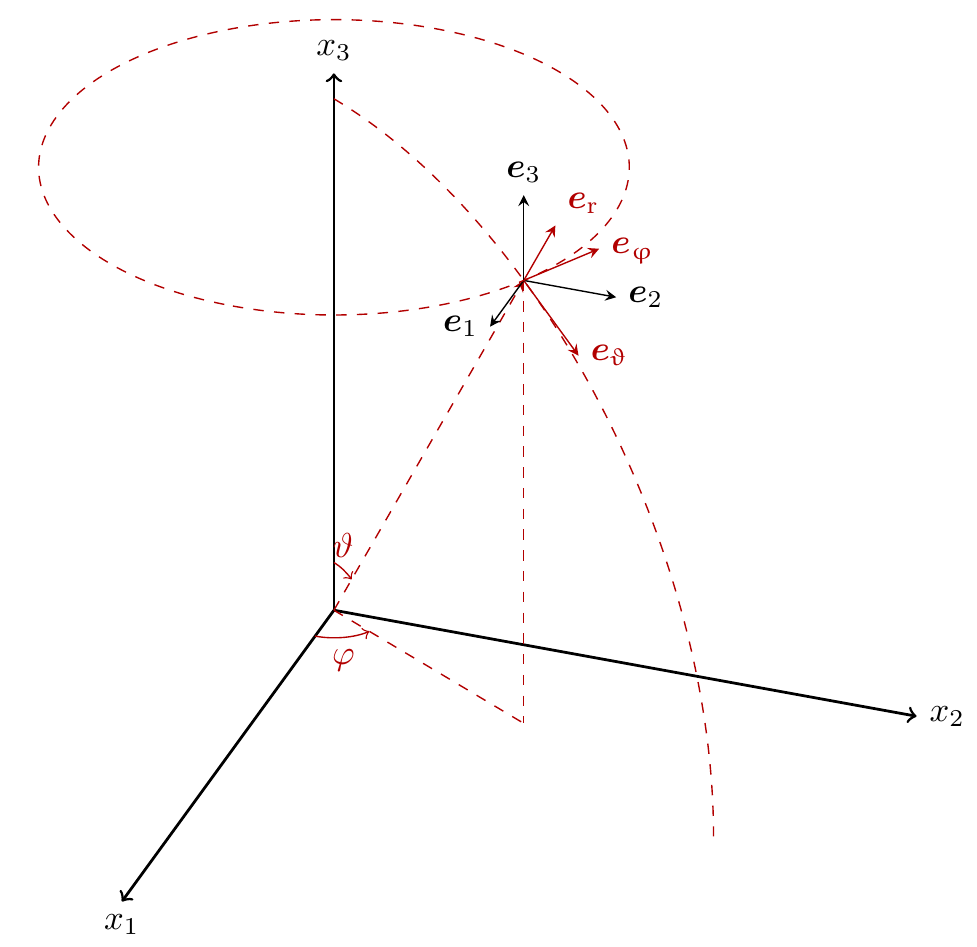}
	\caption{The spherical and Cartesian coordinate system.}
	\label{Fig1:SphericalCoordinateSystem}
\end{figure}

The spherical coordinate system is defined by the transformation $\vec{x}(r,\vartheta,\varphi)=x_i(r,\vartheta,\varphi)\vec{e}_i$, where $\vec{e}_i$ are the standard basis vectors in the Cartesian coordinate system and (see \Cref{Fig1:SphericalCoordinateSystem})
\begin{align}
	x_1 &= r\sin\vartheta\cos\varphi\\
	x_2 &= r\sin\vartheta\sin\varphi\\
	x_3 &= r\cos\vartheta.
\end{align}
The inverse relation is then found to be
\begin{align}
	r &= |\vec{x}|,\quad\text{with}\quad |\vec{x}|=\sqrt{x_1^2+x_2^2+x_3^2}\\
	\vartheta &= \arccos\left(\frac{x_3}{|\vec{x}|}\right)\\
	\varphi &= \operatorname{atan2}(x_2,x_1),
\end{align}
where
\begin{equation}
	\operatorname{atan2}(x_2,x_1) = \begin{cases}
	\arctan(\frac{x_2}{x_1}) & \mbox{if } x_1 > 0\\
	\arctan(\frac{x_2}{x_1}) + \PI & \mbox{if } x_1 < 0 \mbox{ and } x_2 \ge 0\\
	\arctan(\frac{x_2}{x_1}) - \PI & \mbox{if } x_1 < 0 \mbox{ and } x_2 < 0\\
	\frac{\PI}{2} & \mbox{if } x_1 = 0 \mbox{ and } x_2 > 0\\
	-\frac{\PI}{2} & \mbox{if } x_1 = 0 \mbox{ and } x_2 < 0\\
	\text{undefined} & \mbox{if } x_1 = 0 \mbox{ and } x_2 = 0.
	\end{cases}
\end{equation}
Hence, the Jacobian matrix of the spherical transformation is given by
\begin{equation}
	\vec{J}_{\mathrm{s}} = \begin{bmatrix}
		\pderiv{x_1}{r} & \pderiv{x_1}{\vartheta} & \pderiv{x_1}{\varphi}\\
		\pderiv{x_2}{r} & \pderiv{x_2}{\vartheta} & \pderiv{x_2}{\varphi}\\
		\pderiv{x_3}{r} & \pderiv{x_3}{\vartheta} & \pderiv{x_3}{\varphi}
	\end{bmatrix} = \begin{bmatrix}
		\sin\vartheta\cos\varphi & r\cos\vartheta\cos\varphi & -r\sin\vartheta\sin\varphi\\
		\sin\vartheta\sin\varphi & r\cos\vartheta\sin\varphi & r\sin\vartheta\cos\varphi\\
		\cos\vartheta & -r\sin\vartheta & 0
	\end{bmatrix}
\end{equation}
with inverse given by
\begin{equation}
	\vec{J}_{\mathrm{s}}^{-1} = \begin{bmatrix}
		\pderiv{r}{x_1} & \pderiv{r}{x_2} & \pderiv{r}{x_3}\\
		\pderiv{\vartheta}{x_1} & \pderiv{\vartheta}{x_2} & \pderiv{\vartheta}{x_3}\\
		\pderiv{\varphi}{x_1} & \pderiv{\varphi}{x_2} & \pderiv{\varphi}{x_3}
	\end{bmatrix} = \begin{bmatrix}
		\sin\vartheta\cos\varphi & \sin\vartheta\sin\varphi & \cos\vartheta\\
		\frac{1}{r}\cos\vartheta\cos\varphi & \frac{1}{r}\cos\vartheta\sin\varphi & -\frac{1}{r}\sin\vartheta\\
		-\frac{1}{r}\frac{\sin\varphi}{\sin\vartheta} & \frac{1}{r}\frac{\cos\varphi}{\sin\vartheta} & 0
	\end{bmatrix}.
\end{equation}
So for a scalar valued function $\Psi$ the following is obtained (using the chain rule)
\begin{equation}
\label{Eq1:derivativesInSphericalCoordinates}
	\begin{bmatrix}
		\pderiv{\Psi}{r}\\
		\pderiv{\Psi}{\vartheta}\\
		\pderiv{\Psi}{\varphi}
	\end{bmatrix} = \vec{J}_{\mathrm{s}}^{\transpose} 
	\begin{bmatrix}
		\pderiv{\Psi}{x_1}\\
		\pderiv{\Psi}{x_2}\\
		\pderiv{\Psi}{x_3}
	\end{bmatrix}.
\end{equation}
The scale factors in the spherical coordinate system are given by
\begin{equation}
	h_{\mathrm{r}} = \left|\pderiv{\vec{x}}{r}\right| = 1,\quad h_\upvartheta = \left|\pderiv{\vec{x}}{\vartheta}\right| = r,\quad h_\upvarphi = \left|\pderiv{\vec{x}}{\varphi}\right| = r\sin\theta,
\end{equation}
from which the following basis vectors are derived (see \Cref{Fig1:SphericalCoordinateSystem})
\begin{align}
	\vec{e}_{\mathrm{r}} &= \frac{1}{h_{\mathrm{r}}}\pderiv{\vec{x}}{r} = \vec{e}_1\sin\vartheta\cos\varphi + \vec{e}_2\sin\vartheta\sin\varphi + \vec{e}_3\cos\vartheta\\
	\vec{e}_{\upvartheta} &= \frac{1}{h_\upvartheta}\pderiv{\vec{x}}{\vartheta} = \vec{e}_1\cos\vartheta\cos\varphi + \vec{e}_2\cos\vartheta\sin\varphi - \vec{e}_3\sin\vartheta\\
	\vec{e}_{\upvarphi} &= \frac{1}{h_\upvarphi}\pderiv{\vec{x}}{\varphi} = -\vec{e}_1\sin\varphi + \vec{e}_2\cos\varphi.
\end{align}
This can be written in the following matrix form
\begin{equation}
\label{Eq1:XtoSpherical}
	\begin{bmatrix}
		\vec{e}_{\mathrm{r}} & \vec{e}_{\upvartheta} & \vec{e}_{\upvarphi}
	\end{bmatrix} = \vec{J}_{\mathrm{e}}^\transpose\begin{bmatrix}
		\vec{e}_1 & \vec{e}_2 & \vec{e}_3
	\end{bmatrix} = \vec{J}_{\mathrm{e}}^\transpose
\end{equation}
where
\begin{equation}
	\vec{J}_{\mathrm{e}} = \begin{bmatrix}
		\sin\vartheta\cos\varphi & \sin\vartheta\sin\varphi & \cos\vartheta\\
		\cos\vartheta\cos\varphi & \cos\vartheta\sin\varphi & -\sin\vartheta\\
		-\sin\varphi & \cos\varphi & 0
	\end{bmatrix}
\end{equation}
and inverse given by
\begin{equation}
	\vec{J}_{\mathrm{e}}^{-1} = \begin{bmatrix}
		\sin\vartheta\cos\varphi & \cos\vartheta\cos\varphi & -\sin\varphi\\
		\sin\vartheta\sin\varphi & \cos\vartheta\sin\varphi & \cos\varphi\\
		\cos\vartheta & -\sin\vartheta & 0
	\end{bmatrix}.
\end{equation}
So for any vector field
\begin{equation}
	\vec{\Psi} = \Psi_1\vec{e}_1 + \Psi_2\vec{e}_2  + \Psi_3\vec{e}_3 = \Psi_{\mathrm{r}}\vec{e}_{\mathrm{r}} + \Psi_{\upvartheta}\vec{e}_{\upvartheta} + \Psi_{\upvarphi}\vec{e}_{\upvarphi} 
\end{equation}
the following relation is found (by comparing each component)
\begin{equation}
\label{Eq1:SphericalToXfun}
	\begin{bmatrix}
		\Psi_1\\
		\Psi_2\\
		\Psi_3
	\end{bmatrix} = \vec{J}_{\mathrm{e}}^{\transpose} 
	\begin{bmatrix}
		\Psi_{\mathrm{r}}\\
		\Psi_{\upvartheta}\\
		\Psi_{\upvarphi}
	\end{bmatrix},
\end{equation}
and the Jacobian of $\vec{\Psi}$ is given by (using the chain rule)
\begin{align}
\label{Eq1:SphericalToXJacobian}
\begin{split}
	\begin{bmatrix}
		\pderiv{\Psi_1}{x_1} & \pderiv{\Psi_1}{x_2} & \pderiv{\Psi_1}{x_3}\\
		\pderiv{\Psi_2}{x_1} & \pderiv{\Psi_2}{x_2} & \pderiv{\Psi_2}{x_3}\\
		\pderiv{\Psi_3}{x_1} & \pderiv{\Psi_3}{x_2} & \pderiv{\Psi_3}{x_3}
	\end{bmatrix} &= \begin{bmatrix}
		\pderiv{\Psi_1}{r} & \pderiv{\Psi_1}{\vartheta} & \pderiv{\Psi_1}{\varphi}\\
		\pderiv{\Psi_2}{r} & \pderiv{\Psi_2}{\vartheta} & \pderiv{\Psi_2}{\varphi}\\
		\pderiv{\Psi_3}{r} & \pderiv{\Psi_3}{\vartheta} & \pderiv{\Psi_3}{\varphi}
	\end{bmatrix}\vec{J}_{\mathrm{s}}^{-1} \\
	&= \left(\vec{J}_1\Psi_{\mathrm{r}}+\vec{J}_2\Psi_{\upvartheta}+\vec{J}_3\Psi_{\upvarphi} + \vec{J}_{\mathrm{e}}^{\transpose}\begin{bmatrix}
		\pderiv{\Psi_{\mathrm{r}}}{r} & \pderiv{\Psi_{\mathrm{r}}}{\vartheta} & \pderiv{\Psi_{\mathrm{r}}}{\varphi}\\
		\pderiv{\Psi_{\upvartheta}}{r} & \pderiv{\Psi_{\upvartheta}}{\vartheta} & \pderiv{\Psi_{\upvartheta}}{\varphi}\\
		\pderiv{\Psi_{\upvarphi}}{r} & \pderiv{\Psi_{\upvarphi}}{\vartheta} & \pderiv{\Psi_{\upvarphi}}{\varphi}
	\end{bmatrix}\right)\vec{J}_{\mathrm{s}}^{-1}
	\end{split}
\end{align}
where
\begin{equation*}
	\vec{J}_1 = \begin{bmatrix}
		0 & \cos\vartheta\cos\varphi & -\sin\vartheta\sin\varphi\\
		0 & \cos\vartheta\sin\varphi & \sin\vartheta\cos\varphi\\
		0 & -\sin\vartheta & 0
	\end{bmatrix},\quad\vec{J}_2 = \begin{bmatrix}
		0 & -\sin\vartheta\cos\varphi & -\cos\vartheta\sin\varphi\\
		0 & -\sin\vartheta\sin\varphi & \cos\vartheta\cos\varphi\\
		0 & -\cos\vartheta & 0
	\end{bmatrix},\quad\vec{J}_3 = \begin{bmatrix}
		0 & 0 & -\cos\varphi\\
		0 & 0 & -\sin\varphi\\
		0 & 0 & 0
	\end{bmatrix}.
\end{equation*}
Using \Cref{Eq1:derivativesInSphericalCoordinates,Eq1:SphericalToXfun,Eq1:XtoSpherical}, the following formulas are obtained
\begin{align}
	\nabla\Psi &= \pderiv{\Psi}{r}\vec{e}_{\mathrm{r}}+\frac{1}{r}\pderiv{\Psi}{\vartheta}\vec{e}_{\upvartheta}+\frac{1}{r\sin\vartheta}\pderiv{\Psi}{\varphi}\vec{e}_{\upvarphi} \label{Eq1:delScalarSpherical} \\	
	\nabla^2\Psi &= \frac{1}{r^2}\pderiv{}{r}\left(r^2\pderiv{\Psi}{r}\right) + \frac{1}{r^2\sin\vartheta}\pderiv{}{\vartheta}\left(\sin\vartheta\pderiv{\Psi}{\vartheta}\right) +\frac{1}{r^2\sin^2\vartheta}\pderiv[2]{\Psi}{\varphi} \label{Eq1:laplaceScalarSpherical}\\
	\nabla\cdot\vec{\Psi} &= \frac{1}{r^2}\pderiv{(r^2\Psi_{\mathrm{r}})}{r} + \frac{1}{r\sin\vartheta}\pderiv{(\Psi_{\upvartheta}\sin\vartheta)}{\vartheta}+\frac{1}{r\sin\vartheta}\pderiv{\Psi_{\upvarphi}}{\varphi} \label{Eq1:DelDotVecSpherical} \\ 
	\begin{split}\nabla^2\vec{\Psi} &= \left(\nabla^2\Psi_{\mathrm{r}} - \frac{2}{r^2}\Psi_{\mathrm{r}}-\frac{2}{r^2\sin\vartheta}\pderiv{(\Psi_{\upvartheta}\sin\vartheta)}{\vartheta} -\frac{2}{r^2\sin\vartheta}\pderiv{\Psi_{\upvarphi}}{\varphi}\right)\vec{e}_{\mathrm{r}} \\
	&{\hskip1em\relax}+ \left(\nabla^2\Psi_{\upvartheta}-\frac{1}{r^2\sin^2\vartheta}\Psi_{\upvartheta} + \frac{2}{r^2}\pderiv{\Psi_{\mathrm{r}}}{\vartheta} -\frac{2\cos\vartheta}{r^2\sin^2\vartheta}\pderiv{\Psi_{\upvarphi}}{\varphi}\right)\vec{e}_{\upvartheta} \\
	&{\hskip1em\relax}+ \left(\nabla^2\Psi_{\upvarphi} - \frac{1}{r^2\sin^2\vartheta}\Psi_{\upvarphi} + \frac{2}{r^2\sin\vartheta}\pderiv{\Psi_{\mathrm{r}}}{\varphi}+\frac{2\cos\vartheta}{r^2\sin^2\vartheta}\pderiv{\Psi_{\upvartheta}}{\varphi}\right)\vec{e}_{\upvarphi} \end{split}\label{Eq1:LapVecPotSpherical}\\
	\begin{split}
	\nabla\times\vec{\Psi} &=  \frac{1}{r\sin\vartheta}\left(\pderiv{}{\vartheta}\left(\Psi_\upvarphi\sin\vartheta\right) - \pderiv{\Psi_\upvartheta}{\varphi}\right)\vec{e}_{\mathrm{r}} + \frac{1}{r}\left(\frac{1}{\sin\vartheta}\pderiv{\Psi_{\mathrm{r}}}{\varphi} - \pderiv{}{r}\left(r\Psi_\upvarphi\right)\right)\vec{e}_\upvartheta\\
	&{\hskip1em\relax}+ \frac{1}{r}\left(\pderiv{}{r}\left(r\Psi_\upvartheta\right) - \pderiv{\Psi_{\mathrm{r}}}{\vartheta}\right)\vec{e}_\upvarphi.
	\end{split}\label{Eq1:CrossVecPotSpherical}
\end{align}

\section{Linear elasticity}
\label{Sec1:LinearElasticity}
In this section the needed formulas from linear elasticity used in this paper are listed. A more comprehencive introduction to linear elasticity may be found in~\cite{Gould1994itl}. From the displacement field $\vec{u} = u_i\vec{e}_i$ the strain field, $\varepsilon_{ij}$, is defined by
\begin{equation}
	\varepsilon_{ij} = \frac{1}{2}\left(\pderiv{u_i}{x_j} + \pderiv{u_j}{x_i}\right)
\end{equation}
from which the stress field, $\sigma_{ij}$, can be obtained through the constitutive relation\footnote{This representation is often referred to as the Voight notation.} (derived from the generalized Hooke's law) 
\begin{equation}
	\begin{bmatrix}
		\sigma_{11}\\
		\sigma_{22}\\
		\sigma_{33}\\
		\sigma_{23}\\
		\sigma_{13}\\
		\sigma_{12}\\
	\end{bmatrix} = \vec{C}
	\begin{bmatrix}
		\varepsilon_{11}\\
		\varepsilon_{22}\\
		\varepsilon_{33}\\
		2\varepsilon_{23}\\
		2\varepsilon_{13}\\
		2\varepsilon_{12}\\
	\end{bmatrix}\quad\text{with}\quad\vec{C}= \begin{bmatrix}
		K+\frac{4G}{3} & K-\frac{2G}{3} & K-\frac{2G}{3} & 0 & 0 & 0\\
		K-\frac{2G}{3} & K+\frac{4G}{3} & K-\frac{2G}{3} & 0 & 0 & 0\\
		K-\frac{2G}{3} & K-\frac{2G}{3} & K+\frac{4G}{3} & 0 & 0 & 0\\
		0 & 0 & 0 & G & 0 & 0\\
		0 & 0 & 0 & 0 & G & 0\\
		0 & 0 & 0 & 0 & 0 & G
	\end{bmatrix}
\end{equation}
where it has been assumed that the elastic material is isotropic. Note that
\begin{equation}
	\vec{C}^{-1} = \frac{1}{E}\begin{bmatrix}
		1 & -\nu & -\nu & 0 & 0 & 0\\
		-\nu & 1 & -\nu & 0 & 0 & 0\\
		-\nu & -\nu & 1 & 0 & 0 & 0\\
		0 & 0 & 0 & 2(1+\nu) & 0 & 0\\
		0 & 0 & 0 & 0 & 2(1+\nu) & 0\\
		0 & 0 & 0 & 0 & 0 & 2(1+\nu)
	\end{bmatrix}.
\end{equation}
In~\cite[p. 19]{Gould1994itl} the transformation formula for the stress tensor from an arbitrary coordinate system to another can be found. If $\vec{e}_i$ and $\tilde{\vec{e}}_i$ represents the basis vectors of these two coordinate systems and the stress field is known in the first coordinate system, then the stress field in terms of the second coordinate system is found by
\begin{equation}
	\tilde{\sigma}_{ij} = \alpha_{ik}\alpha_{\mathrm{s}l}\sigma_{kl},
\end{equation}
where
\begin{equation}
	\alpha_{ij} = \cos(\tilde{\vec{e}}_i, \vec{e}_j) = \tilde{\vec{e}}_i\cdot\vec{e}_j
\end{equation}
represents the cosine of the angle between the axes corresponding to the vectors $\tilde{\vec{e}}_i$ and $\vec{e}_i$. Letting $\tilde{\vec{e}}_1 = \vec{e}_{\mathrm{r}}$, $\tilde{\vec{e}}_2 = \vec{e}_\upvartheta$ and $\tilde{\vec{e}}_3 = \vec{e}_\upvarphi$ (the basis vectors in the spherical coordinate system), and $\{\vec{e}_1, \vec{e}_2, \vec{e}_3\}$ the standard basis vectors in Cartesian coordinates, one gets (using \Cref{Eq1:XtoSpherical})
\begin{equation}
	[\alpha_{ij}] = \begin{bmatrix}
	\sin\vartheta\cos\varphi & \sin\vartheta\sin\varphi & \cos\vartheta\\
	\cos\vartheta\cos\varphi & \cos\vartheta\sin\varphi & -\sin\vartheta\\
	-\sin\varphi & \cos\varphi & 0
	\end{bmatrix} = \vec{J}_{\mathrm{e}}.
\end{equation}
This yields the relation
\begin{equation}\label{Eq1:StressTransformFromCartToSpherical}
	\begin{bmatrix}
		\sigma_{\mathrm{rr}}\\
		\sigma_{\upvartheta\upvartheta}\\
		\sigma_{\upvarphi\upvarphi}\\
		\sigma_{\upvartheta\upvarphi}\\
		\sigma_{\mathrm{r}\upvarphi}\\
		\sigma_{\mathrm{r}\upvartheta}
	\end{bmatrix} = \vec{D}
	\begin{bmatrix}
		\sigma_{11}\\
		\sigma_{22}\\
		\sigma_{33}\\
		\sigma_{23}\\
		\sigma_{13}\\
		\sigma_{12}
	\end{bmatrix}
\end{equation}
where
\begin{equation*}
	\vec{D} =\begin{bmatrix}
	 \sin^2\vartheta\cos^2\varphi&\sin^2\vartheta\sin^2\varphi&\cos^2\vartheta&\sin2\vartheta\sin\varphi&\sin2\vartheta\cos\varphi&\sin^2\vartheta\sin2\varphi\\ 
	 \cos^2\vartheta\cos^2\varphi&\cos^2\vartheta\sin^2\varphi&\sin^2\vartheta&-\sin2\vartheta\sin\varphi&-\sin2\vartheta\cos\varphi&\cos^2\vartheta\sin2\varphi\\ 
	 \sin^2\varphi&\cos^2\varphi&0&0&0&-\sin2\varphi\\ 
	 -\frac12\cos\vartheta\sin2\varphi&\frac12\cos\vartheta\sin2\varphi&0&-\sin\vartheta\cos\varphi&\sin\vartheta\sin\varphi&\cos\vartheta\cos2\varphi\\ 
	 -\frac12\sin\vartheta\sin2\varphi&\frac12\sin\vartheta\sin2\varphi&0&\cos\vartheta\cos\varphi&-\cos\vartheta\sin\varphi&\sin\vartheta\cos2\varphi\\ 
	 \frac12\sin2\vartheta\cos^2\varphi&\frac12\sin2\vartheta\sin^2\varphi&-\frac12\sin2\vartheta&\cos2\vartheta\sin\varphi&\cos2\vartheta\cos\varphi&\frac12\sin2\vartheta\sin2\varphi
	\end{bmatrix}.
\end{equation*}
The inverse relation is found by inverting the matrix $\vec{D}$, which takes the form
\begin{equation*}
	\vec{D}^{-1} =\begin{bmatrix}
	 \sin^2\vartheta\cos^2\varphi & \cos^2\vartheta\cos^2\varphi & \sin^2\varphi & -\cos\vartheta\sin2\varphi & -\sin\vartheta\sin2\varphi & \sin2\vartheta\cos^2\varphi \\
	 \sin^2\vartheta\sin^2\varphi & \cos^2\vartheta\sin^2\varphi & \cos^2\varphi & \cos\vartheta\sin2\varphi & \sin\vartheta\sin2\varphi & \sin2\vartheta\sin^2\varphi\\
	 \cos^2\vartheta & \sin^2\vartheta & 0 & 0 & 0 & -\sin2\vartheta\\
	 \frac12\sin2\vartheta\sin\varphi & -\frac12\sin2\vartheta\sin\varphi & 0 & -\sin\vartheta\cos\varphi & \cos\vartheta\cos\varphi & \sin\varphi\cos2\vartheta\\
	 \frac12\sin\vartheta\sin2\varphi & -\frac12\sin2\vartheta\cos\varphi & 0 & \sin\vartheta\sin\varphi & -\cos\vartheta\sin\varphi & \cos2\vartheta\cos\varphi\\
	 \frac12\sin^2\vartheta\sin2\varphi & \frac12\cos^2\vartheta\sin2\varphi &-\frac12\sin2\varphi & \cos\vartheta\cos2\varphi & \cos2\varphi\sin\vartheta &\frac12\sin2\vartheta\sin2\varphi
	\end{bmatrix}.
\end{equation*}
Moreover,
\begin{equation}
\label{Eq1:constitutiveRelationSpherical}
	\begin{bmatrix}
		\sigma_{\mathrm{rr}}\\
		\sigma_{\upvartheta\upvartheta}\\
		\sigma_{\upvarphi\upvarphi}\\
		\sigma_{\upvartheta \upvarphi}\\
		\sigma_{\mathrm{r} \upvarphi}\\
		\sigma_{\mathrm{r}\upvartheta}\\
	\end{bmatrix} = \vec{C}
	\begin{bmatrix}
		\varepsilon_{\mathrm{rr}}\\
		\varepsilon_{\upvartheta\upvartheta}\\
		\varepsilon_{\upvarphi\upvarphi}\\
		2\varepsilon_{\upvartheta \upvarphi}\\
		2\varepsilon_{r \upvarphi}\\
		2\varepsilon_{\mathrm{r}\upvartheta}\\
	\end{bmatrix},
\end{equation}
where (cf~\cite[p. 150]{Slaughter2002tlt})
\begin{equation}
\label{Eq1:strainsInSpherical}
\begin{split}
	\varepsilon_{\mathrm{rr}} &= \pderiv{u_{\mathrm{r}}}{r}\\
	\varepsilon_{\upvartheta\upvartheta} &= \frac{1}{r}\left(\pderiv{u_{\upvartheta}}{\vartheta} + u_{\mathrm{r}}\right)\\
	\varepsilon_{\upvarphi\upvarphi} &= \frac{1}{r\sin\vartheta}\left(\pderiv{u_{\varphi}}{\varphi} + u_{\mathrm{r}}\sin\vartheta + u_{\upvartheta}\cos\vartheta\right)\\
	\varepsilon_{\upvartheta\upvarphi} &= \frac{1}{2r}\left(\frac{1}{\sin\vartheta}\pderiv{u_{\upvartheta}}{\varphi} + \pderiv{u_{\upvarphi}}{\vartheta} - u_{\upvarphi}\cot\vartheta\right)\\
	\varepsilon_{\mathrm{r}\upvarphi} &= \frac{1}{2}\left(\frac{1}{r\sin\vartheta}\pderiv{u_{\mathrm{r}}}{\varphi} + \pderiv{u_{\upvarphi}}{r} - \frac{u_{\upvarphi}}{r}\right)\\
	\varepsilon_{\mathrm{r}\upvartheta} &= \frac{1}{2}\left(\frac{1}{r}\pderiv{u_{\mathrm{r}}}{\vartheta} + \pderiv{u_{\upvartheta}}{r} - \frac{u_{\upvartheta}}{r}\right).
\end{split}
\end{equation}
Finally, note that Navier's equation of motion (\Cref{Eq1:navier}) in spherical coordinates are given by (cf.~\cite[p. 189]{Slaughter2002tlt})
\begin{align}
\pderiv{\sigma_{\mathrm{rr}}}{r} + \frac{1}{r}\pderiv{\sigma_{\mathrm{r}\upvartheta}}{\vartheta}+\frac{1}{r\sin\vartheta} \pderiv{\sigma_{\mathrm{r} \upvarphi}}{\varphi} + \frac{1}{r}\left(2\sigma_{\mathrm{r}\mathrm{r}} - \sigma_{\upvartheta\upvartheta} - \sigma_{\upvarphi\upvarphi} + \sigma_{\mathrm{r}\upvartheta}\cot\vartheta\right) +\omega^2\rho_{\mathrm{s}}u_{\mathrm{r}} &= 0\label{Eq1:navierSpherical1}\\
	\pderiv{\sigma_{\mathrm{r}\upvartheta}}{r} + \frac{1}{r}\pderiv{\sigma_{\upvartheta\upvartheta}}{\vartheta}+\frac{1}{r\sin\vartheta} \pderiv{\sigma_{\upvartheta \upvarphi}}{\varphi} + \frac{1}{r}\left[(\sigma_{\upvartheta\upvartheta} - \sigma_{\upvarphi\upvarphi})\cot\vartheta + 3\sigma_{\mathrm{r}\upvartheta} \right] +\omega^2\rho_{\mathrm{s}}u_\upvartheta &= 0\label{Eq1:navierSpherical2}\\
	\pderiv{\sigma_{\mathrm{r}\upvarphi}}{r} + \frac{1}{r}\pderiv{\sigma_{\upvartheta\upvarphi}}{\vartheta}+\frac{1}{r\sin\vartheta} \pderiv{\sigma_{\upvarphi \upvarphi}}{\varphi} + \frac{1}{r}(2\sigma_{\upvartheta\upvarphi}\cot\vartheta + 3\sigma_{\mathrm{r}\upvarphi}) +\omega^2\rho_{\mathrm{s}}u_\upvarphi &= 0. \label{Eq1:navierSpherical3}
\end{align}

\section{Fundamental functions}
Exact solutions for scattering problems on spherical symmetric scatterers are heavily based on the spherical coordinate system defined in \Cref{Sec1:sphericalCoordinates}. Some fundamental functions then naturally arise, and the notation will briefly be presented in the following.

\subsection{Legendre polynomials}
\label{subsec:legendre}
The Legendre polynomials are defined recursively by (cf.~\cite[p. 332]{Abramowitz1965hom})
\begin{equation}
	(n+1)\legendre_{n+1}(x)=(2n+1)x\legendre_n(x)-n\legendre_{n-1}(x)
\end{equation}
starting with $\legendre_0(x) = 1$ and $\legendre_1(x) = x$. From orthogonality property~\cite[\href{http://functions.wolfram.com/05.03.21.0006.01}{05.03.21.0006.01}]{WolframResearch2016m}
\begin{equation}
	\int_{-1}^1 \legendre_m(x)\legendre_n(x)\idiff x = \frac{2}{2n+1}\delta_{mn},
\end{equation}
with $\delta_{mn}$ being the Kronecker delta function, one can do a simple substitution to obtain the following expression
\begin{equation}\label{Eq1:legendreOrthogonality}
	\int_0^\PI \legendre_m(\cos\vartheta)\legendre_n(\cos\vartheta)\sin\vartheta\idiff \vartheta = \frac{2}{2n+1}\delta_{mn}.
\end{equation}
Note the following identity from the expanded Legendre equation
\begin{equation}\label{Eq1:expandedLegendreEquationIdentity}
	\deriv[2]{}{\vartheta} \legendre_n(\cos\vartheta) + \cot\vartheta \deriv{}{\vartheta} \legendre_n(\cos\vartheta) = -n(n+1)\legendre_n(\cos\vartheta).
\end{equation}
The associated Legendre polynomials is a generalization of the Legendre polynomials as they are defined by
\begin{equation}
	\legendre_n^m(x) = (-1)^m(1-x^2)^{\frac{m}{2}}\pderiv[m]{}{x} \legendre_n(x).
\end{equation}
A convenient result of this is the following relation
\begin{equation}
	\legendre_n^1(\cos\vartheta) = \deriv{}{\vartheta} \legendre_n(\cos\vartheta)\label{Eq1:LegendreRelation1}.
\end{equation}
Let $\left\{Q_n^{(j)}\right\}_{j\in\N}$ be a set of functions defined by
\begin{equation}
	Q_n^{(j)}(\vartheta) = \deriv[j]{}{\vartheta}\legendre_n(\cos\vartheta),
\end{equation}
the first four of which are given by
\begin{align}
\begin{split}\label{Eq1:Qs}
	Q_n^{(0)}(\vartheta) &= \legendre_n(\cos\vartheta)\\
	Q_n^{(1)}(\vartheta) &= - \legendre_n'(\cos\vartheta)\sin\vartheta\\
	Q_n^{(2)}(\vartheta) &= - \legendre_n'(\cos\vartheta)\cos\vartheta + \legendre_n''(\cos\vartheta)\sin^2\vartheta\\
	Q_n^{(3)}(\vartheta) &=  \legendre_n'(\cos\vartheta)\sin\vartheta  +\frac32 \legendre_n''(\cos\vartheta)\sin 2\vartheta - \legendre_n'''(\cos\vartheta)\sin^3\vartheta\\
	\end{split}
\end{align}
where the derivatives are found by the recursion relations
\begin{align}
	&(n+1)\legendre_{n+1}'(x)=(2n+1)\left[\legendre_n(x)+x \legendre_n'(x)\right]-n\legendre_{n-1}'(x)\\
	&(n+1)\legendre_{n+1}''(x)=(2n+1)\left[2\legendre_n'(x)+ x\legendre_n''(x)\right]-n\legendre_{n-1}''(x)\\
	&(n+1)\legendre_{n+1}'''(x)=(2n+1)\left[3\legendre_n''(x)+x \legendre_n'''(x)\right]-n\legendre_{n-1}'''(x)
\end{align}
starting with 
\begin{align*}
	&\legendre_0'(x) = 0,\quad \legendre_1'(x) = 1, \quad \legendre_2'(x) = 3x\\
	&\legendre_0''(x) = 0,\quad \legendre_1''(x) = 0, \quad \legendre_2''(x) = 3, \quad \legendre_3''(x) = 15x\\
	&\legendre_0'''(x) = 0,\quad \legendre_1'''(x) = 0, \quad \legendre_2'''(x) = 0, \quad \legendre_3'''(x) = 15, \quad \legendre_4'''(x) = 105x.
\end{align*}
Note that the formulas in \Cref{Eq1:Qs} can be rewritten in the following way
\begin{align}
	Q_n^{(1)}(\vartheta) &= \frac{n}{\sin\vartheta} \left[ \legendre_n(\cos\vartheta)\cos\vartheta - \legendre_{n-1}(\cos\vartheta)\right]\\
	Q_n^{(2)}(\vartheta) &= \frac{n}{\sin^2\vartheta} \left[-\left(n\sin^2\vartheta+1 \right)\legendre_n(\cos\vartheta) -  \legendre_{n-1}(\cos\vartheta)\cos\vartheta\right].
\end{align}
From \Cref{Eq1:expandedLegendreEquationIdentity} the following relations can be obtained
\begin{align}\label{Eq1:expandedLegendreEquationIdentity2}
	Q_n^{(2)}(\vartheta) &= - Q_n^{(1)}(\vartheta)\cot\vartheta -n(n+1)Q_n^{(0)}(\vartheta)\\
	Q_n^{(3)}(\vartheta) &= - Q_n^{(2)}(\vartheta)\cot\vartheta + Q_n^{(1)}(\vartheta)\cot^2\vartheta + (-n^2-n+1)Q_n^{(1)}(\vartheta).\label{Eq1:expandedLegendreEquationIdentity3}
\end{align}

\subsection{Spherical Bessel and Hankel functions}
\label{subsec:sphericalBesselAndHankel}
The Bessel functions of the first kind can be defined by~\cite[p. 360]{Abramowitz1965hom}
\begin{equation}
	\besselJ_\upsilon(x) = \sum_{m=0}^\infty \frac{(-1)^m}{m! \GAMMA(m+\upsilon+1)}\left(\frac{x}{2}\right)^{2m+\upsilon},
\end{equation}
while the Bessel functions of the second kind are defined by
\begin{equation}
	\besselY_\upsilon(x) = \frac{\besselJ_\upsilon(x) \cos(\upsilon\PI)-\besselJ_{-\upsilon}(x)}{\sin(\upsilon\PI)},
\end{equation}
where
\begin{equation}
	\besselY_n(x) = \lim_{\upsilon\to n} \besselY_\upsilon(x)
\end{equation}
whenever $n\in\Z$ (cf.~\cite[p. 358]{Abramowitz1965hom}). These definitions may be used to define the \textit{spherical Bessel functions}. The spherical Bessel functions of the first kind are defined by (cf.~\cite[p. 437]{Abramowitz1965hom})
\begin{equation}
	\besselj_n(x) = \sqrt{\frac{\PI}{2x}}\besselJ_{n+\frac{1}{2}}(x)
\end{equation}
and the second kind are defined by
\begin{equation}
	\bessely_n(x) = \sqrt{\frac{\PI}{2x}}\besselY_{n+\frac{1}{2}}(x).
\end{equation}
Some important limits of the spherical Bessel function of the first kind at the origin are~\cite[\href{http://functions.wolfram.com/03.21.20.0016.01}{03.21.20.0016.01} and \href{http://functions.wolfram.com/03.21.20.0017.01}{03.21.20.0017.01}]{WolframResearch2016m}
\begin{align}
	&\lim_{x\to 0} \besselj_0(x) = 1,\quad \lim_{x\to 0} \besselj_n(x) = 0 \quad\forall n\in\N^*\\
	&\lim_{x\to 0} \deriv{}{x}\besselj_1(x) = \frac{1}{3},\quad \lim_{x\to 0} \deriv{}{x}\besselj_n(x) = 0 \quad\forall n\in\N\setminus\{1\}\\
	&\lim_{x\to 0} \deriv[2]{}{x}\besselj_0(x) = -\frac{1}{3},\quad\lim_{x\to 0} \deriv[2]{}{x}\besselj_2(x) = \frac{2}{15}, \quad\lim_{x\to 0} \deriv[2]{}{x}\besselj_n(x) = 0 \quad\forall n\in\N\setminus\{0,2\}.
\end{align}
From this the following limits are obtained
\begin{equation}\label{Eq1:BesselLimits}
	\lim_{x\to 0^+} \frac{\besselj_n(x)}{x} = \begin{cases} \infty & n = 0\\
	\frac{1}{3} & n = 1\\
	0 & n > 1
	\end{cases}
\end{equation}
and
\begin{equation}\label{Eq1:BesselLimits2}
	\lim_{x\to 0^+} \frac{\besselj_n(x)}{x^2} = \begin{cases} \infty & n = 0,1\\
	\frac{1}{15} & n = 2\\
	0 & n > 2.
	\end{cases}
\end{equation}
A couple of convenient identities involving the derivatives of the spherical Bessel functions are given by~\cite[\href{http://functions.wolfram.com/03.21.20.0007.01}{03.21.20.0007.01} and \href{http://functions.wolfram.com/03.21.20.0007.01}{03.21.20.0008.01}]{WolframResearch2016m}
\begin{align}\label{Eq1:BesselDerivIdentity1}
	\deriv{}{x}Z_n^{(i)}(x) &=  Z_{n-1}^{(i)}(x) -\frac{n+1}{x}Z_n^{(i)}(x)\\
	\deriv{}{x}Z_n^{(i)}(x) &= \frac{n}{x}Z_n^{(i)}(x) - Z_{n+1}^{(i)}(x)\label{Eq1:BesselDerivIdentity2}
\end{align}
for $i=1,2$. By combining these two formulas, one can compute higher order derivatives. For example
\begin{equation}\label{Eq1:BesselDerivIdentity3}
	\deriv[2]{}{x}Z_n^{(i)}(x) =  \left[\frac{n(n-1)}{x^2}-1\right] Z_n^{(i)}(x) + \frac{2}{x}Z_{n+1}^{(i)}(x).
\end{equation}
The spherical Hankel functions of the first and second kind can now be expressed by
\begin{equation}
	\hankel^{(1)}_n(x) = \besselj_n(x) +  \imag \bessely_n(x)
\end{equation}
and 
\begin{equation}
	\hankel^{(2)}_n(x) = \besselj_n(x) -  \imag \bessely_n(x).
\end{equation}
respectively. Two important limits for spherical Hankel functions are~\cite[p. 25]{Ihlenburg1998fea}
\begin{align}
	\lim_{x\to\infty} x\euler^{-\imag x} \hankel^{(1)}_n(x) = \imag^{-n-1}\label{Eq1:sphericalHankelLimit}\\
	\lim_{x\to\infty} x\euler^{\imag x} \hankel^{(2)}_n(x) = \imag^{n+1}
\end{align}
One can trivially show that the \Cref{Eq1:BesselDerivIdentity1,Eq1:BesselDerivIdentity2,Eq1:BesselDerivIdentity3} holds for spherical Hankel functions as well
\begin{align}
	\deriv{}{x}\hankel^{(i)}_n(x) &=  \hankel^{(i)}_{n-1}(x) -\frac{n+1}{x}\hankel^{(i)}_n(x)\\
	\deriv{}{x}\hankel^{(i)}_n(x) &= \frac{n}{x}\hankel^{(i)}_n(x) - \hankel^{(i)}_{n+1}(x)\label{Eq1:HankelDerivIdentity2}\\
	\deriv[2]{}{x}\hankel^{(i)}_n(x) &= \left[\frac{n(n-1)}{x^2}-1\right] \hankel^{(i)}_n(x) + \frac{2}{x}\hankel^{(i)}_{n+1}(x),
\end{align}
for $i=1,2$.
\section{The incident wave}
\label{Sec1:incidentWave}
The coefficients $F_n^{(1)}$ and $F_n^{(2)}$ in \Cref{Eq1:IncidentWaveConds} may be computed by using the orthogonality property of the Legendre polynomials in \Cref{Eq1:legendreOrthogonality}. In fact, any square integrable function $\Psi(\vartheta)$ on the interval $[0,\PI]$ can be written as (see~\cite[p. 27]{Ihlenburg1998fea})
\begin{equation}
	\Psi(\vartheta) = \sum_{n=0}^\infty \Psi_n \legendre_n(\cos\vartheta)
\end{equation}
where
\begin{equation}\label{Eq1:H_n}
	\Psi_n = \frac{2n+1}{2}\int_0^\PI \Psi(\vartheta) \legendre_n(\cos\vartheta)\sin\vartheta\idiff\vartheta.
\end{equation}
For example, a plane wave traveling along the $x_3$-axis can be expanded as~\cite[10.1.47]{Abramowitz1965hom}
\begin{equation}\label{Eq1:planeWave}
	p_{\mathrm{inc}}(\vec{x},\omega) = P_{\mathrm{inc}}(\omega)\euler^{\imag k_1 x_3} = P_{\mathrm{inc}}(\omega)\euler^{\imag k_1 r\cos\vartheta} = P_{\mathrm{inc}}(\omega)\sum_{n=0}^\infty (2n+1)\imag^n j_n(k_1 r)\legendre_n(\cos\vartheta)
\end{equation}
such that
\begin{align}
	F_n^{(1)} = P_{\mathrm{inc}}(\omega) (2n+1)\imag^n j_n(k_1 R_{0,1})\\
	F_n^{(2)} = P_{\mathrm{inc}}(\omega) (2n+1)\imag^n k_1 j_n'(k_1 R_{0,1}).
\end{align}
Another example of an incident wave satisfying the axisymmetry property, is a wave due to a point source located at $\vec{x}_{\mathrm{s}} = -r_{\mathrm{s}}\vec{e}_3$. The incident wave can then be expressed with the fundamental solution of the Helmholtz equation
\begin{equation}
	p_{\mathrm{inc}}(\vec{x},\omega) = P_{\mathrm{inc}}(\omega)\frac{r_{\mathrm{s}}}{|\vec{x}_{\mathrm{s}}-\vec{x}|}\euler^{\imag k_1 |\vec{x}_{\mathrm{s}}-\vec{x}|},\quad |\vec{x}_{\mathrm{s}}-\vec{x}| = \sqrt{r^2+2r_{\mathrm{s}}r\cos\vartheta + r_{\mathrm{s}}^2}.
\end{equation}
By a simple substitution $v=\cos\vartheta$ in \Cref{Eq1:H_n} one gets
\begin{align}\label{Eq1:coeffsG}
\begin{split}
	F_n^{(1)} &= P_{\mathrm{inc}}(\omega)\frac{2n+1}{2}r_{\mathrm{s}}\int_{-1}^1 \frac{\euler^{\imag k_1 q(v)}}{q(v)}\legendre_n(v)\idiff v\\
	F_n^{(2)} &= P_{\mathrm{inc}}(\omega)\frac{2n+1}{2}r_{\mathrm{s}}\int_{-1}^1 \left(R_{0,1}+r_{\mathrm{s}}v\right)\frac{\euler^{\imag k_1 q(v)}}{q^3(v)}\left[\imag k_1 q(v) - 1\right]\legendre_n(v)\idiff v
	\end{split}
\end{align}
where
\begin{equation*}
	q(v) = \sqrt{R_{0,1}^2+2r_{\mathrm{s}}R_{0,1}v+r_{\mathrm{s}}^2}.
\end{equation*}
One can obtain simple expressions for some of these coefficients, for example
\begin{equation*}
	F_0^{(1)} = P_{\mathrm{inc}}(\omega)\sinc\left(k_1R_{0,1}\right)\euler^{\imag k_1 r_{\mathrm{s}}}.
\end{equation*}
But in general one needs to use a numerical routine to evaluate the integrals in \Cref{Eq1:coeffsG}.

\section*{References}
\bibliographystyle{TK_CM} 

\bibliography{references}

\end{document}